\pdfoutput=1 
\documentclass[10pt,aps,nofootinbib,prd,preprintnumbers,twocolumn]{revtex4-1}
\usepackage{amsmath,amssymb}
\usepackage[english]{babel}
\usepackage{boldline}
\usepackage{braket}
\usepackage{float}
\usepackage{hyperref}
\hypersetup{colorlinks=true}
\usepackage{tikz}
\usetikzlibrary{arrows}
\usetikzlibrary{calc}
\usetikzlibrary{decorations.markings}
\usetikzlibrary{math}

\allowdisplaybreaks

\newcommand{\aglproj}{\hat{\Pi}_{\text{A}}}
\newcommand{\aL}{a(L)}
\newcommand{\aR}{a(R)}
\newcommand{\EL}{\hat{E}_L}
\newcommand{\ER}{\hat{E}_R}
\newcommand{\Nleft}{\hat{N}_L}
\newcommand{\Nright}{\hat{N}_R}
\newcommand{\UL}{\hat{U}_L}
\newcommand{\UR}{\hat{U}_R}
\newcommand{\Lpp}[1][]{\mathcal{L}^{++}_{ #1}}
\newcommand{\Lpm}[1][]{\mathcal{L}^{+-}_{ #1}}
\newcommand{\Lmp}[1][]{\mathcal{L}^{-+}_{ #1}}
\newcommand{\Lmm}[1][]{\mathcal{L}^{--}_{ #1}}
\newcommand{\Sin}{\mathcal{S}_{\text{in}}}
\newcommand{\Sout}{\mathcal{S}_{\text{out}}}
\newcommand{\Sinpp}{\Sin^{++} {}}
\newcommand{\Sinpm}{\Sin^{+-} {}}
\newcommand{\Sinmp}{\Sin^{-+} {}}
\newcommand{\Sinmm}{\Sin^{--} {}}
\newcommand{\Soutpp}{\Sout^{++} {}}
\newcommand{\Soutpm}{\Sout^{+-} {}}
\newcommand{\Soutmp}{\Sout^{-+} {}}
\newcommand{\Soutmm}{\Sout^{--} {}}
\newcommand{\Hpp}{\mathcal{H}^{++}}
\newcommand{\Hmm}{\mathcal{H}^{--}}
\newcommand{\N}[1][]{\mathcal{N}_{ #1}}
\newcommand{\Na}{\mathcal{N}_a}
\newcommand{\Nb}{\mathcal{N}_b}
\newcommand{\Ni}{\mathcal{N}_i}
\newcommand{\NL}{\mathcal{N}_L}
\newcommand{\Nl}{\mathcal{N}_l}
\newcommand{\No}{\mathcal{N}_o}
\newcommand{\NR}{\mathcal{N}_R}
\newcommand{\Nq}{\mathcal{N}_{\psi}}
\newcommand{\Nj}{\mathcal{N}_j}
\newcommand{\inverseRoot}[1][]{ \frac{1}{\sqrt{ #1}}}

\newcommand{\midarrow}{\tikz \draw[-triangle 45] (0,0) -- +(.1,0);}
\newcommand{\midreversearrow}{\tikz \draw[-triangle 45] +(.1,0) -- (0,0);}

\newcommand{\LoopDraw}[4]{
  \draw[#1] ({-0.4 * #3}, - #4) -- (0, - #4);
  \draw[#2] ({+0.4 * #3}, - #4) -- (0, - #4);
}

\newcommand{\loopsize}{1.25}

\newcommand{\LoopPicture}[2]{
  \begin{tikzpicture}
    \begin{scope}[thick]
      \LoopDraw{#1}{#2}{\loopsize}{0}
      \path ({-0.5 * \loopsize}, -0.15) -- ({+0.5 * \loopsize}, -0.15);
    \end{scope}
  \end{tikzpicture}
}

\newcommand{\LoopOpPicture}[3]{
  \begin{tikzpicture}[every node/.style={inner sep=0,outer sep=0}]
    \begin{scope}
      \draw[#1, very thick] ({-0.5 * \loopsize}, 0) -- (0, 0);
      \draw[#2, very thick] ({+0.5 * \loopsize}, 0) -- (0, 0);
      \draw (0,0) node[yshift= 0.05cm] {$\widehat{ } $};
      \draw (0,0) node[yshift=-0.25cm] {#3};
      \useasboundingbox (0,-0.4);
    \end{scope}
  \end{tikzpicture}
}

\newcommand{\SinDraw}[4]{
  \path ({-0.5 * #3}, -#4) -- ({+0.5 * #3}, -#4);
  \draw[#1, thick] ({-0.35 * #3}, -#4) -- ({-0.15 * #3}, -#4);
  \filldraw[#2, thick, fill=white, draw=black] ({-0.15 * #3}, -#4) circle [radius={0.125}];
}

\newcommand{\stringlinewidth}{1.25}
\newcommand{\quarkcircleradius}{0.06125}

\newcommand{\SinPicture}[2]{
  \begin{tikzpicture}
    \begin{scope}[thick]
      \SinDraw{#1}{#2}{\stringlinewidth}{0}
    \end{scope}
  \end{tikzpicture}
}

\newcommand{\SinOpPicture}[3]{
  \begin{tikzpicture}[every node/.style={inner sep=0,outer sep=0}]
    \begin{scope}[very thick]
      \draw [#1] ({-1 * \stringlinewidth}, 0) -- (0,0);
      \filldraw[#2, thick, fill=white, draw=black] (0, 0) circle [radius={\quarkcircleradius * \stringlinewidth}];
      \draw (0,0) node[yshift=-0.3cm] {#3};
      \draw (0,0) node[yshift= 0.15cm] {$\widehat{ } $};
    \end{scope}
  \end{tikzpicture}
}

\newcommand{\SinProjPicture}[2]{
  \begin{tikzpicture}
    \begin{scope}[thick]
      \SinDraw{#1}{#1}{\stringlinewidth}{-0.1}
      \SinDraw{#2}{#2}{\stringlinewidth}{+0.2};
  \end{scope}
\end{tikzpicture}
}

\newcommand{\SoutDraw}[4]{
  \path ({-0.5 * \stringlinewidth}, -#4) -- ({+0.5 * \stringlinewidth}, -#4);
  \draw[#2, thick] ({+0.35 * \stringlinewidth}, -#4) -- ({+0.15 * \stringlinewidth}, -#4);
  \filldraw[#1, thick, fill=white, draw=black] ({+0.15 * \stringlinewidth}, -#4) circle [radius={0.125}];
}

\newcommand{\SoutPicture}[2]{
  \begin{tikzpicture}
    \begin{scope}[thick]
      \SoutDraw{#1}{#2}{\stringlinewidth}{0}
    \end{scope}
  \end{tikzpicture}
}

\newcommand{\SoutOpPicture}[3]{
  \begin{tikzpicture}[every node/.style={inner sep=0,outer sep=0}]
    \begin{scope}[very thick]
      \draw [#2] ({1.0 * \stringlinewidth}, 0) -- (0,0);
      \filldraw[#1, thick, fill=white, draw=black] (0, 0) circle [radius={\quarkcircleradius * \stringlinewidth}];
      \draw (0,0) node[yshift=-0.3cm] {#3};
      \draw (0,0) node[yshift= 0.15cm] {$\widehat{ } $};
    \end{scope}
  \end{tikzpicture}
}

\newcommand{\SoutProjPicture}[2]{
  \begin{tikzpicture}
    \begin{scope}[thick]
      \SoutDraw{#1}{#1}{\stringarcradius}{-0.1}
      \SoutDraw{#2}{#2}{\stringarcradius}{+0.2};
  \end{scope}
\end{tikzpicture}
}

\newcommand{\hadronsize}{2}

\newcommand{\HadronDraw}[2]{
  \path ({-0.5 * \loopsize}, 0) -- ({+0.5 * \loopsize}, 0);
  \draw[#1] ({-0.15 * \loopsize}, -0.05) -- ({+0.15 * \loopsize}, -0.05);
  \draw[#1] ({-0.15 * \loopsize}, +0.05) -- ({+0.15 * \loopsize}, +0.05);
      \filldraw[#1, fill=white, draw=black] ({-0.15 * \loopsize}, -#2) circle [radius=0.0625 * \hadronsize];
      \filldraw[#1, fill=white, draw=black] ({+0.15 * \loopsize}, -#2) circle [radius=0.0625 * \hadronsize];
}

\newcommand{\HadronPicture}[1]{
\begin{tikzpicture}
  \begin{scope}[thick]
    \HadronDraw{#1}{0}
  \end{scope}
\end{tikzpicture}
}

\newcommand{\HadronOpPicture}[2]{
  \begin{tikzpicture}
    \begin{scope}[very thick]
      \HadronDraw{#1}{0};
      \draw (0,0) node[yshift=-0.3cm] {#2};
      \draw (0,0) node[yshift= 0.15cm] {$\widehat{ } $};
    \end{scope}
  \end{tikzpicture}
}

\begin{document}

\title{Loop, String, and Hadron Dynamics in SU(2) Hamiltonian Lattice Gauge Theories}

\author{Indrakshi Raychowdhury}
\email{iraychow@umd.edu}
\affiliation{Maryland Center for Fundamental Physics and Department of Physics,
University of Maryland, College Park, Maryland 20742, USA}

\author{Jesse R.~Stryker}
\email[]{stryker@uw.edu}
\affiliation{Institute for Nuclear Theory, University of Washington, Seattle, Washington 98195, USA}

\date{\today}

\preprint{INT-PUB-19-059}
\preprint{UMD-PP-019-08}

\begin{abstract}
  The question of how to efficiently formulate Hamiltonian gauge theories is experiencing renewed interest due to advances in building quantum simulation platforms.
  We introduce a reformulation of an SU(2) Hamiltonian lattice gauge theory---a loop-string-hadron (LSH) formulation---that describes dynamics directly in terms of its loop, string, and hadron degrees of freedom, while alleviating several disadvantages of quantumly simulating the Kogut-Susskind formulation.
  This LSH formulation transcends the local loop formulation of $d+1$-dimensional lattice gauge theories by incorporating staggered quarks, furnishing the algebra of gauge-singlet operators, and being used to reconstruct dynamics between states that have Gauss's law built in to them.
  LSH operators are then factored into products of ``normalized'' ladder operators and diagonal matrices, priming them for classical or quantum information processing.
  Self-contained expressions of the Hamiltonian are given up to $d=3$.
  The LSH formalism makes little use of structures specific to SU(2) and its conceptual clarity makes it an attractive approach to apply to other non-Abelian groups like SU(3).
\end{abstract}


\maketitle

\section{Introduction}

Non-Abelian gauge symmetries play a fundamental role in modeling the interactions observed in Nature.
The Standard Model is a relativistic quantum field theory with local $\text{U}(1) \times \text{SU}(2) \times \text{SU}(3)$ gauge symmetry.
The $\text{U}(1) \times \text{SU}(2)$ symmetry describes the electroweak sector, while the $\text{SU}(3)$ color symmetry describes QCD \cite{gross.wilczekUltravioletBehavior73,politzerReliablePerturbative73,gell-mannEightfoldWay61}.
Gauge symmetry also plays an important role in understanding the theory of high-temperature superconductors.
Nonrelativistic and dynamical SU(2) gauge fields emerge in the low-energy effective theory of doped and undoped Mott insulators in their spin-liquid phase, which models the physics of high-temperature superconductivity \cite{leursNonabelianGauge07,lee.nagaosa.eaDopingMott06}.

Non-Abelian gauge theories such as QCD are also known for admitting asymptotic freedom and becoming strongly coupled at low energies.
Predicting their low-energy properties requires nonperturbative calculations, and these are rarely tractable analytically.
Lattice quantum field theory, namely lattice QCD \cite{wilsonConfinementQuarks74}, has proven to be a successful nonperturbative approach to studying gauge theories numerically.
It has enabled novel \emph{ab initio} calculations in a variety of applications, such as
heavy quark physics \cite{bowler.kenway.eaHeavyBaryon96},
low-lying hadron masses \cite{pacs-cscollaboration.aoki.eaFlavorLattice09},
QCD thermodynamics \cite{bazavov.bhattacharya.eaEquationState09,aoki.borsanyi.eaQCDTransition09},
baryon number fluctuations \cite{borsanyi.fodor.eaFreezeOutParameters13},
and weak matrix elements \cite{nplqcdcollaboration.tiburzi.eaDoubleBeta17,nplqcdcollaboration.savage.eaProtonProtonFusion17}.

Lattice field theory calculations are usually performed by importance sampling the functional integral in Euclidean space (imaginary time).
These computational feats have predominantly characterized static or equilibrium properties at zero chemical potential \cite{kronfeldTwentyFirstCentury12}.
However, lattice QCD calculations with non-zero baryon chemical potential, with a topological $\theta$-term, or in real (Minkowskian) time are generally hampered by exponentially hard sign problems.
The fact that these scenarious can be so much harder is an artifact of the way they are formulated for simulation on classical machines, i.e., the path integral and the breakdown of Monte Carlo methods when applied to it.
A class of problems that especially stands to benefit from new methods is real-time dynamics.

The Hamiltonian formulation of gauge theories, which requires fixing a timelike direction, seems most natural for describing intrinsically real-time processes.
Hamiltonian lattice gauge theory \cite{kogut.susskindHamiltonianFormulation75} was formulated shortly following the action-based formulation.
Despite its more physical description of dynamics in terms of electric fluctuations, Hamiltonian lattice gauge theory has remained rather unexplored.
This is because the exponential growth of Hilbert space means the resources needed for simulation rapidly overwhelm any classical computing architecture.

In the 1980s, it was proposed that computers based on quantum mechanical degrees of freedom ought to be better suited for simulating quantum many-body systems \cite{feynmanSimulatingPhysics82}, such as a gauge theory.
The idea is that degrees of freedom of the system under study be mapped onto those of the quantum computer, and unitary operations are done on it to mimic time evolution.
In this scenario, it seems far more natural to express theories with Hamiltonians and Hilbert spaces rather than functional integrals and classical field configurations.

The arrival of functional quantum devices \cite{arute.arya.eaQuantumSupremacy19} thus creates an urgent need for a thorough grasp of Hamiltonian lattice gauge theory and how its structure can be related to that of quantum architectures.
Several proposals or steps toward proposals for quantum-simulating lattice gauge theories have been made in recent years
\cite{banerjee.dalmonte.eaAtomicQuantum12,banuls.cichyReviewNovel20,banuls.blatt.eaSimulatingLattice19}.
So far, most of them have been for simpler models like $\mathbb{Z}_2$ gauge theories \cite{zohar.farace.eaDigitalQuantum17,banuls.blatt.eaSimulatingLattice19,schweizer.grusdt.eaFloquetApproach19} or U(1) gauge theories in $1+1$ dimensions \cite{rico.pichler.eaTensorNetworks14,muschik.heyl.eaWilsonLattice17,klco.dumitrescu.eaQuantumclassicalComputation18,davoudi.hafezi.eaAnalogQuantum20}, including the first digital quantum simulation of the Schwinger model on a small lattice \cite{martinez.muschik.eaRealtimeDynamics16}.
Such simulations are instructive, but generalizing to non-Abelian gauge groups and multidimensional space is necessary to address the important problems where classical computers fall short.
Work on these generalizations is underway
\cite{%
  byrnes.yamamotoSimulatingLattice06,
  zohar.cirac.eaColdAtomQuantum13,
  tagliacozzo.celi.eaSimulationNonAbelian13,
  banerjee.bogli.eaAtomicQuantum13,
  wieseUltracoldQuantum13,
  wieseQuantumSimulating14,
  mezzacapo.rico.eaNonAbelianSU15,
  nuqscollaboration.lamm.eaGeneralMethods19,
  nuqscollaboration.alexandru.eaGluonField19,
  klco.savage.eaSUNonAbelian20%
}
(see also Refs. \cite{banuls.cichyReviewNovel20,banuls.blatt.eaSimulatingLattice19} and references therein), but the state of these studies is even less mature due to the significant practical complications involved with non-Abelian interactions.

In this paper we revisit and reformulate a non-Abelian lattice gauge theory---SU(2) gauge theory in $d+1$ dimensions with one flavor of staggered quarks---ultimately putting it into an explicit form to which (classical or) quantum algorithms can be readily applied.
The main theoretical contributions of this work include a thorough introduction to a loop-string-hadron (LSH) formulation of SU(2) lattice gauge theory, which uses local loops, strings, and hadrons as dynamical variables.
The derivations provide the detail needed to adapt the framework to similar SU(2) theories.
Furthermore, the exposition on the Hamiltonian contains the details required to develop comprehensive simulation algorithms.
In a companion work \cite{raychowdhury.strykerSolvingGauss20}, we provide a mapping of the LSH formalism to qubits along with quantum circuit solutions to all the constraints, representing a major advance toward implementing verifiably-gauge-invariant states and quantum error mitigation.
Generalizing the present LSH formalism for SU(3) would be a key step toward future quantum simulations of lattice QCD.

This LSH formulation is the result of working with strictly SU(2)-invariant operators and is an extension of the Schwinger boson (prepotential) formulation of lattice gauge theory
\cite{
  schwingerAngularMomentum52,
  mathurHarmonicOscillator05,
  mathurLoopStates06,
  mathurLoopApproach07,
  anishetty.mathur.eaPrepotentialFormulation10,
  mathur.raychowdhury.eaSUIrreducible10,
  raychowdhuryPrepotentialFormulation13,
  anishetty.raychowdhurySULattice14,
  anishetty.sreerajMassGap18%
}.
The non-Abelian Gauss law that usually appears as a constraint is made intrinsic, meaning the local excitations are physical and even intuitive.
The price paid is the introduction of an Abelian Gauss law that must be enforced instead, and the introduction of additional lattice links.
These are not fundamental hurdles because i) the Abelian constraints are simpler to work with and, ii) if the Abelian constraints are also solved, then the gauge-invariant Hilbert space is covered much more efficiently than it would be in a Kogut-Susskind formulation.
(Addressing the latter point is the subject of ongoing work.)
Importantly, by making the operator structure so explicit, algorithms can start being applied to simulating dynamics and compared against any other proposals made for non-Abelian simulations.

The organization of this paper is as follows:
Section \ref{sec:KSFormulation} reviews key points of the Kogut-Susskind Hamiltonian formulation.
In Sec.\,\ref{sec:SBFormulation}, we briefly review how that framework is expressed using Schwinger bosons.
In Sec.\,\ref{sec:LSHFormulation}, we describe the LSH formulation in one spatial dimension in detail.
This includes the LSH operators and their algebra, the Hamiltonian and Gauss's law, definition of an orthonormal LSH basis, and complete specification of LSH matrix elements in that basis.
In Secs.\,\ref{sec:LSHMultidim} and \ref{sec:LSH3dim}, we generalize to $2+1$ and $3+1$ dimensions respectively.
Finally, Sec.\,\ref{sec:LSH-KSComparison} compares the LSH formalism against a conventional framework.

\section{\label{sec:KSFormulation}Kogut-Susskind formulation}
In this section we review the basic setup of Hamiltonian lattice gauge theory.
Subsequently, we discuss aspects this framework in the context of digital quantum simulation.
Here and throughout we use the Einstein summation convention.

\subsection{Basic properties}

SU(2) lattice gauge theory is formulated in terms of matrix-valued link operators $\hat{U}$ and their conjugate electric fields $\hat{E}$.
In this work we consider $d$-dimensional Cartesian lattices (meaning square, cubic, or hypercubic), with sites $x$ and links $(x,j)$ that emanate from $x$ and terminate at $x+e_j$.
By the latter we are setting a convention for how a link operator $\hat{U}(x,j)$ associated with $(x,j)$ behaves under local gauge transformations $\Omega(x)$:
\begin{equation}
  \hat{U} (x,j) \rightarrow \Omega^{\dagger} (x) \hat{U} (x,j) \Omega (x+e_j) \ ,
\end{equation}
where $\hat{U}$ and $\Omega$ are taken to be in the fundamental representation.
Accordingly, we refer to the $x$ side of $(x,j)$ as the ``left'' end of the link, the $x+e_j$ side as the ``right'' end of the link, and the link's orientation as going ``from $x$ to $x+e_j$.''

On the link ends live left and right chromoelectric fields $\hat{E}^{\text{a}}_{L/R}$ ($\mathrm{a}=1,2,3$), which generate gauge transformations at their respective sides:
\begin{subequations}
\label{eq:CCR}
\begin{align}
  [ \EL^{\text a}(x,i) \ , \ \hat{U}_{\alpha \beta}(y,j) ]  &= - T^{\text a}_{\alpha \gamma} \hat{U}_{\gamma \beta} (y,j) \ \delta_{ij} \ \delta_{y,x} \ , \\
  [ \ER^{\text a}(x,i) \ , \ \hat{U}_{\alpha \beta}(y,j) ]  &= + \hat{U}_{\alpha \gamma} (y,j) T^{\text a}_{\gamma \beta} \ \delta_{ij} \ \delta_{y,x-e_i} ,
\end{align}
\end{subequations}
with $2T^{\text{a}}=\sigma^{\text a}$ being the Pauli matrices and greek indices running over $\{1,2\}$.
The canonical commutation relations (\ref{eq:CCR}) are more succinctly expressed by only displaying generator indices,
\begin{subequations}
\begin{align}
  [ \EL^{\text a} \ , \ \hat{U} ]  &=  - T^{\text a} \hat{U} \ , \\
  [ \ER^{\text a} \ , \ \hat{U} ]  &=  + \hat{U} T^{\text a} \ ,
\end{align}
\end{subequations}
and understanding that only same-link commutators may be nonvanishing.
The left and right electric fields each form SU(2) Lie algebras and they commute among each other:
\begin{align}
  [ \EL^{\text a} \ , \ \EL^{\text b} ] &= i \epsilon^{\text{abc}} \EL^{\text c} \ , \\*
  [ \ER^{\text a} \ , \ \ER^{\text b} ] &= i \epsilon^{\text{abc}} \ER^{\text c} \ , \\*
  \label{eq:LAndRCommute} [ \EL^{\text a} \ , \ \ER^{\text b} ] &= 0 \ .
\end{align}
While left and right gauge transformations commute, they are not strictly independent---being related by parallel transport \cite{sharatchandraLocalObservables82};
a consequence of this is that the gauge-invariant Casimirs on either side must be equal:
\begin{equation}
  \label{eq:KSCasimir} \hat{E}^2 \equiv \EL^{\text a}\EL^{\text a}= \ER^{\text a}\ER^{\text a} \ .
\end{equation}

By the same token, elements of the special unitary link operator matrices also commute but are not strictly independent of each other:
\begin{align}
  \label{eq:linkOpElementsCommute} [ \hat{U}_{\alpha\beta}, \hat{U}_{\gamma\delta}] = [ \hat{U}_{\alpha\beta}, (\hat{U}_{\gamma\delta})^{\dagger}] &= 0 \ , \\*
  \label{eq:linkOpsAreSU} \hat{U}_{22} = \hat{U}_{11}^{\dagger} \ , \qquad \hat{U}_{21} &= -\hat{U}_{12}^{\dagger} \ .
\end{align}
Any pair of $\hat{U}$ or $\hat{E}$ operators from different links, however, is truly independent.

In addition to the $\hat{U}$ and $\hat{E}$ gauge variables living on links, a fundamental matter field $\hat{\psi}^{\alpha}$ ($\alpha=1,2$) will live on lattice sites.
The matter field commutes with the $\hat{U}$ and $\hat{E}$ fields, and its behavior under local gauge transformations is
\begin{equation}
  \hat{\psi}(x) \rightarrow \Omega^\dagger (x) \hat{\psi}(x) .
\end{equation}
This transformation is generated by the charge density,
\begin{align}
  \hat{\rho}^{\text a}(x) &= \hat{\psi}^\dagger(x) T^{\text a} \hat{\psi}(x) \ , \\
  [ \hat{\rho}^{\text a} , \hat{\psi}^{\alpha} ] &= - T^{\text a}_{\alpha \beta} \hat{\psi}^{\beta} \ .
\end{align}
These charge densities again form an SU(2) Lie algebra.

Full generators of local gauge transformations (applicable to any of the fields above) are given by the Gauss law operators $\hat{G}^{\text a}$:
\begin{equation}
  \label{eq:gaussOps} \hat{G}^{\mathrm a}(x)=\sum_{i=1}^{d} (\hat{E}^{\mathrm a}_{L,i}(x)+\hat{E}^{\mathrm a}_{R,i}(x)) + \hat{\psi}^{\dagger}_{\alpha}(x)T^{\mathrm a}_{\alpha\beta}   \hat{\psi}_\beta(x) \ .
\end{equation}
Gauge invariance of the theory means all the $\hat{G}^{\text a}(x)$ commute with the Hamiltonian.

A lattice Hamiltonian for SU(2) gauge bosons coupled to one flavor of staggered fermionic matter, in units of the lattice spacing, may now be formulated as \cite{kogut.susskindHamiltonianFormulation75}
\begin{equation}
  \hat{H} = \hat{H}_E + \hat{H}_B + \hat{H}_M + \hat{H}_I \ ,\\
\end{equation}
with
\begin{subequations}
  \label{eq:KSHam}
  \begin{align}
    \hat{H}_E &= \frac{g_0^2}{2}  \sum_{(x,i)}  \hat{E}^{2}(x,i) \ , \\
    \label{eq:KSHamB} \hat{H}_B &= \frac{1}{g_0^2} \sum_{x} \sum_{i<j} \text{tr} \left[ 2- \hat{U}_{\Box}^{(ij)} (x) - \hat{U}_{\Box}^{(ij)}(x)^\dagger  \right] \ , \\
    \hat{H}_M &= m_0 \sum_{x} (-)^x \hat{\psi}^{\dagger}(x) \hat{\psi}(x) \ , \\
    \hat{H}_I &= \sum_{(x,i)}\hat{\psi}^{\dagger}(x)\hat{U}(x,i)\hat{\psi}(x+e_i) + \text{H.c.}
  \end{align}
\end{subequations}
Above, $g_0$ and $m_0$ are the bare coupling and bare mass;
the magnetic energy $\hat{H}_B$ is formed from gauge-invariant traces of plaquette operators,
\begin{equation}
  \hat{U}_{\square}^{(ij)}(x) \equiv \hat{U}(x,i) \hat{U}(x+e_i,j)  \hat{U}^{\dagger}(x+e_j,i)  \hat{U}^{\dagger} (x,j) \ ,
  \label{eq:plaquetteDef}
\end{equation}
with $\text{tr}[\hat{U}_{\square}^\dagger ] = \text{tr} [ \hat{U}_{\square} ]$ for SU(2); 
and the alternating sign $(-)^{x}\equiv(-1)^{\sum_i x_i}$ in the mass energy reflects the staggered fermion prescription.

\subsection{Hilbert space of states}

The Hamiltonian generates dynamics among states that satisfy Gauss's law, i.e.,
\begin{equation}
  \hat{G}^{\text a}(x)\ket{\text{phys}} = 0 \quad \text{for all $x$, a.}
\end{equation}
Dynamics of this sort is usually described using redundant but local degrees of freedom.
Alternatively, these states can be described in terms of arbitrary gauge-invariant Wilson loop and Wilson line operators acting on the staggered strong-coupling vacuum ($g_0,m_0 \rightarrow \infty$).
Characterizing states this way, however, leads to a highly overcomplete basis;
mutually dependent loops satisfy Mandelstam constraints
\cite{
  mandelstamChargemonopoleDuality79,
  lollYangMillsTheory93,
  watsonSolutionSU94%
}, which are nonlocal and notoriously hard to solve.
A central aim of the remainder of this work is to encode dynamics using local degrees of freedom that both satisfy Gauss's law and are orthogonal, starting from doublets of harmonic oscillators \cite{mathurHarmonicOscillator05}.

Since the $\hat{G}^{\text a}(x)$ operators are expressed in terms of electric fields, this makes it natural to diagonalize electric operators.
A typical complete set of commuting observables (CSCO) on a link is given by
\begin{equation}
  \label{eq:KS-CSCO} \left\{ \EL^{\text a} \EL^{\text a}, \ \EL^{3} \ , \ER^{\text a} \ER^{\text a} \ , \ER^{3} \right\} \ .
\end{equation}
This leads to an irreducible representation or ``irrep'' basis.
One could truncate the gauge field Hilbert space (while preserving gauge invariance) by keeping all states up to some cutoff on the Casimirs ($\hat{E}_{L/R}^{\text a} \hat{E}_{L/R}^{\text a}$), e.g., $j_{\text{max}} (j_{\text{max}} + 1)$ with $j=0,1/2,...,j_{\text{max}}$ labeling the ``angular momentum'' or irrep of SU(2).

The fermionic Fock spaces are naturally finite dimensional and require no truncation.

\subsection{Practical considerations}
Compared to compact Abelian gauge groups, there are several aspects of the Kogut-Susskind formulation that make SU(N) groups especially formidable for simulation.
\begin{itemize}
  \item \emph{Noncommuting constraints}.---%
    In U(1) lattice gauge theories, the constraints are simultaneously diagonalizable.
    This means it is possible to choose a basis where each basis ket is definitely in the allowed subspace or definitely in the disallowed subspace.
    But in SU(N) lattice gauge theory the Gauss law constraints form an algebra, $[\hat{G}^{\text a}(x) , \hat{G}^{\text b}(x)]=i f^{\text{abc}} \hat{G}^{\text c}(x)$, so simultaneously diagonalizing all constraints is impossible.
    Then the basis kets that would be represented by and measured on a quantum device would not be meaningful by themselves.
  \item \emph{Asymmetric quantum numbers}.---%
    For compact U(1) theories, the eigenstates of electric fields are characterized by a single integer quantum number.
    In that eigenbasis, link operators simply raise or lower the quantum number by one unit.
    In SU(2) theories one typically diagonalizes $\hat{E}^{\text a} \hat{E}^{\text a}$ and $\hat{E}^{\text 3}$ at every side of every link, yielding local $\ket{j,m}$ structures on all link ends.
    Every irrep $j$ has a different dimensionality, and these irreps are mixed by the action of link operators according to
\begin{align}
  \qquad \hat{U}^{ }_{M N} & \ket{j,m}_L \otimes \ket{j,n}_R = \nonumber \\
  & \sum_{j', m',n'} \braket{\tfrac{1}{2}, M; j, m| j', m'} \braket{j', n'| \tfrac{1}{2}, N; j, n} \times \nonumber  \\
  & \qquad \quad \times \sqrt{\frac{\dim\, j }{\dim \, j'}} \ket{j', m'}_L \otimes \ket{j', n'}_R \label{eq:linkOpAction}
\end{align}
\cite{zohar.burrelloFormulationLattice15}.
    [For SU(2), nonvanishing contributions on the right-hand side come from $j'=j\pm 1/2$.]
    Representing these mixings using qubit (or qudit) registers seems awfully forced and unnatural.
  \item \emph{Group-specific coefficients}.---%
    The action of link operators (\ref{eq:linkOpAction}) more generally involves group-dependent Clebsch-Gordon coefficients.
    In principle, designing simulation protocols specific to SU(2) and to SU(3) is not unreasonable and should even be expected.
    But if SU(2) and SU(3) were first expressed in a common framework then one could expect optimizations found for the former to better translate to the latter.
  \item \emph{Gauge redundancy in noisy simulations}.---%
    Local gauge constraints mean basis states are largely wasted representing disallowed states.
    State vectors in non-error-corrected simulations will wander away from the exponentially small space of allowed states.
    Moreover, nontrivial gauge-invariant states are very specific linear combinations of conventional irrep basis states;
    if the computational basis represents the irrep basis, then any single-qubit error could potentially spoil gauge invariance.
\end{itemize}
All of these disadvantages provide the impetus for exploring alternative frameworks.

\section{\label{sec:SBFormulation}Schwinger boson formulation}

The Schwinger boson or ``prepotential'' formulation of lattice gauge theory
\cite{
  mathurHarmonicOscillator05,
  mathurLoopStates06,
  mathurLoopApproach07,
  anishetty.mathur.eaPrepotentialFormulation10,
  raychowdhuryPrepotentialFormulation13%
}
provides an alternative, equivalent framework to that of Kogut and Susskind.
Below we review key points of the Schwinger boson formulation and describe how it alleviates some but not all the obstacles identified in the previous section.

\subsection{Basic properties}

\begin{figure}
  \begin{center}
  \begin{tikzpicture}[scale=2.0, thick]
    \begin{scope}[every node/.style={sloped,allow upside down}]
      \filldraw [fill=black, draw=black] (-1,0) circle [radius=0.02]
      node[left] {$x$}
      node[below right] {$L$}
      node[above right] {$\left[ \begin{array}{c} a_1(L) \\ a_2(L) \end{array} \right]$};
      \draw (-1,0) -- node {\midarrow} (1,0);
      \filldraw [fill=black, draw=black] (1,0) circle [radius=0.02]
      node[right] {$x+e_i$}
      node[below left] {$R$}
      node[above left] {$\left[ \begin{array}{c} a_1(R) \\ a_2(R) \end{array} \right]$};
    \end{scope}
  \end{tikzpicture}
  \end{center}
  \caption{\label{fig:genericLink}
    A link $(x,i)$ from a Cartesian lattice starting at $x$ and terminating at $x+e_i$.
    We denote the starting side ``left'' ($L$) and the terminal side ``right'' ($R$).
    Link ends each host a bosonic oscillator doublet, here indicated by $a(L)=(a_1(L),a_2(L))$ at $L$ and $a(R)=(a_1(R),a_2(R))$ at $R$.
    The arrow indicates the link orientation, pointing from $L$ to $R$.
  }
\end{figure}
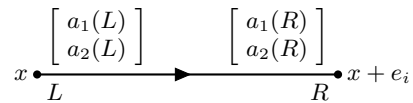

This formulation replaces $\hat{E}$ and $\hat{U}$ with bilinears of harmonic oscillator doublets $\hat{a}_\alpha(L/R)$ (the Schwinger bosons or prepotentials) at the left ($L$) and right ($R$) ends of each link $(x,i)$, as shown in Fig. \ref{fig:genericLink}.

The SU(2) electric fields on a link are realized by
\begin{align}
\label{eq:sbElectric}
  \hat{E}_{L/R}^{\mathrm a} &\equiv \hat{a}^{\dagger}(L/R) T^{\mathrm a} \hat{a}(L/R) \ .
\end{align}
Beyond the SU(2)$_{L/R}$ electric fields, there also exist U(1) generators:
\begin{align}
  \hat{N}_{L/R} &= \hat{a}^{\dagger}(L/R) \cdot \hat{a}(L/R) \ .
\end{align}
The requirement that left and right Casimirs be equal leads to an Abelian Gauss law (AGL) relating these U(1) generators along each link:
\begin{equation}
  \Nleft(x,i) \ket{\text{phys}} = \Nright(x+e_i,i) \ket{\text{phys}} \ .
  \label{eq:AGL}
\end{equation}
This constraint neutralizes undesired U(1) transformations that are a symmetry of the definitions (\ref{eq:sbElectric}), but are not a part of the target theory.
It is useful to define Hermitian projectors for the AGL-satisfying subspace within the tensor product space of all Schwinger boson modes:
\begin{equation}
  \aglproj(x,i) = \int_{-\pi}^{\pi} \frac{d\phi}{2\pi} e^{i\phi\left(\Nright (x+e_i,i)-\Nleft (x,i) \right)} \ .
\end{equation}

Turning to the SU(2) link operators, these are given by
\begin{align}
  \hat{U} &\equiv \UL \UR \ , \label{eq:KSLinkOp}
\end{align}
\vspace{-\baselineskip}
\begin{subequations}
  \label{eq:sbLinks}
\begin{align}
  \UL &\equiv \frac{1}{\sqrt{\hat{N}_L+1}} \left( \begin{array}{cc} \hat{a}^\dagger_2({L}) & \hat{a}_1(L) \\ -\hat{a}^\dagger_1(L) & \hat{a}_2(L) \end{array} \right) \ , \label{eq:sbUL} \\
  \UR &\equiv   \left( \begin{array}{cc} \hat{a}^\dagger_1(R) & \hat{a}^\dagger_2(R) \\ -\hat{a}_2(R) & \hat{a}_1(R)\end{array} \right) \frac{1}{\sqrt{\hat{N}_R+1}} \ . \label{eq:sbUR}
\end{align}
\end{subequations}
With the above definitions, the relations \eqref{eq:CCR}--\eqref{eq:LAndRCommute} follow.
One can also show that \eqref{eq:linkOpElementsCommute} and \eqref{eq:linkOpsAreSU} are satisfied on the AGL Hilbert space,
\begin{alignat}{4}
  \aglproj [ \hat{U}_{\alpha\beta}, \hat{U}_{\gamma\delta}] \aglproj &= \aglproj [ \hat{U}_{\alpha\beta}, \hat{U}^{\dagger}_{\gamma\delta}] \aglproj & &= 0 \ , \\
  \aglproj ( \hat{U}_{22} - \hat{U}_{11}^{\dagger} ) \aglproj &= \aglproj ( \hat{U}_{21} + \hat{U}_{12}^{\dagger} ) \aglproj & &= 0 \ .
\end{alignat}
And finally, the Schwinger boson construction of $\hat{U}$ is indeed a special unitary matrix on the AGL Hilbert space:
\begin{align}
  \aglproj \hat{U}^\dagger \hat{U} \aglproj  &= \aglproj  \hat{U}\hat{U}^\dagger \aglproj \nonumber \\*
  &= \aglproj \left( \begin{array}{cc} 1  & 0 \\ 0 & 1  \end{array} \right) \aglproj \ ,\\
  \aglproj \det(\hat{U}) \aglproj &= \aglproj \left( \hat{U}_{11} \hat{U}_{22} - \hat{U}_{12} \hat{U}_{21} \right) \aglproj \nonumber \\
  &= \aglproj (1) \aglproj = \aglproj \ .
\end{align}

\subsection{Hilbert space of states}
Allowed states in the Schwinger boson framework are characterized similarly to the Kogut-Susskind formulation.
Physically permissible wave functions must be annihilated by the Schwinger boson implementations of $\hat{G}^{\text{a}}(x)$, and the same reasons to diagonalize electric operators continue to apply.

Where the two diverge is in the local Hilbert space structure and choice of a complete set of commuting observables.
Instead of having aggregate link Hilbert spaces, the gauge field Hilbert space is built from local harmonic oscillators: two modes at the left and right ends of every link.
The natural choice of a CSCO for such Hilbert spaces is
\begin{equation} 
  \label{eq:SB-CSCO} \left\{ \hat{N}_1(L), \ \hat{N}_2(L), \ \hat{N}_1(R), \ \hat{N}_2(R) \right\}
\end{equation}
for each link.
This choice is equivalent to (\ref{eq:KS-CSCO}), but the spectrum of quantum numbers is different.

Truncating the Kogut-Susskind theory at some representation $j_{\text{max}}$ is equivalent to truncating all Schwinger boson occupation numbers to $j_{\text{max}}$.

\subsection{Practical considerations}

The Schwinger boson formulation offers the following advantages:
\begin{itemize}
  \item \emph{Symmetric quantum numbers}.---%
    All quantum numbers are on the same footing, being integer bosonic occupation numbers.
    Now it is obvious how one could represent these quantum numbers with binary registers.
    It is also obvious how to truncate the electric field [a uniform cutoff on all the occupation numbers in (\ref{eq:SB-CSCO})].
  \item \emph{Non-group-specific matrix elements}.---%
    The link operator is expressed in terms of simple harmonic oscillator ladder operators, and Clebsch-Gordon coefficients are implicit in the various rescaling factors carried along by them.
    In this sense, the elementary degrees of freedom are group agnostic.
    [Of course, in going from SU(2) to SU(3), one needs SU(3) irreducible Schwinger bosons as described in \cite{anishetty.mathur.eaIrreducibleSU09}.]
\end{itemize}
These features ought to be favorable for developing algorithms in this framework.

What remains to be addressed is the non-Abelian constraints, and redundancy of states.
The former is addressed starting with the following observation:
Local gauge transformations act site locally, with Schwinger bosons and matter all transforming identically.
This enables one to construct site-local intertwining operators automatically invariant under the action of the local generators;
these can be identified as segments of all possible SU(2)-invariant excitations hosted by a site (like a section of a Wilson loop).
Using these, one can construct an SU(2)-invariant Hilbert space locally at each site.
For pure gauge theory, the resulting local ``loop states'' \cite{mathurLoopStates06,mathurLoopApproach07,anishetty.raychowdhurySULattice14,raychowdhuryLowEnergy19} are characterized by integer-valued loop quantum numbers directly related to the angular momentum flux $j$.
Truncating the Kogut-Susskind theory at some representation $j_{\text{max}}$ is equivalent to truncating all Schwinger boson occupation numbers to $j_{\text{max}}$, and that is equivalent to truncating local loop numbers at $2j_{\text{max}}/(2d-1)$.

A drawback of the loop basis is that it is overcomplete.
Finding the complete and orthogonal gauge-invariant Hilbert space requires solving the Mandelstam constraints, which becomes increasingly complicated in higher dimensions and with higher cutoff.
These issues have been discussed in great detail in earlier works on the prepotential formulation of pure gauge theory.
A central objective of the loop-string-hadron framework below will be to give a complete and local description of gauge-invariant dynamics with minimal redundancy, equipped with fundamental matter, and adaptable to any number of spatial dimensions.

\section{\label{sec:LSHFormulation}Loop-string-hadron formulation: One dimension}
The SU(2)-invariant excitations at a site are parts of flux loops, parts of meson strings, or hadrons.
We now derive a loop-string-hadron formulation starting from prepotentials that has non-Abelian gauge invariance built into it.
We start in $1+1$ dimensions, where the essential features of coupling to matter---which was not previously a part of the prepotential framework---already appear.

\begin{figure}[!t]
  \centering
  \begin{tikzpicture}[scale=2.0, thick]
    \begin{scope}
      \draw (-1.5,0) node[left] {$\cdots$} -- (1.5,0) node[right] {$\cdots$};
      \draw (-1.1,0) -- node {\midarrow} (0,0);
      \draw (0.25,0) -- node {\midarrow} (1,0);
      \foreach \x in {-1,0,1}{
        \filldraw [fill=white, draw=black] (\x,0) circle [radius=0.125]
        node {$\psi$}
        node[xshift=-11, yshift=-6] {$i$}
        node[xshift=11, yshift=-6] {$o$};
      };
      \draw [densely dashed, draw=black!25] (-0.5, -0.25) rectangle (0.5, 0.25);
      \draw[densely dotted, draw=black!25] (-0.5,-0.25) -- (-1,-1.25);
      \draw[densely dotted, draw=black!25] ( 0.5,-0.25) -- ( 1,-1.25);
      \draw[densely dotted, draw=black!25] ( 0.5, 0.25) -- ( 1,-0.5);
      \draw[densely dotted, draw=black!25] (-0.5, 0.25) -- (-1,-0.5);
      \draw [densely dashed,fill=white, draw=black!25] (-1,-1.25) rectangle (1, -0.5);
      \draw (-1,-1)
      -- node[above] {$\left[ \begin{array}{c} a_1(R) \\ a_2(R) \end{array} \right]$\ \ \ \ }
      (0,-1)
      node[yshift=-5, xshift=-20] {$i$}
      node[yshift=-5, xshift=20] {$o$}
      node[above] {$\left[ \begin{array}{c} \psi_1 \\ \psi_2 \end{array} \right]$}
      -- node[above] {\ \ \ \ $\left[ \begin{array}{c} a_1(L) \\ a_2(L) \end{array} \right]$}
      (1,-1);
      \fill (0,-1) circle[fill=black, radius=.03];
    \end{scope}
  \end{tikzpicture}
  \caption{\label{fig:1dLattice}
    Pictorial representation of a 1D lattice with matter.
    Every site on the 1D lattice is associated with a fermionic doublet $\psi=(\psi_1, \psi_2)$.
    Bosonic doublets $a(L)=(a_1(L),a_2(L))$ and $a(R)=(a_1(R),a_2(R))$ are associated with link ends attached to any site along directions $o$ and $i$, respectively.
  }
\end{figure}
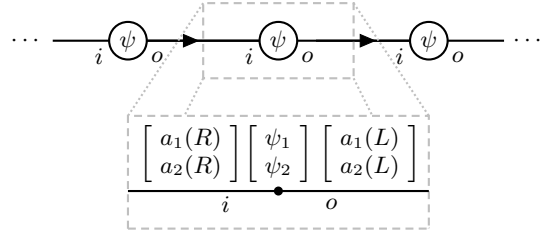

In $1+1$ dimensions, the Kogut-Susskind Hamiltonian (\ref{eq:KSHam}) reduces to
\begin{equation}
  \hat{H} = \hat{H}_E + \hat{H}_I + \hat{H}_M \ .
\end{equation}
Each site $x$ of this lattice is connected to  one incoming link along direction $i$ and one outgoing link along direction $o$, as in Fig. \ref{fig:1dLattice}.
Within the prepotential framework, Schwinger bosons $\hat{a}_\alpha(L)$ ($\hat{a}_\alpha(R)$) are attached to the link along the direction $o$ ($i$).
A staggered fermion field $\hat{\psi} = (\hat{\psi}_1 ,\hat{\psi}_2)$ lives on the sites themselves.

The site-local doublets shown in the box in Fig. \ref{fig:1dLattice} can contract in many possible ways to form SU(2) singlets.
It follows that SU(2) invariance can be made manifest by passing from Schwinger boson and quark operators to using only their SU(2)-invariant combinations.
The gauge theory will be expressed entirely in terms of the dynamics generated by all such operators.

\subsection{\label{subsec:SU2Invariants}SU(2) singlets: Loop, string, and hadron operators}

The complete set of SU(2) invariants at a one-dimensional (1D) site of a spatial lattice is obtained by constructing all possible singlet tensors out the available doublets and their conjugates.
It is a special feature of SU(2) that fundamental doublets are unitarily equivalent to antifundamentals:
if $f$ transforms like a fundamental, then $\tilde{f}$ given by
\begin{align}
  \epsilon &\equiv i \sigma_{y} = \left( \begin{tabular}{rr} 0 & 1 \\ $-1$ & 0 \end{tabular} \right) \ ,\\
  \tilde{f}_{\alpha} & \equiv \epsilon_{\alpha \beta} f_{\beta} \ ,
\end{align}
transforms in the conjugate representation.
This equivalence implies $\tilde{a}^{\dagger}_{\alpha}(L/R) \equiv \epsilon_{\alpha \beta} a^{\dagger}_{\beta}(L/R)$ gives another set of doublets to work with.

Using the available tensors, the complete set of nonvanishing singlets is listed below in \eqref{eq:1dLoopOps}--\eqref{eq:1dNumOps}:
\begin{itemize}
  \item \emph{Pure gauge loop operators}.---$\mathcal{L}^{\sigma, \sigma'}$:
    \begin{subequations}
      \label{eq:1dLoopOps}
      \begin{align}
        \label{eq:L++} \Lpp &= \aR^\dagger_{\alpha} \aL^\dagger_\beta \epsilon_{\alpha\beta}\\
        \label{eq:L--} \Lmm &= \aR_{\alpha} \aL_\beta \epsilon_{\alpha\beta} = (\Lpp)^\dagger \\
        \label{eq:L+-} \Lpm &= \aR^\dagger_{\alpha}\aL_\beta \delta_{\alpha\beta} \\
        \label{eq:L-+} \Lmp &= \aR_{\alpha}\aL^\dagger_\beta \delta_{\alpha\beta} = (\Lpm)^\dagger
      \end{align}
    \end{subequations}
  \item \emph{Incoming string operators}.---$\Sin^{\sigma, \sigma'}$:
    \begin{subequations}
      \label{eq:1dSinOps}
      \begin{align}
        \label{eq:Sin++} \Sinpp &= \aR^\dagger_{\alpha} \psi^\dagger_\beta \epsilon_{\alpha\beta}\\
        \label{eq:Sinmm} \Sinmm &= \aR_{\alpha}\psi_\beta \epsilon_{\alpha\beta} = (\Sinpp)^\dagger \\
        \label{eq:Sin+-} \Sinpm &= \aR^\dagger_{\alpha}\psi_\beta \delta_{\alpha\beta} \\
        \label{eq:Sin-+} \Sinmp &= \aR_{\alpha}\psi^\dagger_\beta \delta_{\alpha\beta} = (\Sinpm)^\dagger
      \end{align}
    \end{subequations}
  \item \emph{Outgoing string operators}.---$\Sout^{\sigma, \sigma'}$:
    \begin{subequations}
      \label{eq:1dSoutOps}
      \begin{align}
        \label{eq:Sout++} \Soutpp &= \psi^\dagger_\alpha \aL^\dagger_\beta \epsilon_{\alpha\beta} \\*
        \label{eq:Sout--} \Soutmm &= \psi_{\alpha} \aL_\beta \epsilon_{\alpha\beta}= (\Soutpp)^\dagger \\*
        \label{eq:Sout+-} \Soutpm &= \psi^\dagger_\alpha \aL_\beta \delta_{\alpha\beta}  \\*
        \label{eq:Sout-+} \Soutmp &= \psi_\alpha \aL^\dagger_\beta \delta_{\alpha\beta} = (\Soutpm)^\dagger
      \end{align}
    \end{subequations}
  \item \emph{Hadron operators}.---$\mathcal{H}^{\sigma, \sigma}$:
   \begin{subequations}
     \label{eq:1dHadronOps}
     \begin{align}
       \label{eq:H++} \Hpp &= -\frac{1}{2!}\psi^\dagger_{\alpha}\psi^\dagger_\beta \epsilon_{\alpha\beta} \\*
       \label{eq:H--} \Hmm &= \,\,\,\,\,\frac{1}{2!}\psi_{\alpha}\psi_{\beta} \epsilon_{\alpha\beta} = (\Hpp)^\dagger
     \end{align}
   \end{subequations}
    [Baryons and mesons are the same for SU(2).]
  \item \emph{Gauge flux, quark number operators}.---$\mathcal{N}_{L/R}$, $\Nq$:
    \begin{subequations}
      \label{eq:1dNumOps}
      \begin{align}
        \NL &= a(L)^\dagger_{\alpha}a(L)_\alpha \\
        \NR &= a(R)^\dagger_{\alpha}a(R)_\alpha \\
        \Nq &= \psi^\dagger_{\alpha}\psi_\alpha
      \end{align}
    \end{subequations}
\end{itemize}
These invariants exhaust all possible singlet bilinears and they are referred to as LSH operators.
They obey a closed operator algebra, which will be necessary to establish since the original $E$, $U$, and $\psi$ variables have been replaced.

Before giving the complete algebra, it is helpful to first build some intuition for these operators.
One can visualize LSH operators in terms of creation and annihilation of the gauge and matter degrees of freedom appearing in their definitions.
Below, this is done using line segments for gauge flux, circles for quarks, and solid (dotted) lines for creation (annihilation) actions:
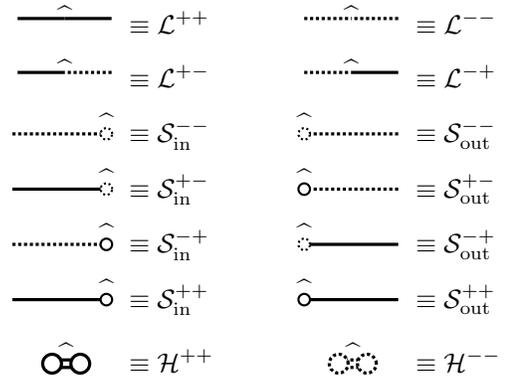
\begin{alignat*}{4}
  \vcenter{\hbox{\LoopOpPicture{solid}{solid}{$ $}}} &\equiv \Lpp\qquad\quad  & \vcenter{\hbox{\LoopOpPicture{densely dotted}{densely dotted}{$ $}}} &\equiv \Lmm \\
  \vcenter{\hbox{\LoopOpPicture{solid}{densely dotted}{$ $}}} &\equiv \Lpm & \vcenter{\hbox{\LoopOpPicture{densely dotted}{solid}{$ $}}} &\equiv \Lmp \\
  \vcenter{\hbox{\SinOpPicture{densely dotted}{densely dotted}{$ $}}} &\equiv \Sinmm & \vcenter{\hbox{\SoutOpPicture{densely dotted}{densely dotted}{$ $}}} &\equiv \Soutmm \\
  \vcenter{\hbox{\SinOpPicture{solid}{densely dotted}{$ $}}} &\equiv \Sinpm & \vcenter{\hbox{\SoutOpPicture{solid}{densely dotted}{$ $}}} &\equiv \Soutpm \\
  \vcenter{\hbox{\SinOpPicture{densely dotted}{solid}{$ $}}} &\equiv \Sinmp & \vcenter{\hbox{\SoutOpPicture{densely dotted}{solid}{$ $}}} &\equiv \Soutmp \\
  \vcenter{\hbox{\SinOpPicture{solid}{solid}{$ $}}} &\equiv \Sinpp & \vcenter{\hbox{\SoutOpPicture{solid}{solid}{$ $}}} &\equiv \Soutpp \\
  \vcenter{\hbox{\HadronOpPicture{solid}{$ $}}} &\equiv \Hpp & \vcenter{\hbox{\HadronOpPicture{densely dotted}{$ $}}} &\equiv \Hmm
\end{alignat*}
The placement of solid and dotted lines is in direct correspondence with the superscripts on the LSH operators.

The simplest examples from these are $\Lpp$ and $\Lmm$, which create or destroy an SU(2)-invariant flux line passing through the site.
By contrast, the mixed-type operators $\mathcal{L}^{\pm \mp}$ deform a flux line flowing out one side to instead flow out the other;
physically, this corresponds to changing the direction flux emanates from a single quark.
The hadron operators $\Hpp$ and $\Hmm$ create or annihilate a hadron, consistent with the Pauli principle (at most two quarks present). 

The actions of string operators are more subtle.
For example, $\Sinpp$ will create the ``right'' end of a meson string, provided no quark is initially present.
Alternatively, if a quark is already present in the form of an out string, the strings ends actually join and leave behind independent hadron and loop flux excitations.
The variety of actions $\Sinpp$ and other string operators can have will be summarized later on.

Further intuition for how LSH operators interact with each other is also gained from and made mathematically precise by now looking at their algebra.
The algebra of LSH operators at any 1D lattice site \eqref{eq:1dLoopOps}--\eqref{eq:1dNumOps} is tabulated in two parts.
Table \ref{tab:bigCommTable} lists commutators of operators where at most one operator from the pair has fermionic statistics.
Table \ref{tab:bigAnticommTable} lists anticommutators of operators that both have fermionic statistics.
In addition to these, Table \ref{tab:bigLoopCommTable} contains the operators and algebra needed for $d>1$ that will be discussed in Sec.\,\,\ref{sec:gluonicLoopOps};
they are displayed now so they can be referred to alongside Tables \ref{tab:bigCommTable} and \ref{tab:bigAnticommTable}.
The 1D LSH algebra is sufficient to completely describe the dynamics of the theory, to be discussed below.

\begin{table*}[!tbhp]
  \renewcommand{\arraystretch}{1.2}
  \centering
  \begin{tabular}{c V{3} ccc V{3} cccc V{3} cc}
&$[\cdot, \NR ]$&$[\cdot, \NL ]$&$[\cdot, \Nq ]$&$[\cdot, \Lmm ]$&$[\cdot, \Lmp ]$&$[\cdot, \Lpm ]$&$[\cdot, \Lpp ]$&$[\cdot , \Hpp]$&$[\cdot , \Hmm]$ \\
\hlineB{3}
$[\NR , \cdot ]$&$ 0 $&$ 0 $&$ 0 $&$ -\Lmm $&$ -\Lmp $&$+\Lpm $&$+\Lpp $&$ 0 $&$ 0 $ \\  
$[\NL , \cdot ]$&$ 0 $&$ 0 $&$ 0 $&$ -\Lmm $&$+\Lmp $&$ -\Lpm $&$+\Lpp $&$ 0 $&$ 0 $ \\  
$[\Nq , \cdot ]$&$ 0 $&$ 0 $&$ 0 $&$ 0 $&$ 0 $&$ 0 $&$ 0 $&$ 2\Hpp $&$ -2 \Hmm $ \\
\hlineB{3}
$[\Lpp , \cdot ]$&$ -\Lpp $&$ -\Lpp $&$ 0 $&$-\NL - \NR -2$&$ 0 $&$ 0 $&$ 0 $&$ 0 $&$ 0 $ \\  
$[\Lpm , \cdot ]$&$ -\Lpm $&$+\Lpm $&$ 0 $&$ 0 $&$ \NR - \NL $&$ 0 $&$ 0 $&$ 0 $&$ 0 $ \\  
$[\Lmp , \cdot ]$&$+\Lmp $&$ -\Lmp $&$ 0 $&$ 0 $&$ 0 $&$\NL - \NR $&$ 0 $&$ 0 $&$ 0 $ \\  
$[\Lmm , \cdot ]$&$+\Lmm $&$+\Lmm $&$ 0 $&$ 0 $&$ 0 $&$ 0 $&$\NL + \NR +2$&$ 0 $&$ 0 $ \\
\hlineB{3}
$[\Sinpp , \cdot ]$&$ -\Sinpp $&$ 0 $&$ -\Sinpp $&$ -\Soutpm $&$ +\Soutpp $&$ 0 $&$ 0 $&$ 0 $&$-\Sinpm $ \\  
$[\Sinpm , \cdot ]$&$ -\Sinpm $&$ 0 $&$ +\Sinpm $&$ -\Soutmm $&$ -\Soutmp $&$ 0 $&$ 0 $&$ -\Sinpp $&$ 0 $ \\  
$[\Sinmp , \cdot ]$&$ +\Sinmp $&$ 0 $&$ -\Sinmp $&$ 0 $&$ 0 $&$ +\Soutpm $&$ +\Soutpp $&$ 0 $&$ +\Sinmm$ \\  
$[\Sinmm , \cdot ]$&$ +\Sinmm $&$ 0 $&$ +\Sinmm $&$ 0 $&$ 0 $&$ -\Soutmm $&$ +\Soutmp $&$ +\Sinmp $&$ 0 $ \\ 
\hlineB{3}
$[\Soutpp , \cdot ]$&$ 0 $&$ -\Soutpp $&$ -\Soutpp $&$ -\Sinmp $&$ 0 $&$ +\Sinpp $&$ 0 $&$ 0 $&$ +\Soutmp$ \\  
$[\Soutmp , \cdot ]$&$ 0 $&$ -\Soutmp $&$ +\Soutmp $&$ -\Sinmm $&$ 0 $&$ -\Sinpm $&$ 0 $&$ +\Soutpp$&$ 0 $ \\  
$[\Soutpm , \cdot ]$&$ 0 $&$ +\Soutpm $&$ -\Soutpm $&$ 0 $&$ +\Sinmp $&$ 0 $&$+\Sinpp $&$ 0 $&$ -\Soutmm$ \\  
$[\Soutmm , \cdot ]$&$ 0 $&$ +\Soutmm $&$ +\Soutmm $&$ 0 $&$ -\Sinmm $&$ 0 $&$ +\Sinpm $&$ -\Soutpm $&$ 0 $ \\ 
\hlineB{3}
$[\Hmm , \cdot ]$&$ 0 $&$ 0 $&$ 2\Hmm$&$ 0 $&$ 0 $&$ 0 $&$ 0 $&$1-\Nq$&$ 0 $ \\  
$[\Hpp , \cdot ]$&$ 0 $&$ 0 $&$ -2\Hpp$&$ 0 $&$ 0 $&$ 0 $&$ 0 $&$ 0 $&$\Nq -1$ \\  
  \end{tabular}
  \caption{\label{tab:bigCommTable} Commutator algebra for the loop, string, and hadron operators at a matter site.}
  \begin{tabular}{c V{3} cccc V{3} cccc}
    \multicolumn{1}{l}{$\quad$ }\\
&$\{ \cdot, \Sinpp\}$&$\{ \cdot, \Sinpm\}$&$\{ \cdot, \Sinmp\}$&$\{ \cdot, \Sinmm\}$&$\{ \cdot, \Soutpp\}$&$\{ \cdot, \Soutpm\}$&$\{ \cdot, \Soutmp\}$&$\{ \cdot, \Soutmm\}$ \\
\hlineB{3}
$\{\Sinpp , \cdot \}$&$ 0 $&$ 0 $&$ -2\Hpp $&$ 2+\NR-\Nq$&$ 0 $&$ 0 $&$ +\Lpp $&$ -\Lpm $ \\  
$\{\Sinpm , \cdot \}$&$ 0 $&$ 0 $&$ \NR + \Nq $&$ -2\Hmm $&$ +\Lpp $&$ +\Lpm $&$ 0 $&$ 0 $ \\  
$\{\Sinmp , \cdot \}$&$ -2\Hpp $&$ \NR + \Nq $&$ 0 $&$ 0 $&$ 0 $&$ 0 $&$ +\Lmp $&$ +\Lmm $ \\  
$\{\Sinmm , \cdot \}$&$ 2 +\NR-\Nq$&$ -2\Hmm $&$ 0 $&$ 0 $&$ -\Lmp $&$ +\Lmm $&$ 0 $&$ 0 $ \\ 
\hlineB{3}
$\{\Soutpp , \cdot \}$&$ 0 $&$ +\Lpp $&$ 0 $&$ -\Lmp $&$ 0 $&$ 2\Hpp $&$ 0 $&$ 2+\NL-\Nq $\\   
$\{\Soutpm , \cdot \}$&$ 0 $&$ +\Lpm $&$ 0 $&$ +\Lmm $&$ 2\Hpp $&$ 0 $&$ \NL + \Nq $&$ 0 $ \\   
$\{\Soutmp , \cdot \}$&$ +\Lpp $&$ 0 $&$ +\Lmp $&$ 0 $&$ 0 $&$ \NL + \Nq $&$ 0 $&$ 2\Hmm $\\   
$\{\Soutmm , \cdot \}$&$ -\Lpm $ &$ 0 $&$ +\Lmm $&$ 0 $&$ 2+\NL - \Nq$&$ 0 $&$ 2\Hmm $&$ 0 $ \\   
  \end{tabular}
  \caption{\label{tab:bigAnticommTable} Anticommutator algebra for incoming and outgoing string operators at a matter site.
  }
    \setlength{\tabcolsep}{1pt}
    \resizebox{\textwidth}{!}{%
   \begin{tabular}{c V{3} cccc V{3} cccc V{3} cccc}
     \multicolumn{1}{l}{$\quad$ }\\
&$[\cdot, \Lpp[ij] ]$&$[\cdot, \Lpm[ij] ]$&$[\cdot, \Lmp[ij] ]$&$[\cdot, \Lmm[ij] ]$&$[\cdot, \Lpp[jk] ]$&$[\cdot, \Lpm[jk] ]$&$[\cdot, \Lmp[jk] ]$&$[\cdot, \Lmm[jk] ]$&$[\cdot, \Lpp[ki] ]$&$[\cdot, \Lpm[ki] ]$&$[\cdot, \Lmp[ki] ]$&$[\cdot, \Lmm[ki] ]$ \\
\hlineB{3}
$[\Lpp[ij] , \cdot ]$&$ 0 $&$ 0 $&$ 0 $&\scriptsize{$ -\mathcal{N}_{i} - \mathcal{N}_{j} - 2 $}&$ 0 $&$ 0 $&$ +\Lpp[ki] $&$ +\Lmp[ki] $&$ 0 $&$ +\Lpp[jk] $&$ 0 $&$ +\Lpm[jk] $ \\
$[\Lpm[ij] , \cdot ]$&$ 0 $&$ 0 $&$ \mathcal{N}_{i} – \mathcal{N}_{j} $&$ 0 $&$ -\Lpp[ki] $&$ +\Lmp[ki] $&$ 0 $&$ 0 $&$ 0 $&$ -\Lmp[jk] $&$ 0 $&$ +\Lmm[jk] $ \\
$[\Lmp[ij] , \cdot ]$&$ 0 $&$ \mathcal{N}_{j} – \mathcal{N}_{i} $&$ 0 $&$ 0 $&$ 0 $&$ 0 $&$ -\Lpm[ki] $&$ +\Lmm[ki] $&$ -\Lpp[jk] $&$ 0 $&$ +\Lpm[jk] $&$ 0 $ \\
$[\Lmm[ij] , \cdot ]$&\scriptsize{$ \mathcal{N}_{i} + \mathcal{N}_{j} + 2 $}&$ 0 $&$ 0 $&$ 0 $&$ -\Lpm[ki] $&$ -\Lmm[ki] $&$ 0 $&$ 0 $&$ -\Lmp[jk] $&$ 0 $&$ -\Lmm[jk] $&$ 0 $ \\
\hlineB{3}
$[\Lpp[jk] , \cdot ]$&$ 0 $&$ +\Lpp[ki] $&$ 0 $&$ +\Lpm[ki] $&$ 0 $&$ 0 $&$ 0 $&\scriptsize{$ -\mathcal{N}_{j} – \mathcal{N}_{k} - 2 $}&$ 0 $&$ 0 $&$ +\Lpp[ij] $&$ +\Lmp[ij] $ \\
$[\Lpm[jk] , \cdot ]$&$ 0 $&$ -\Lmp[ki] $&$ 0 $&$ +\Lmm[ki] $&$ 0 $&$ 0 $&$ \mathcal{N}_{j} – \mathcal{N}_{k} $&$ 0 $&$ -\Lpp[ij] $&$ +\Lmp[ij] $&$ 0 $&$ 0 $ \\
$[\Lmp[jk] , \cdot ]$&$ -\Lpp[ki] $&$ 0 $&$ +\Lpm[ki] $&$ 0 $&$ 0 $&$ \mathcal{N}_{k} – \mathcal{N}_{j} $&$ 0 $&$ 0 $&$ 0 $&$ 0 $&$ -\Lpm[ij] $&$ +\Lmm[ij] $ \\
$[\Lmm[jk] , \cdot ]$&$ -\Lmp[ki] $&$ 0 $&$ -\Lmm[ki] $&$ 0 $&\scriptsize{$ \mathcal{N}_{j} + \mathcal{N}_{k} +2 $}&$ 0 $&$ 0 $&$ 0 $&$ -\Lpm[ij] $&$ -\Lmm[ij] $&$ 0 $&$ 0 $ \\
\hlineB{3}
$[\Lpp[ki] , \cdot ]$&$ 0 $&$ 0 $&$ +\Lpp[jk] $&$ +\Lmp[jk] $&$ 0 $&$ +\Lpp[ij] $&$ 0 $&$ +\Lpm[ij] $&$ 0 $&$ 0 $&$ 0 $&\scriptsize{$ -\mathcal{N}_{k} - \mathcal{N}_{i} - 2 $} \\
$[\Lpm[ki] , \cdot ]$&$ -\Lpp[jk] $&$ +\Lmp[jk] $&$ 0 $&$ 0 $&$ 0 $&$ -\Lmp[ij] $&$ 0 $&$ +\Lmm[ij] $&$ 0 $&$ 0 $&$ \mathcal{N}_{k} – \mathcal{N}_{i} $&$ 0 $ \\
$[\Lmp[ki] , \cdot ]$&$ 0 $&$ 0 $&$ -\Lpm[jk] $&$ +\Lmm[jk] $&$ -\Lpp[ij] $&$ 0 $&$ +\Lpm[ij] $&$ 0 $&$ 0 $&$ \mathcal{N}_{i} – \mathcal{N}_{k} $&$ 0 $&$ 0 $ \\
$[\Lmm[ki] , \cdot ]$&$ -\Lpm[jk] $&$ -\Lmm[jk] $&$ 0 $&$ 0 $&$ -\Lmp[ij] $&$ 0 $&$ -\Lmm[ij] $&$ 0 $&\scriptsize{$ \mathcal{N}_{k} + \mathcal{N}_{i} + 2 $}&$ 0 $&$ 0 $&$ 0 $ \\
 \end{tabular}
 }
 \caption{\label{tab:bigLoopCommTable} Commutator algebra for the loop operators at a gluonic (pure gauge) vertex. ($d\geq 2$)}
\end{table*}

The commutation relations in Table \ref{tab:bigCommTable} have a number of qualitative features:
\begin{itemize}
  \item The $[\mathcal{N},\cdot]$ rows and $[\cdot,\mathcal{N}]$ columns express simply how LSH operators change gauge flux or fermion density.
  \item The $[\mathcal{H},\mathcal{L}]$ and $[\mathcal{L},\mathcal{H}]$ sectors express the independence of exciting hadrons and exciting gauge flux.
  \item The $[\mathcal{S},\mathcal{L}]$ sectors express how loop operators can deform outgoing (incoming) string operators into incoming (outgoing) string operators.
  \item The $[\mathcal{S},\mathcal{H}]$ sectors express how hadron operators can change the behavior of $\Sin$ and $\Sout$ operators.
\end{itemize}

String operators inherit fermionic statistics and naturally obey anticommutation relations due to their linearity in fermionic fields.
Qualitative patterns can also be found in Table \ref{tab:bigAnticommTable}:
\begin{itemize}
  \item The $\{\Sin,\Sin\}$ and $\{\Sout,\Sout\}$ sectors express both the Pauli exclusion principle as well as the fact that certain combinations of string operators acting from the same side are equivalent to hadron creation or annihilation.
  \item The $\{\Sin,\Sout\}$ and $\{\Sout,\Sin\}$ sectors express how string operators acting on both sides without changing net quark number should be thought of as a loop action.
\end{itemize}

The closure of the operator algebra confirms the completeness that was asserted for the singlets in \eqref{eq:1dLoopOps}-\eqref{eq:1dNumOps}.

\subsection{\label{subsec:LSHHam}Gauss laws and translation of the Hamiltonian}
The loop-string-hadron operators introduced above are sufficient to express the Hamiltonian for SU(2) gauge bosons coupled to one flavor of staggered fermions.
This is all that is necessary to define dynamics, since the algebra of operators is known.
In this section, all the pieces of the Hamiltonian are reconstructed from their LSH equivalents, leaving everything expressed in terms of SU(2)-invariant operators alone.

By working solely with SU(2) singlets, the only gauge constraints that will have to be enforced ``by hand'' are the Abelian Gauss laws (\ref{eq:AGL}):
\begin{equation}
  (\NR(x+1) - \NL(x)) \ket{\text{phys}} = 0 \ .
  \label{eq:1dAGL}
\end{equation}
This was always the case in the Schwinger boson formulation, but now the on-site non-Abelian Gauss law is solved at the operator level in the Hamiltonian.
Importantly, the constraints all commute.
Also note that these AGL constraints retain the same form they had in pure gauge loop formulations
\cite{
  mathurLoopApproach07,
  anishetty.mathur.eaPrepotentialFormulation10,
  raychowdhuryPrepotentialFormulation13%
}.
These constraints can be solved too, but for now the map will be given just for passing to the SU(2)-invariant variables \eqref{eq:1dLoopOps}--\eqref{eq:1dNumOps}.
\footnote{It is well known that completely solving Gauss's law in 1D space is trivial.  Doing so destroys locality and does not generalize to multidimensional space.}

The electric energy measures the gauge flux running along a link.
The quadratic Casimirs are expressed in terms of LSH number operators as
\begin{equation}
  \hat{E}_{L/R}^{\mathrm a}(x) \hat{E}_{L/R}^{\mathrm a}(x) = \tfrac{1}{2} \mathcal{N}_{L/R} (x) \left(\tfrac{1}{2} \mathcal{N}_{L/R} (x) + 1 \right)
\end{equation}
To form the system's electric energy, all link ends are put on the same footing by taking
\begin{align}
    \hat{H}_E \rightarrow \frac{g_0^2}{4} \sum_{x} & \left[ \tfrac{1}{2} \NR(x) \left(\tfrac{1}{2} \NR(x) + 1 \right)  \right. \nonumber \\
    & \left. + \tfrac{1}{2} \NL(x) \left(\tfrac{1}{2} \NL(x) + 1 \right) \right] \ .
    \label{eq:LSH1dHE}
\end{align}
Note that $\NR(x)$ and $\NL(x)$ are on either side of some site $x$, rather than opposite ends of a link.

The staggered mass terms are given quite simply in terms of the quark number operators, $\Nq = \psi^\dagger \psi$, so the mass energy is just
\begin{equation}
  \hat{H}_M \rightarrow m_0 \sum_x (-)^x \Nq (x) \ .
  \label{eq:LSH1dHM}
\end{equation}

The hopping terms from $\hat{H}_I$ can create, destroy, break, or glue together meson strings, so their expressions naturally involve the local string operators.
To translate a hopping term, the sites at each end of a link can be considered separately. 
Recall from \eqref{eq:KSLinkOp} and \eqref{eq:sbLinks} that link operators were given in terms of Schwinger bosons by
\begin{align*}
  \hat{U}(x,i) &= \UL(x) \UR(x+e_i)\ , \\*
  \UL(x,i) &= \left. \frac{1}{\sqrt{\Nleft+1}}
  \left(
    \begin{array}{rr}
      \hat{a}_2^\dagger(L) & \hat{a}_1 (L) \\
      -\hat{a}_1^\dagger(L) & \hat{a}_2(L)
  \end{array}\right) \right|_{x,i} \ , \\
  \UR(x,i) &= \left. \left(
    \begin{array}{rr}
      \hat{a}_1^\dagger(R) & \hat{a}_2^\dagger(R) \\
      -\hat{a}_2(R) &  \hat{a}_1(R)
    \end{array}
  \right) \frac{1}{\sqrt{\Nright+1}} \right|_{x,i} \ .
\end{align*}
Using the separate $\hat{U}_{L/R}$ factors at 1D sites, it follows that
\begin{subequations}
\begin{align}
  \hat{\psi}^\dagger(x) \UL(x) &= \frac{1}{\sqrt{\NL(x) +1}} \left(\begin{matrix} \Soutpp(x) , & \Soutpm(x) \end{matrix}\right) \ , \\*
  \UR(x) \hat{\psi}(x) &= \left(\begin{matrix}\Sinpm(x)\\ \Sinmm(x)\end{matrix}\right)\frac{1}{\sqrt{\NR(x)+1}} \ .
\end{align}
\end{subequations}
Thus, the translation of the interaction into LSH operators is given by
\begin{align}
  \hat{H}_I \rightarrow & \sum_x \frac{1}{\sqrt{\NL(x)+1}} \left[ \sum_{\sigma=\pm} \Sout^{+,\sigma}(x) \Sin^{\sigma,-}(x+1) \right] \times \nonumber \\*
  & \qquad \qquad \qquad \qquad \times \frac{1}{\sqrt{\NR(x+1)+1}} \quad + \text{H.c.} \label{eq:LSH1dInteraction}
\end{align}

The entire Hamiltonian \eqref{eq:LSH1dHE}-\eqref{eq:LSH1dInteraction} is now expressed solely in terms of the SU(2) singlets from \eqref{eq:1dLoopOps}-\eqref{eq:1dNumOps}.

\subsection{\label{sec:1dDynamics}An orthonormal loop-string-hadron basis and operator factorization}

To describe dynamics in a way useful for computational algorithms, it is helpful to set up a basis.
It would seem natural to use as a CSCO the operators $\{ \NR, \NL, \Nq\}$ since these naturally appeared in the algebra, and to then express the Hamiltonian in terms of their quantum numbers.
However, these may not be the most desirable due to the fact that these are constrained by the possible excitations LSH operators can create.
(For example, $\Nq=1$ while $\NR=\NL=0$ is not gauge invariant.)
As will be shown below, one can instead enumerate states more directly in terms of SU(2)-invariant LSH excitations---leading to a loop-string-hadron basis.
In this way, only allowed on-site states are ever represented.

A second practical issue to be addressed concerns operator factorization.
The Hamiltonian was expressed in terms of LSH operators, but in an orthonormal basis these operators change state normalization in addition to changing quantum numbers.
Factorizing these two behaviors has the benefits of making the matrix elements of any operator completely evident and also setting the stage for a Wigner-Jordan transformation.
This factorization will be done for convenience with respect to a LSH basis (though the factorization itself is basis independent). 

\subsubsection{\label{sec:LSHOnSite}On-site Hilbert space construction}
Until this point, the LSH constructions have been built on underlying harmonic oscillator operators, but there was no need to choose a basis.
The formal tools introduced will now be used to construct a basis of SU(2)-invariant excitations in which to express the action of the Hamiltonian.
This is done by first defining ``on-site'' bases and later stitching these together to construct lattice states.

An on-site Hilbert space for the 1D lattice has three apparent degrees of freedom corresponding to the original occupation numbers (\ref{eq:1dNumOps}), i.e., $n_L$, $n_R$, and $n_{\psi}$.
But as remarked above these are constrained by the possible excitations generated by LSH operators.

A more physical on-site basis consists of states $\ket{n_l,n_i,n_o}$ with a loop quantum number $n_l$ and quark quantum numbers $n_i$, $n_o$ that describe strictly SU(2)-invariant gauge and matter excitations.
Such a ``loop-string-hadron basis'' of unnormalized kets, denoted by a double-bar ket $\ket{| \ \ }$, can be defined as follows.
\begin{subequations}
  \label{eq:onSiteBasisDef}
  \begin{align}
    \ket{|n_l, n_i=0, n_o=0} &\equiv (\Lpp{})^{n_l} \ket{0} \ ,\\
    \ket{|n_l, n_i=0, n_o=1} &\equiv (\Lpp{})^{n_l} \Soutpp \ket{0} \ ,\\
    \ket{|n_l, n_i=1, n_o=0} &\equiv (\Lpp{})^{n_l} \Sinpp \ket{0} \ ,\\
    \ket{|n_l, n_i=1, n_o=1} &\equiv (\Lpp{})^{n_l} \Hpp \ket{0} \ ,
  \end{align}
\end{subequations}
where
\begin{equation}
  n_i=0,1 \qquad n_o=0,1  \qquad n_l=0,1,2,\cdots \ ,
  \label{eq:LSH1dQuantumNumbersRange}
\end{equation}
$\ket{0}$ is the local vacant state annihilated by any LSH operator carrying at least one minus sign, and $\braket{0|0}=1$.
Note that $n_i$ and $n_o$ indicate quark content, but not necessarily strings;
exactly one of these equaling 1 implies the presence of a flux string, but both equaling 1 means they are paired up into a hadron.
Furthermore, one must take care to remember that the quark numbers are properly handled as ordered fermionic occupation numbers.
The states above uniquely enumerate all SU(2)-invariant excitations that can be hosted by a site.
\footnote{
  Though its utility is limited, it is straightforward to give one unifying expression valid for all states:
$      \ket{|n_l,n_i,n_o} =  (\Lpp)^{n_l} \left[\Pi_{00} + \Pi_{01} + \Pi_{10} +(1/2)\Lmm \right] (\Sinpp)^{n_i} (\Soutpp)^{n_o} \ket{0} $,
  where
    $\Pi_{00} = \Hmm \Hpp$,
    $\Pi_{01} = \Lmp \Lpm$,
    $\Pi_{10} = \Lpm \Lmp$,
  and
    $( 1/2) \Lmm (\Sinpp)^{n_i} (\Soutpp)^{n_o} \ket{0} = \delta_{n_i,1} \delta_{n_o,1}\Hpp \ket{0}$.
}

The norms of $\ket{|n_l,n_i,n_o}$ can be derived by repeated use of the operator algebra.
These types of calculations are described in Appendix \ref{app:norm}.
The result is that a normalized basis is given by
\begin{equation}
  \label{eq:normalSiteBasis}
  \ket{n_l,n_i,n_o} = \frac{\ket{|n_l,n_i,n_o}}{\sqrt{n_l! \ (n_l+1+(n_i \oplus n_o ))!}}  \ ,
\end{equation}
where $\oplus$ denotes addition modulo 2.

Before reexpressing the Hamiltonian, the SU(2)-invariant LSH quantum numbers will have to be related to the prepotential quantum numbers.
This relationship can be inferred from
\begin{subequations}
  \begin{align}
    \Nq \ket{n_l, n_i, n_o} &= (n_i + n_o) \ket{n_l, n_i, n_o} \ , \\*
    \NL \ket{n_l, n_i, n_o} &= (n_l + n_o(1-n_i)) \ket{n_l, n_i, n_o} \ , \\*
    \NR \ket{n_l, n_i, n_o} &= (n_l + n_i(1-n_o)) \ket{n_l, n_i, n_o} \ .
  \end{align}%
  \label{eq:quantumNumberConversion}%
\end{subequations}
These imply that the following act as number operators on the $\ket{n_l, n_i, n_o}$ states:
\begin{subequations}
  \begin{align}
    \Ni&\equiv \tfrac{1}{2}\left[ \Nq + \NR -\NL \right] \ ,\\*
    \No&\equiv \tfrac{1}{2}\left[ \Nq + \NL -\NR \right] \ ,\\*
    \Nl&\equiv \tfrac{1}{2}\left[ \NL+\NR-\Nq + \tfrac{1}{2}\left( \Nq^2 - (\NL-\NR)^2\right) \right] \ .
  \end{align}%
  \label{eq:LSHNumberOps}%
\end{subequations}
(Note again that $\mathcal{N}_{R/L}$ belong to a common site, not opposite ends of a link.)
The relations (\ref{eq:quantumNumberConversion}) can now be promoted to operator identities to be inserted in the Hamiltonian:
\begin{subequations}
  \begin{align}
    \Nq &= \Ni + \No \ , \\
    \NL &= \Nl + \No(1-\Ni) \ , \\
    \NR &= \Nl + \Ni(1-\No) \ .
  \end{align}
  \label{eq:loopStringToSHONumbers}
\end{subequations}

To summarize, the LSH basis characterizes local states by counting units of loop flux passing through a site, and keeping track of quark species present.
A lone ``out quark'' ($n_o=1$) or a lone ``in quark'' ($n_i=1$) is shorthand for indicating the type of string present, while completely full orbitals just signify a gauge-invariant hadron.
The LSH quantum numbers $\{ \Nl, \Ni, \No\}$ are equivalent to allowed combinations of the $\{ \NR, \NL, \No\}$ quantum numbers, but have the benefit of being unconstrained over their ranges (\ref{eq:LSH1dQuantumNumbersRange}).

\subsubsection{Global Hilbert space construction in one dimension}
While the loop-string-hadron formulation largely focuses on characterizing site-local excitations, the dynamics ultimately couples sites and is expressed using states of the lattice as a whole.
The global Hilbert space consists of the excitations coming from all sites: one bosonic loop mode and two fermionic quark modes each.
However, the global space can only be viewed as a tensor product space of all the local modes to the extent that fermionic statistics are accounted for.
One can account for fermionic statistics with binary occupation numbers if the associated basis states have a prescription for how the fermionic operators are ordered.
The ordered product of operators can then be applied to some fixed reference state that satisfies the Abelian Gauss law and any other desired symmetries.

In the loop-string-hadron framework, the lattice ``vacant'' state $\ket{0}$ (not to be confused with a qubit computational basis state) is characterized as a state devoid of any excitations,
\begin{equation}
  \Ni(x) \ket{0} = \No(x) \ket{0} = \Nl(x) \ket{0} = 0 \quad \text{for all $x$} \ .
\end{equation}
It is annihilated by any $\mathcal{L}^{\pm \pm}$, $\mathcal{S}^{\pm \pm}$, or $\mathcal{H}^{\pm \pm}$ carrying at least one minus sign.
One can construct the entire Hilbert space by using $\ket{0}$ as a reference state.

Another reference state would be the staggered strong-coupling vacuum $\ket{v}$, which is the true vacuum at $g_0,m_0 \rightarrow \infty$.
The staggered strong-coupling vacuum is characterized by having vanishing electric fields and full fermion orbitals on odd sites;
$\ket{v}$ can be related to $\ket{0}$ by applying to it $\Hpp$ from every even site.

For all the other lattice basis states, it is necessary to fix a convention for fermion ordering.
A site-local ordering was already chosen earlier, so all that is necessary is to order sites.
The convention we choose is that sites receive excitations from greatest $x$ down to least;
the associated expressions would then have $\mathcal{S}_{\text{in}/\text{out}}^{++}$'s written with $x$ increasing from left to right.
On a lattice with an even number of sites $L_x$, these states are denoted by
\begin{equation*}
  \begin{split}
    \left| n_l(0),n_i(0),n_o(0); n_l(1),n_i(1),n_o(1);  \cdots \right.\\
    \quad \left. \cdots ; n_l(L_x-1),n_i(L_x-1),n_o(L_x-1) \right>
      &
  \end{split}
\end{equation*}
For example, the (normalized) staggered strong-coupling vacuum of a four site lattice is given by
\begin{align}
  \ket{v} &= \ket{\ 0,0,0;\ 0,1,1;\ 0,0,0;\ 0,1,1} \nonumber \\
  &= \left[ \tfrac{1}{2}\Lmm(1)\Sinpp(1)\Soutpp(1) \right] \times \nonumber \\
  & \quad \times \left[ \tfrac{1}{2}\Lmm(3)\Sinpp(3)\Soutpp(3) \right] \ket{0} \nonumber \\
  &= \Hpp(1)\Hpp(3)\ket{0} \ . 
  \label{eq:fourSiteSCV}
\end{align}
Of course, the ordering is especially important for states that actually have on-site net fermionic excitations.
An an example of this would be a basis state describing a meson string between sites $x=0$ and $x=1$, which is given by
\begin{align*}
    \ket{\text{meson}} &= \ket{0,0,1;\ 0,1,0;\ 0,0,0;\ 0,1,1} \\
    &= \frac{1}{2} \Soutpp(0) \Sinpp(1) \Hpp(3) \ket{0} \ ,
\end{align*}
as opposed to $\frac{1}{2} \Sinpp(1) \Sinpp(0) \Hpp(3) \ket{0}$ with the opposite ordering.
This state appears after one application of $\hat{H}_I$ to the staggered strong-coupling vacuum $\ket{v}$.

We can summarize the characterization of basis states with the following rule:
Local quarks are created going from greatest $x$ down to least, and with $\Soutpp(x)$ always acting before $\Sinpp(x)$.

Working with the full lattice, the Abelian Gauss law is imposed for physical states.
In the $\ket{n_l, n_i, n_o}$ basis, this translates to
\begin{equation}
  \left[ n_l + n_o (1-n_i) \right]_{x} = \left[ n_l + n_i(1-n_o) \right]_{x+1}.
\end{equation}

\subsubsection{Operator factorization}
It was remarked at the beginning of this section that LSH operators in the Hamiltonian change quantum numbers as well as state normalization.
The on-site operators can now be factored in order to isolate the two behaviors, at which point matrix elements with respect to the LSH basis can be read off trivially.

Pertaining to the loop quantum number $n_l$, we introduce \emph{normalized ladder operators}, $\Lambda^{+}$ and $\Lambda^{-}$:
\begin{equation}
  \Lambda^{\pm} \equiv \mathcal{L}^{\pm \pm} \frac{1}{\sqrt{(\Nl + \tfrac{1}{2} \pm \tfrac{1}{2})(\Nl + \tfrac{3}{2} \pm \tfrac{1}{2} + (\Ni \oplus \No))}}
  \label{eq:lambdas}
\end{equation}
Here a ``normalized operator'' refers to any operator $\mathcal{O}$ such that nonvanishing eigenvalues of $\mathcal{O}^\dagger \mathcal{O}$ are unity.
The significance of $\Lambda^{\pm}$ is that their nonvanishing matrix elements in the LSH basis are all unity:
\begin{equation}
  \bra{n_l^\prime , n_i^\prime, n_o^\prime } \Lambda^{\pm} \ket{n_l,n_i,n_o} = \delta_{n_l^\prime,n_l \pm 1} \delta_{n_i^\prime ,n_i} \delta_{n_o^\prime ,n_o} \ .
\end{equation}
Hence, they move states up and down the ladder of $n_l$ without changing normalization, except for the possibility of annihilation at the bottom.
The ladder operators were constructed in (\ref{eq:lambdas}) to make factoring $\Lpp$ and $\Lmm$ trivial.

As for the quark quantum numbers, these are affected by the string operators (and the mixed-type loop operators $\mathcal{L}^{\pm, \mp}$).
The string operators were found to obey fermionlike anticommutation relations, but they are not canonically normalized.
This motivates introducing SU(2)-invariant fermionic modes $\chi_i,\chi_o$ to describe them, with
\begin{alignat}{4}
  \{ \chi_{q'} \, , \, \chi_{q} \} = \{ \chi_{q'}^\dagger \, , \, \chi_{q}^\dagger \} &= 0 \ , & & \quad (q = i,o) \\
  \{ \chi_{q'} \, , \, \chi_{q}^\dagger \} & = \delta_{q'q} \ . & & \quad (q = i,o)
\end{alignat}
These also qualify as normalized ladder operators.
Because string operators can affect loop numbers, it will prove helpful to also introduce the following shorthand \emph{conditional ladder operators}:
\begin{subequations}
  \label{eq:lambdaPowers}
  \begin{align}
    (\Lambda^\pm )^{\mathcal{N}_{q}} &\equiv (1-\mathcal{N}_{q}) + \Lambda^\pm \mathcal{N}_{q} \ ,  &(q=i,o)\\
    (\Lambda^\pm )^{1-\mathcal{N}_{q}} &\equiv \Lambda^\pm (1 - \mathcal{N}_{q})  + \mathcal{N}_{q} \ . &(q=i,o)
  \end{align}
\end{subequations}
Each term in these operator exponentials projects on to one or the other eigenspace of $\mathcal{N}_q$ and is followed by a corresponding loop ladder action or lack thereof.

The SU(2)-invariant quark modes $\chi_i$ and $\chi_o$ are also helpful for characterizing global basis states.
One can express any of the LSH basis states by simply acting all the $\chi^\dagger_{i/o}$'s on $\ket{0}$ with the same rule for ordering as before---there is no need for string and $\Lmm$ operators or factors of $1/2$ like those in (\ref{eq:fourSiteSCV}).

Equipped with the normalized ladder operators, all loop and string operators can be factorized as shown in Table \ref{tab:LSHOpFactorizations}.
It is straightforward to show that these operator factorizations completely reproduce the LSH algebra.
The factorizations are all given in a canonical form, with diagonal scaling operators sitting on the right and normalized ladder operators following them.
Acting an LSH operator on a ket $\ket{n_l, n_i, n_o}$, the numerical value of its matrix element can just be read off, and the resultant quantum numbers are easily deduced from the ladder operator content.

\begin{table}[!thb]
  \centering
  \begin{tabular}{m{8.5cm}}
    \hline
    \begin{center}\textsc{Loop-string-hadron operator factorizations} \end{center}\\
    \hline
    {
      \begin{subequations}
        \label{eq:LSHEffectiveOps}
        \begin{align}
          \Lpp &= \ \ \Lambda^+ \sqrt{(\Nl + 1) (\Nl + 2 + (\Ni \oplus \No))} \label{eq:Lppfactored} \\
          \Lmm &= \ \ \Lambda^- \sqrt{\Nl (\Nl + 1 + (\Ni \oplus \No))} \label{eq:Lmmfactored} \\
          \Lpm &= - \chi_i^\dagger \ \chi_o \label{eq:Lpmfactored} \\
          \Lmp &= \ \ \chi_i \ \chi_o^\dagger \label{eq:Lmpfactored} \\
          \Sinpp &= \ \ \chi_i^\dagger \ (\Lambda^+)^{\No} \quad \sqrt{\Nl + 2 - \No} \label{eq:Sinppfactored} \\
          \Sinmm &= \ \ \chi_i \ (\Lambda^-)^{\No} \quad \sqrt{\Nl + 2(1 - \No)} \label{eq:Sinmmfactored} \\
          \Soutpp &= \ \ \chi_o^\dagger \ (\Lambda^+)^{\Ni} \quad \sqrt{\Nl + 2 - \Ni} \label{eq:Soutppfactored} \\
          \Soutmm &= \ \ \chi_o \ (\Lambda^-)^{\Ni} \quad \sqrt{\Nl + 2(1 - \Ni)} \label{eq:Soutmmfactored} \\
          \Sinmp &= \ \ \chi_o^\dagger \ (\Lambda^-)^{1-\Ni} \sqrt{\Nl+2\Ni} \label{eq:Sinmpfactored} \\
          \Sinpm &= \ \ \chi_o  \ (\Lambda^+)^{1-\Ni}\sqrt{\Nl+1+\Ni} \label{eq:Sinpmfactored} \\
          \Soutpm &= \ \ \chi_i^\dagger \ (\Lambda^-)^{1-\No} \sqrt{\Nl+2\No} \label{eq:Soutpmfactored} \\
          \Soutmp &= \ \ \chi_i \ (\Lambda^+)^{1-\No} \sqrt{\Nl+1+\No} \label{eq:Soutmpfactored}\\
          \Hpp &= \ \ \chi_i^\dagger \chi_o^\dagger \label{eq:Hppfactored} \\
          \Hmm &= - \chi_i \chi_o \label{eq:Hmmfactored}
        \end{align}
      \end{subequations}
    } \\
    \hline
  \end{tabular}
  \caption{Factorization of all SU(2) invariant operators into canonically normalized fermionic modes times a loop ladder operator times a function of number operators. The operator exponentials are conditional ladder operators defined in (\ref{eq:lambdaPowers}).}
  \label{tab:LSHOpFactorizations}
\end{table}

\subsubsection{Wigner-Jordan transform for one dimension}
Using fermionic operators can be convenient analytically, but computation models usually assume native operations that commute for different sites.
In classical lattice QCD, Grassman variables are avoided because the quark fields can be integrated out of the functional integral analytically.
Quantum simulation, however, frequently involves choosing a fermionic mapping onto commuting computational degrees of freedom.

Qubits are two-state systems with a ``computational basis'' often denoted with states $\ket{0}$ and $\ket{1}$, but most computation models do not regard these as having a fermionic character.
For example, the ``raising'' operators $\ket{1}\bra{0}$ for distinct qubits commute with each other.
The bottom line is that for applications the Hamiltonian will need to be converted to spin operators at some point.
For one-dimensional systems with localized interactions, the Wigner-Jordan transformation maps fermionic modes into spin operators rather cleanly.

The fermionic modes $\chi_i(x)$, $\chi_o(x)$ for $x=0,\ldots,L_x-1$ express physical (SU(2)-invariant) quark degrees of freedom that dynamically couple to each other through the hopping terms.
However, it turns out that the $\chi_i$'s and $\chi_o$'s, in fact, decouple from each other.
The operator-factorized Hamiltonian will be discussed below, but to see this decoupling one only needs the string operators from the hopping terms in (\ref{eq:LSH1dInteraction}).
With the factorizations in (\ref{eq:LSHEffectiveOps}), the fermionic content of $\Sout^{+\sigma}(x) \Sin^{\sigma -}(x+1)$ terms takes the form
\begin{align*}
  \Soutpp(x) \Sinpm(x+1) & \sim \chi_o^\dagger(x) \chi_o(x+1) \cdots \ , \\
  \Soutpm(x) \Sinmm(x+1) & \sim \chi_i^\dagger(x) \chi_i(x+1) \cdots \ . 
\end{align*}
The decoupling is now manifest.

Knowing this, we relabel the fermionic modes using $\Psi_k$ for $k=0,\ldots,2L_x -1$, with the map
\begin{equation*}
  \Psi_k = \begin{cases} \chi_i(k), \qquad 0 \leq k \leq L_x - 1 \\ \chi_o(k-L_x), \qquad L_x \leq k \leq 2L_x - 1\end{cases} \ .
\end{equation*}
The Wigner-Jordan transformation converts the $\Psi_k$ into spin operators via
\begin{equation}
  \Psi_k \equiv \sigma^{+}_{k} \prod_{k' = 0}^{k-1} Z_{k'} \ .
\end{equation}
Assuming open boundary conditions, all fermionic couplings are then nearest-neighbor in the $x$ coordinate as well as the $k$ label, and as a result the Wigner-Jordan transformation has no leftover Pauli-$Z$ strings.
The couplings in the hopping terms will all take the form $\sigma_k^\pm \sigma_{k+1}^\mp$:
\begin{align}
  \chi_i^\dagger(x) \chi_i (x+1) &\rightarrow \sigma_{x}^- \sigma_{x+1}^+ \ , \\
  \chi_o^\dagger(x) \chi_o (x+1) &\rightarrow \sigma_{L_x +x}^- \sigma_{L_x+x+1}^+ \ .
\end{align}
Hence, on the 1D open lattice only, it is possible to essentially replace anticommuting $\chi$'s and $\chi^\dagger$'s with commuting $\sigma^-$'s and $\sigma^+$'s in the operator factorizations.

\subsection{Dynamics of loop-string-hadron states}
The terms of the Hamiltonian presented in Sec. \ref{subsec:LSHHam} were expressed in terms of site-local loop-string-hadron operators.
The Hamiltonian will now be reexpressed once more using the operator factorizations from above, with the final result expediting the process of calculating matrix elements in the LSH basis.
Subsequently, a graphical method is given for determining how states are mixed by terms in the Hamiltonian.

Starting with the electric Hamiltonian, the Casimirs continue to be diagonal as they always were.
Using the conversion (\ref{eq:loopStringToSHONumbers}) from prepotential to LSH number operators, we have
\begin{align}
  \hat{H}_E = \frac{g_0^2}{4} \sum_{x} & \left\{ \left[ \tfrac{1}{2} \left( \Nl + \No(1-\Ni) \right) \right]_{x} \right. \times \nonumber \\*
    & \quad \times \left[\tfrac{1}{2} \left( \Nl + \No(1-\Ni) \right) + 1 \right]_{x} \nonumber \\*
    & \ \ + \left[ \tfrac{1}{2} \left( \Nl + \Ni(1-\No) \right) \right]_{x} \times \nonumber \\*
    & \qquad \times \left. \left[ \tfrac{1}{2} \left( \Nl + \Ni(1-\No) \right) + 1 \right]_{x} \right\} \ .
\end{align}

The mass Hamiltonian is also diagonal and given simply by
\begin{align}
  \hat{H}_M &= m_{0} \sum_x (-)^x (\Ni(x) + \No(x)) \ .
  \label{eq:HMLoopString} 
\end{align}
  
And lastly, the interaction $\hat{H}_I$ in terms of SU(2) invariants was originally given as (\ref{eq:LSH1dInteraction}), with the off-diagonal part of a hopping term being
$\sum_{\sigma=\pm} \sigma \Sout^{+,\sigma}(x) \Sin^{\sigma,-}(x+1)$.
Using the operator factorizations (\ref{eq:LSHEffectiveOps}), these hopping terms are given by
\begin{subequations}
\label{eq:hoppingLoopString}
\begin{align}
  &\begin{split}
    &\Soutpp(x) \Sinpm (x+1) \\
    &=     \left[ \chi_o^\dagger \right]_{x} \left[ \chi_o \right]_{x+1} \times \\
    &\quad\ \ \left[ (1-\Ni)+\Lambda^+ \Ni \right]_{x} \left[ \Ni +  \Lambda^+ (1-\Ni)  \right]_{x+1} \times\\
    &\quad\ \ \left[ \sqrt{\Nl-\Ni+2} \right]_{x} \left[ \sqrt{\Nl-(1-\Ni)+2} \right]_{x+1} \ ,
  \end{split}\\*
  &\begin{split}
    &\Soutmm(x) \Sinmp (x+1) \\
    &=     \left[ \chi_o \right]_{x} \left[ \chi_o^\dagger \right]_{x+1} \times \\
    &\quad\ \ \left[ (1-\Ni) + \Lambda^- \Ni \right]_{x} \left[ \Ni +  \Lambda^- (1-\Ni)  \right]_{x+1} \times\\
    &\quad\ \ \left[ \sqrt{\Nl+2(1-\Ni)} \right]_{x} \left[ \sqrt{\Nl + 2\Ni} \right]_{x+1} \ ,
  \end{split}\\*
  &\begin{split}
    &\Soutpm(x) \Sinmm (x+1) \\
    &=     \left[ \chi_i^\dagger \right]_{x} \left[ \chi_i \right]_{x+1} \times \\
    &\quad\ \ \left[ \No + \Lambda^- (1-\No) \right]_{x} \left[ (1-\No)+\Lambda^- \No \right]_{x+1} \times\\
    &\quad\ \ \left[ \sqrt{\Nl + 2\No} \right]_{x} \left[ \sqrt{\Nl + 2(1-\No)} \right]_{x+1} \ ,
  \end{split}\\*
  &\begin{split}
    &\Soutmp(x) \Sinpp (x+1) \\
    &=     \left[ \chi_i \right]_{x} \left[ \chi_i^\dagger  \right]_{x+1} \times \\
    &\quad\ \ \left[ \No + \Lambda^+ (1-\No) \right]_{x} \left[ (1-\No)+\Lambda^+ \No \right]_{x+1} \times\\
    &\quad\ \ \left[ \sqrt{\Nl + \No +1} \right]_{x} \left[ \sqrt{\Nl + (1-\No)+1} \right]_{x+1} \ .
  \end{split}
\end{align}
\end{subequations}
To complete $\hat{H}_I$, one also needs the diagonal ``outer'' factors that sandwich these.
By (\ref{eq:loopStringToSHONumbers}),
\begin{equation}
  \label{eq:denominatorsLoopString}
  \frac{1}{\sqrt{\mathcal{N}_{L/R}+1}} = \frac{1}{\sqrt{\Nl + \mathcal{N}_{o/i}\left(1-\mathcal{N}_{i/o}\right) + 1} } \ .
\end{equation}
The above expressions in terms of diagonalized scaling operators and normalized ladder operators are everything one needs to immediately express the action of the Hamiltonian in the LSH basis.

\begin{figure}[!thb]
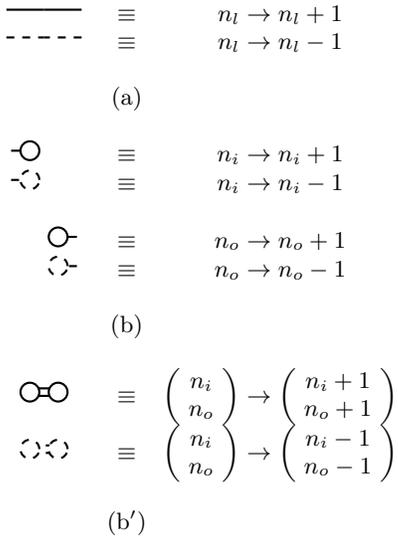

\begin{center}
\textsc{Graphical quantum number instructions} \\
\[
\begin{array}{ccc}
  \LoopPicture{solid}{solid} &\equiv& n_l \rightarrow n_l + 1 \\
  \LoopPicture{dashed}{dashed} &\equiv& n_l \rightarrow n_l - 1 \\
  \\
  & \text{(a)} & \\
  \\
  \SinPicture{solid}{solid} &\equiv& n_i \rightarrow n_i+1 \\
  \SinPicture{densely dashed}{dashed} &\equiv& n_i \rightarrow n_i-1 \\
  \\
  \SoutPicture{solid}{solid} &\equiv& n_o \rightarrow n_o+1 \\
  \SoutPicture{dashed}{densely dashed} &\equiv& n_o \rightarrow n_o-1 \\
  \\
  & \text{(b)} & \\
  \\
  \HadronPicture{solid} &\equiv&  \left( \begin{array}{c} n_i \\ n_o \end{array} \right) \rightarrow \left( \begin{array}{c} n_i+1 \\ n_o+1 \end{array} \right) \\
  \HadronPicture{dashed} &\equiv&  \left( \begin{array}{c} n_i \\ n_o \end{array} \right) \rightarrow \left( \begin{array}{c} n_i-1 \\ n_o-1 \end{array} \right) \\
  \\
  & \text{(b$^\prime$)} & \\
  \\
\end{array}
\]
\end{center}
\caption{\label{fig:picac}
Pictorial representation of changes in quantum numbers between initial and final states, which represent (a) flux creation and annihilation, (b) quark creation and annihilation, and (b$^\prime$) hadron creation and annihilation (a composite action).}
\end{figure}

The actions of the loop-string-hadron operators are easier to intuit given the fact they are 1-sparse in the LSH basis, i.e., any of the $\mathcal{L}$, $\mathcal{S}$, or $\mathcal{H}$ operators acting on a basis state either turns it into another basis state or annihilates it.
They do not expand into linear combinations like link operators do in irrep bases [cf. Eq.\,(\ref{eq:linkOpAction})].
To express all possible actions of the LSH operators in terms of quantum numbers, we introduce a pictorial mapping shown in Fig. \ref{fig:picac} that associates pictures with changes in quantum numbers of the basis states.

A summary of these pictorial actions on quantum numbers is as follows:
\begin{itemize}
  \item Solid (dashed) line: Increment (decrement) $n_l$ by one unit.
  \item Solid (dashed) in quark: Increment (decrement) $n_i$ by one unit.
  \item Solid (dashed) out quark: Increment (decrement) $n_o$ by one unit.
  \item Solid (dashed) hadron: Increment (decrement) both $n_i$ and $n_o$ by one unit.
\end{itemize}
If the resulting quantum numbers are forbidden, this corresponds to annihilation of the basis state.
Note also that these graphical rules use symbols that are related to, but distinct from, the basis-independent operator pictures introduced in Sec.\,\ref{subsec:SU2Invariants}.

As a simple example of their usage, the fact that $\Lpp \ket{n_l,n_i,n_o} \propto \ket{n_l+1,n_i,n_o}$ means the action of $\Lpp$ is represented by a single solid line:
\begin{align*}
  \Lpp \ket{n_l,n_i,n_o} &\propto \ket{n_l+1,n_i,n_o} \\*
  \\*
  \raisebox{3pt}{$\Rightarrow \quad \Lpp =$} \vcenter{\hbox{\LoopOpPicture{solid}{solid}{$ $}}} \quad & \sim \quad \begin{array}{c} \LoopPicture{solid}{solid} \end{array}
\end{align*}

In general, however, the operators may have composite actions, so the instructions are composed vertically along with an ordering to them.
This is summarized as follows:
\begin{itemize}
  \item Effect the changes indicated by each instruction, going from top to bottom.
  \item The state is annihilated if at any step the quantum numbers are forbidden.
\end{itemize}
Consider $\Sinpp$ for example.
Using the factorization (\ref{eq:Sinppfactored}) and conditional ladder operators (\ref{eq:lambdaPowers}), one can write $\Sinpp \sim \chi_i^\dagger (1- \mathcal{N}_o) + \chi_i^\dagger \Lambda^+ \mathcal{N}_o$.
Acting on a basis state, at most one of these terms can be non zero.
Each term tries to raise $n_i$, while only one can raise $n_l$.
These behaviors are diagrammatically summarized by
\begin{align*}
  \raisebox{1pt}{$ \Sinpp = $} \vcenter{\hbox{\SinOpPicture{solid}{solid}{$ $}}} \quad &\sim \begin{array}{c} \SoutProjPicture{solid}{densely dotted} \\ \SinPicture{solid}{solid} \end{array} + \begin{array}{c} \SoutPicture{densely dotted}{densely dotted} \\ \LoopPicture{solid}{solid} \\ \HadronPicture{solid} \end{array} \ ,
\end{align*}
where the first term creates an in string on a quarkless site, while the second connects an in string to an already-existing out string to form a baryon and gauge flux line.
We similarly represent and describe the actions of all loop-string-hadron operators pictorially in Table \ref{tab:opactions}.

\subsection{Summary of matter sites}
To conclude this section, we summarize the results and what their significance is in 1 and higher dimensions.

Prepotentials were used to construct a closed algebra of manifestly SU(2)-invariant LSH operators, and it was shown how to translate the Hamiltonian into them.
These operators were then used to construct a LSH basis in which every possible combination of quantum numbers (consistent with the Abelian Gauss law) describes a unique set of on-site excitations.
For future applications, all LSH operators were then factored for convenience on that basis and the Hamiltonian was again reexpressed in a more explicit form.

For $d=1$ and with open boundary conditions, one can essentially just forget about Fermi statistics and replace singlet-quark operators with spin operators.
In higher dimensions this will no longer be the case.
However, the local bases and operator factorizations will carry over to ``matter sites'' in $d>1$, so the main feature that gets lost is really just simplicity of the Wigner-Jordan transformation.

\newcolumntype{L}{>{\arraybackslash}m{9cm}}
\newcommand{\cartooncaption}[1]{\text{\scriptsize{#1}}}
\begin{table*}[!htb]
  \centering
  \textsc{Graphical representation of operators}
  \[
    \begin{array}{c|ccc}
    \hline
    \text{LSH operator} & \multicolumn{3}{c}{\text{Physical description; Graphical action on state $\ket{n_l,n_i,n_o}_x$}} \\
    \hline\hline
\renewcommand{\arraystretch}{.7}%
    \\[-0.20cm]
    \Lpp(x)\equiv \vcenter{\hbox{\LoopOpPicture{solid}{solid}{$x$}}} & \multicolumn{3}{c}{\begin{array}{c} \vcenter{\hbox{\LoopPicture{solid}{solid}}} \\ \cartooncaption{Create unit of gauge flux.} \end{array}} \\
    \hline
    \\[-0.20cm]
    \Lmm(x)\equiv \vcenter{\hbox{\LoopOpPicture{densely dotted}{densely dotted}{$x$}}} & \multicolumn{3}{c}{\begin{array}{c} \vcenter{\hbox{\LoopPicture{densely dotted}{densely dotted}}} \\ \cartooncaption{Destroy unit of gauge flux.} \end{array}} \\
    \hline
    \\[-0.20cm]
    \Lpm(x)\equiv \vcenter{\hbox{\LoopOpPicture{solid}{densely dotted}{$x$}}} & \multicolumn{3}{c}{\begin{array}{c} \vcenter{\hbox{\LoopPicture{solid}{densely dotted}}} \\ \cartooncaption{Change matter-sourced flux direction. ($d>1$)} \end{array}} \\
    \hline
    \\[-0.20cm]
    \Lmp(x)\equiv \vcenter{\hbox{\LoopOpPicture{densely dotted}{solid}{$x$}}} & \multicolumn{3}{c}{\begin{array}{c} \vcenter{\hbox{\LoopPicture{densely dotted}{solid}}} \\ \cartooncaption{Change matter-sourced flux direction. ($d>1$)} \end{array}} \\
    \hline
    \\[-0.20cm]
    \Sinpp(x) \equiv \vcenter{\hbox{\SinOpPicture{solid}{solid}{$x$}}} & \begin{array}{c} \SoutProjPicture{solid}{densely dotted} \\ \SinPicture{solid}{solid} \\ \cartooncaption{Create string to left.} \end{array} & + & \begin{array}{c} \SoutPicture{densely dotted}{densely dotted} \\ \LoopPicture{solid}{solid} \\ \HadronPicture{solid} \\ \cartooncaption{Join strings, detaching quark pair.} \end{array}   \\
    \hline
    \\[-0.20cm]
    \Sinmm(x) \equiv \vcenter{\hbox{\SinOpPicture{densely dotted}{densely dotted}{$x$}}} & \begin{array}{c} \SoutProjPicture{solid}{densely dotted} \\ \SinPicture{densely dotted}{densely dotted} \\ \cartooncaption{Destroy string to left.} \end{array} & + & \begin{array}{c} \HadronPicture{densely dotted} \\ \LoopPicture{densely dotted}{densely dotted} \\ \SoutPicture{solid}{solid} \\ \cartooncaption{
        Extract left string from loop flux + hadron.
    } \end{array}  \\
    \hline
    \\[-0.20cm]
    \Sinpm(x) \equiv \vcenter{\hbox{\SinOpPicture{solid}{densely dotted}{$x$}}} & \begin{array}{c} \HadronPicture{densely dotted} \\ \SinPicture{solid}{solid} \\ \cartooncaption{Replace one quark from} \\ \cartooncaption{a pair with incoming flux.} \end{array} & + & \begin{array}{c} \SinProjPicture{solid}{densely dotted} \\ \SoutPicture{densely dotted}{densely dotted} \\ \LoopPicture{solid}{solid} \\ \cartooncaption{Replace a meson-string end with gauge flux.} \end{array} \\
    \hline
    \\[-0.20cm]
    \Sinmp(x) \equiv \vcenter{\hbox{\SinOpPicture{densely dotted}{solid}{$x$}}} & \begin{array}{c} \SinProjPicture{solid}{densely dotted} \\ \LoopPicture{densely dotted}{densely dotted} \\ \SoutPicture{solid}{solid}\\ \cartooncaption{Cut a flux tube from left.} \end{array} & + & \begin{array}{c}  \SinPicture{densely dotted}{densely dotted} \\ \HadronPicture{solid} \\ \cartooncaption{
    Neutralize incoming flux by completing a pair.
  } \end{array}  \\
    \hline
    \\[-0.20cm]
    \Soutpp(x) \equiv \vcenter{\hbox{\SoutOpPicture{solid}{solid}{$x$}}} & \begin{array}{c} \SinProjPicture{solid}{densely dotted} \\ \SoutPicture{solid}{solid} \\ \cartooncaption{Create string to right.} \end{array} & + & \begin{array}{c} \SinPicture{densely dotted}{densely dotted} \\ \LoopPicture{solid}{solid} \\ \HadronPicture{solid} \\ \cartooncaption{Join strings, detaching quark pair.} \end{array}   \\
    \hline
    \\[-0.20cm]
    \Soutmm(x) \equiv \vcenter{\hbox{\SoutOpPicture{densely dotted}{densely dotted}{$x$}}} & \begin{array}{c} \SinProjPicture{solid}{densely dotted} \\ \SoutPicture{densely dotted}{densely dotted} \\ \cartooncaption{Destroy string to right.} \end{array} & + & \begin{array}{c} \HadronPicture{densely dotted} \\ \LoopPicture{densely dotted}{densely dotted} \\ \SinPicture{solid}{solid} \\ \cartooncaption{Extract right string from loop flux + hadron.} \end{array}  \\
    \hline
    \\[-0.20cm]
    \Soutpm(x) \equiv \vcenter{\hbox{\SoutOpPicture{densely dotted}{solid}{$x$}}} & \begin{array}{c} \SoutProjPicture{solid}{densely dotted} \\ \LoopPicture{densely dotted}{densely dotted} \\ \SinPicture{solid}{solid}\\ \cartooncaption{Cut a flux tube from right.} \end{array} & + & \begin{array}{c}  \SoutPicture{densely dotted}{densely dotted} \\ \HadronPicture{solid} \\ \cartooncaption{Neutralize outgoing flux by completing a pair.} \end{array} \\
    \hline
    \\[-0.20cm]
    \Soutmp(x) \equiv \vcenter{\hbox{\SoutOpPicture{solid}{densely dotted}{$x$}}} & \begin{array}{c} \HadronPicture{densely dotted} \\ \SoutPicture{solid}{solid} \\ \cartooncaption{Replace one quark from} \\ \cartooncaption{a pair with outgoing flux.} \end{array} & + & \begin{array}{c} \SoutProjPicture{solid}{densely dotted} \\ \SinPicture{densely dotted}{densely dotted} \\ \LoopPicture{solid}{solid} \\ \cartooncaption{Replace a meson-string end with gauge flux.} \end{array} \\
    \hline
    \\[-0.20cm]
    \Hpp(x)\equiv \vcenter{\hbox{\HadronOpPicture{solid}{$x$}}} & \multicolumn{3}{c}{\begin{array}{c} \vcenter{\hbox{\HadronPicture{solid}}} \\ \cartooncaption{Create a hadron.}  \end{array}} \\
    \hline
    \\[-0.20cm]
    \Hmm(x)\equiv \vcenter{\hbox{\HadronOpPicture{densely dotted}{$x$}}} & \multicolumn{3}{c}{\begin{array}{c} \vcenter{\hbox{\HadronPicture{densely dotted}}} \\ \cartooncaption{Destroy a hadron.}  \end{array}} \\
    \hline
  \end{array}
\]
\caption{Graphical representation of the (1D) loop-string-hadron operators.
Left column:
Pictorial representations of the operators.
Right column:
The operators' actions on local LSH states $\ket{n_l,n_i,n_o}_x$, in terms of the graphical rules in Sec. \ref{sec:1dDynamics}.
The graphical instructions (right) indicate how states get mapped, whereas the operator symbols (left) are just alternatives to the LSH operator themselves.
Not shown are the pure number factors required by (\ref{eq:LSHEffectiveOps}).}
  \label{tab:opactions}
\end{table*}

\section{\label{sec:LSHMultidim}Loop-string-hadron formulation: Multiple dimensions}
The prepotential formulation of pure SU(2) gauge theory on Cartesian lattices was studied in great detail in Refs.~\cite{mathurHarmonicOscillator05,
  mathurLoopStates06,
  mathurLoopApproach07,
  anishetty.mathur.eaPrepotentialFormulation10,
  raychowdhuryPrepotentialFormulation13%
}.
While it yields a local loop basis, on a Cartesian lattice that basis is overcomplete and consequently associated with a local form of Mandelstam constraints.
Solving these constraints is involved and becomes increasingly difficult in higher dimensions.

More recently, a virtual ``point splitting'' of lattice sites on square lattices \cite{raychowdhuryLowEnergy19,anishetty.sreerajMassGap18} was found to be quite fruitful because it bypasses the Mandelstam constraints and casts all constraints of the theory into the Abelian form of (\ref{eq:AGL}).
Below, the point splitting method is reviewed and how this development generalizes to higher dimensions is explained.
We additionally describe how to couple to matter in higher dimensions, giving a complete suite for describing SU(2) lattice gauge theory coupled to one flavor of staggered quarks.

\subsection{Virtual point splitting: Two dimensions}
\begin{figure}[!tbh]
\begin{center}
  \begin{tikzpicture}[scale=1.5, thick]
    %
    %
    \begin{scope}
      \draw (-1,0) node[above right] {$\bar{1}$} -- (1,0) node[above left] {$1$};
      \draw (0,-1)  node[below] {(a)}node[above right] {$\bar{2}$} -- (0,1) node[below right] {$2$};
      \filldraw [fill=white, draw=black] (0,0) circle [radius=0.125] node {$\psi$};
      \draw (-0.2,-0.2) node {$x$};
    \end{scope}
  \end{tikzpicture}
    %
    %
  \begin{tikzpicture}
    \draw (0,0) node {$\quad\Rightarrow\quad$};
    \useasboundingbox (0,-2);
  \end{tikzpicture}
    %
    %
  \begin{tikzpicture}[scale=1., thick]
    \begin{scope}
      \draw (-1,0) node[above right] {$\bar{1}$} -- (0,0);
      \draw (0,-1) node[above right] {$\bar{2}$} node[below] {(b)} -- (0,0);
      \draw (1.5,1.5) -- (2.5,1.5) node[above left] {$1$};
      \draw (1.5,1.5) -- (1.5,2.5) node[below right] {$2$};
      \draw (0,0) node[right] {$\bar{x}^{\prime}$} node [shift={(0.1,0.3)}] {\scriptsize $\bar{3}$}
      -- (1.5,1.5) node[below] {$x^{\prime}$} node [shift={(-0.3,-0.1)}] {\scriptsize $3$};
      \filldraw [fill=white, draw=black] (0.75,0.75) circle [radius=0.2]
      node {$\psi$} node [shift={(-0.35,-0.15)}] {\scriptsize $o$} node [shift={(0.15,0.35)}] {\scriptsize $i$};
      \draw (1.1,0.4) node {$x$};
      \filldraw [fill=gray, draw=gray] (0,0) circle [radius=0.06125];
      \filldraw [fill=gray, draw=gray] (1.5,1.5) circle [radius=0.06125];
    \end{scope}
  \end{tikzpicture}
\end{center}
\caption{(a) A site $x$ from a 2D lattice. (b) Virtual point splitting divides $x$ into the pair $x', \bar{x}'$, with matter living on the central site $x$.}
\label{fig:2dPointSplit}
\end{figure}
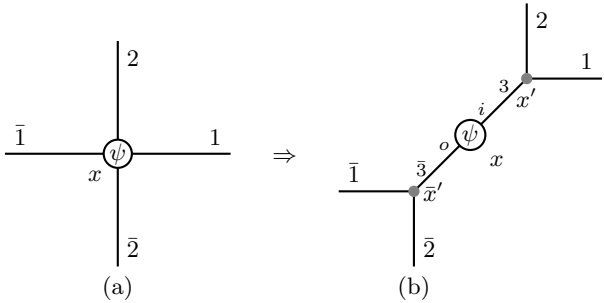

Virtual splitting of a site from a square lattice involves formally dividing each four-point vertex into a pair of three-point vertices with one shared virtual leg, as depicted in Fig.\ \ref{fig:2dPointSplit}.
It is notationally convenient to split the site by pairing the $+e_j$ ($-e_j$) directions together, to label the attached link ends 1 and 2 ($\bar{1}$ and $\bar{2}$), and to label their common vertex $x'$ ($\bar{x}'$).
As for the internal link, it is further broken into two links with an intermediate vertex that will accommodate matter.
This extra division is not needed for pure gauge theory.
Point splitting the two-dimensional (2D) square lattice results topologically in a hexagonal lattice.
One can now formulate prepotentials on this virtual hexagonal lattice as in (\ref{eq:sbElectric}) and (\ref{eq:KSLinkOp}).

The virtual links can carry gauge flux, but the flux through them is not actually counted toward $\hat{H}_E$.
The utility of the links really lies in the fact that a three point-vertex has no ambiguities in how nonintersecting SU(2) flux lines are routed through it.
Four-point vertices do suffer from such an ambiguity, and this is responsible for redundant states on the square lattice that normally have to be removed via Mandelstam constraints.
The formal hexagonal lattice still harbors redundancy, but dealing with it is significantly easier:  the relevant constraint is just another Abelian Gauss law for virtual links.

As for plaquettes, the elementary loops are indeed hexagonal plaquettes corresponding to six link operators in pure gauge theory.

For more discussion on the original pure gauge version, see Ref. \cite{raychowdhuryLowEnergy19}.

The matter field living at site $x$ is now situated between two virtual links as shown in Fig.\ \ref{fig:2dPointSplit}.
The virtual matter vertex is locally identical to a 1D lattice site (cf. Fig. \ref{fig:1dLattice}), so the on-site SU(2)-invariant operators and local Hilbert space for $x$ are handled as in one dimension.
Because hosting matter divides each virtual link into two, the plaquette operators end up involving eight sites.

The other trivalent virtual sites $x'$ and $\bar{x}'$ are gluonic vertices on the same footing as in pure gauge theory \cite{raychowdhuryLowEnergy19}, which we now review.

\subsection{\label{sec:gluonicLoopOps}SU(2) invariants: Loop operators at gluonic vertices}
At any gluonic site $x_g$ ($x'$ and $\bar{x}'$ vertices for the hexagonal lattice), links emerge in three directions and can be labeled with integers $p,q,r$ such that $p<q<r$:
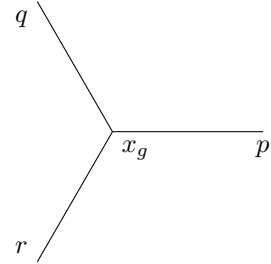
\begin{equation*}
  \begin{tikzpicture}
    \begin{scope}
      \draw ( 0, 0) -- ( 2, 0) node [below] {$p$};
      \draw ( 0, 0) -- ( {2 * cos(120)}, {2 * sin(120)} ) node [below left] {$q$};
      \draw ( 0, 0) -- ( {2 * cos(240)}, {2 * sin(240)} ) node [above left] {$r$};
      \draw ( 0, 0) node [below right] {$x_g$};
    \end{scope}
  \end{tikzpicture}
\end{equation*}
For the 2D lattice we only ever need $(p,q,r)=(1,2,3)$ ($(p,q,r)=(\bar{1},\bar{2},\bar{3})$) at the $x'$ ($\bar{x}'$) sites like in Fig. \ref{fig:2dPointSplit}, but more $(p,q,r)$ combinations will be used in three dimensions.

The attached link ends are associated with Schwinger bosons $\hat{a}^\alpha(x_g,i)$ for $i=p,q,r$.
From these doublets, one can form the complete set of SU(2) invariants at $x_g$ as given in \eqref{eq:gluonicLoopOps}--\eqref{eq:gluonicNumOps}.
\begin{itemize}
  \item \emph{Pure gauge loop operators:}.---$\mathcal{L}_{ij}^{\sigma, \sigma'}$:
\begin{subequations}
    \label{eq:gluonicLoopOps}
\begin{align}
  \Lpp[ij] &\equiv \hat{a}^\dagger_{\alpha}(i) \hat{\tilde{a}}^{\dagger}_\alpha(j)\\
  \Lpm[ij] &\equiv \hat{a}^\dagger_{\alpha}(i) \hat{a}_{\alpha}(j) \\
  \Lmp[ij] &\equiv \hat{a}_{\alpha}(i) \hat{a}^\dagger_{\alpha}(j) = (\Lpm[ij])^\dagger \\
  \Lmm[ij] &\equiv \hat{a}_{\alpha}(i) \hat{\tilde{a}}_\alpha(j) = (\Lpp[ij])^\dagger
\end{align}
\end{subequations}
  \item \emph{Gauge flux number operators}.---$\N[j]$:
    \begin{equation}
    \label{eq:gluonicNumOps}
    \N[j] = \hat{a}^\dagger_{\alpha}(j) \hat{a}_{\alpha}(j) \ .
      \end{equation}
\end{itemize}
Above, $i$ and $j$ are distinct direction indices from the set $\{p,q,r\}$.
It is easily seen from (\ref{eq:gluonicLoopOps}) that the $\mathcal{L}^{\sigma,\sigma^\prime}_{ij}$ are redundant in their link labels.
For example, $\Lpp[12] = -\Lpp[21]$.
More generally, this interdependence is summarized by
\begin{equation}
  \mathcal{L}^{\sigma,\sigma'}_{ij} = - \sigma \sigma' \mathcal{L}^{\sigma' ,\sigma}_{ji}
\end{equation}
To avoid this redundancy, we will usually deal only with the ``cyclic'' pairs $ij=(pq,qr,rs)$.

As with matter sites, the loop operators associated with gluonic vertices form a closed commutator algebra, displayed in Table \ref{tab:bigLoopCommTable}.

\subsection{\label{sec:vertex} Vertex factors and contractions}
Completely migrating from the $E$, $U$, and $\psi$ variables to LSH variables is greatly aided by furnishing a dictionary to translate spatially extended, composite operators.
Since the LSH formalism isolates on-site degrees of freedom, such operators are formed by multiplying together LSH operators from the traversed vertices.

For example, when tracing out a plaquette operator ($d\geq2$) or any other spatial Wilson line or loop, one would ordinarily encounter vertex contractions in four possible forms:
$U_{\alpha \beta} V_{\beta \gamma}$, $U_{\alpha \beta}^{\dagger} V_{\beta \gamma}^{\dagger}$, $U_{\alpha \beta} V_{\beta \gamma}^{\dagger}$, or $U_{\alpha \beta}^\dagger V_{\beta \gamma}$ (with $U$ and $V$ being link operators attached to a given vertex).
In the LSH framework, the four types of vertex contractions are naturally identified with \emph{vertex factor} matrices.
The four possible vertex contractions are expressed in \eqref{eq:RLvertex}-\eqref{eq:LLvertex}, which show the appropriate factor to assign to a vertex depending on how the links are oriented relative to the ``path'' being traced by the Wilson line:
\begin{itemize}
\item $RL$-type traversal:
\begin{align}
  &\begin{tikzpicture}[scale=2.0]
    \begin{scope}[very thick, every node/.style={sloped,allow upside down}]
      \draw[-*] (0,1) -- node {\midarrow} node[above] {$a=\binom{a_1}{a_2}$} node[below] {$\UR(a)$} (1,1);
      \draw (1,1) -- node {\midarrow}  node[above] {$b=\binom{b_1}{b_2}$} node[below] {$\UL(b)$} (2,1);
      \draw (1,1.5) node {$\rightarrow$ \small{path of Wilson line} $\rightarrow$};
    \end{scope}
  \end{tikzpicture} \nonumber \\
  \UR(a) \UL(b) =& \frac{1}{\sqrt{\Nb +1}} \left( \begin{matrix} \Lpp[ab] & \Lpm[ab]\\ - \Lmp[ab] & \Lmm[ab] \end{matrix} \right)\frac{1}{\sqrt{\Na +1}}
  \label{eq:RLvertex}
\end{align}
\item $LR$-type traversal:
\begin{align}
  &\begin{tikzpicture}[scale=2.0]
    \begin{scope}[very thick, every node/.style={sloped,allow upside down}]
      \draw (1,1.25) node {$\rightarrow$ \small{path of Wilson line} $\rightarrow$};
      \draw[-*] (0,1) -- node {\midreversearrow} node[below] {$\UL^\dagger(a)$} (1,1);
      \draw (1,1) -- node {\midreversearrow}  node[below] {$\UR^\dagger(b)$} (2,1);
    \end{scope}
  \end{tikzpicture} \nonumber \\
  \UL^\dagger(a) \UR^\dagger(b) =& \frac{1}{\sqrt{\Nb +1}} \left( \begin{matrix} -\Lmm[ab] & -\Lmp[ab]\\ \Lpm[ab] & -\Lpp[ab] \end{matrix} \right)\frac{1}{\sqrt{\Na +1}}
  \label{eq:LRvertex}
\end{align}
\item $RR$-type traversal:
\begin{align}
  &\begin{tikzpicture}[scale=2.0]
    \begin{scope}[very thick, every node/.style={sloped,allow upside down}]
      \draw[-*] (0,1) -- node {\midarrow} node[below] {$\UR(a)$} (1,1);
      \draw (1,1) -- node {\midreversearrow}  node[below] {$\UR^\dagger(b)$} (2,1);
      \draw (1,1.25) node {$\rightarrow$ \small{path of Wilson line} $\rightarrow$};
    \end{scope}
  \end{tikzpicture} \nonumber \\*
  \UR(a) \UR^\dagger(b) =& \frac{1}{\sqrt{\Nb +1}} \left( \begin{matrix} \Lpm[ab] & -\Lpp[ab]\\ \Lmm[ab] & \Lmp[ab] \end{matrix} \right)\frac{1}{\sqrt{\Na +1}}
  \label{eq:RRvertex}
\end{align}
\item $LL$-type traversal:
\begin{align}
  &\begin{tikzpicture}[scale=2.0]
    \begin{scope}[very thick, every node/.style={sloped,allow upside down}]
      \draw[-*] (0,1) -- node {\midreversearrow} node[below] {$\UL^\dagger(a)$} (1,1);
      \draw (1,1) -- node {\midarrow} node[below] {$\UL(b)$} (2,1);
      \draw (1,1.25) node {$\rightarrow$ \small{path of Wilson line} $\rightarrow$};
    \end{scope}
  \end{tikzpicture} \nonumber \\
  \UL^\dagger(a) \UL(b) =& \frac{1}{\sqrt{\Nb +1}} \left( \begin{matrix} \Lmp[ab] & -\Lmm[ab]\\ \Lpp[ab] & \Lpm[ab] \end{matrix} \right)\frac{1}{\sqrt{\Na +1}}
  \label{eq:LLvertex}
\end{align}
\end{itemize}
In the graphics, the symbols $a$ and $b$ are used to refer to the harmonic oscillator doublets encountered when ``flowing in'' to and ``flowing out'' of a vertex, respectively.
Therefore, they make use of the following singlets:
\begin{alignat*}{4}
  \Lpp[ab] &= a^\dagger \cdot \epsilon \cdot b^\dagger \qquad & \Lmm[ab] &= a \cdot \epsilon \cdot b \\
  \Lpm[ab] &= a^\dagger \cdot b & \Lmp[ab] &= a \cdot  b^\dagger \\
  \Na &= a^\dagger \cdot a & \Nb &= b^\dagger \cdot b \ .
\end{alignat*}
The vertex factor matrices can be transformed into each other by using the $\epsilon$ matrix;
schematically,
\begin{align*}
  (LR) &= (-\epsilon) (RL) (-\epsilon) \ , \\
  (RR) &= (RL) (-\epsilon) \ , \\
  (LL) &= (-\epsilon) (RL) \ ,
\end{align*}
allowing all four types to be expressed in terms of one matrix and suitable contractions with $\epsilon$.

To get a feel for how the vertex factors are used, consider an elementary plaquette in 2D pure gauge theory that follows the path $x\rightarrow x+e_2 \rightarrow x+e_1+e_2 \rightarrow x+e_1 \rightarrow x$.
[This corresponds to $U^{(21)}_{\square}(x)$ from \eqref{eq:plaquetteDef}.]
By multiplying all vertex factors together going around the path and tracing over the leftover gauge indices, the associated loop takes the schematic form
\begin{equation*}
\text{tr} \left( \left[ \mathcal{V}_{12} \right]_{x} 
\left[
  \mathcal{V}_{\bar{2}\bar{3}} \mathcal{V}_{31}
\right]_{x+e_2}
\left[
  \mathcal{V}_{\bar{1}\bar{2}}
\right]_{x+e_1+e_2}
\left[
   \mathcal{V}_{23} \mathcal{V}_{\bar{3}\bar{1}}
 \right]_{x+e_1}
 \right) \ ,
\end{equation*}
for appropriately chosen vertex factor matrices $\mathcal{V}$.
Plaquette and Wilson loop operators will be constructed explicitly in Sec. \ref{sec:2dHam} below.

To form hopping terms and general meson strings in the LSH framework, one additionally needs vertex factors at matter sites to form the string ends:
\begin{align} 
  \hat{\psi}^\dagger(x) \UL(x) &= \frac{1}{\sqrt{\NL(x) +1}} \left(\begin{matrix} \Soutpp(x) , & \Soutpm(x) \end{matrix}\right) \\*
  \hat{\psi}^\dagger(x) \UR^{\dagger}(x) &= \frac{1}{\sqrt{\NR(x) +1}} \left(\begin{matrix} \Sinmp(x) , & \Sinpp(x) \end{matrix}\right) \\*
  \UR(x) \hat{\psi}(x) &= \left(\begin{matrix}\Sinpm(x)\\ \Sinmm(x)\end{matrix}\right)\frac{1}{\sqrt{\NR(x)+1}} \\*
  \UL^{\dagger}(x) \hat{\psi}(x) &= \left(\begin{matrix}\Soutmm(x)\\ \Soutmp(x)\end{matrix}\right)\frac{1}{\sqrt{\NL(x)+1}}
\end{align}
The full meson string operator is then a path-ordered product of pure-glue vertex factors, sandwiched between two appropriate string ends.
Elementary matrix multiplication of all such factors leaves no uncontracted group indices.

\subsection{\label{sec:2dHam}The 2D Hamiltonian}
The Hamiltonian for two dimensions will now be translated into loop-string-hadron operators.
The essential difference from $d=1$ will be the presence of magnetic energy $\hat{H}_B$.

The electric energy $\hat{H}_E$ is the same as a square lattice, in the sense that contributions from all the $1$- and $2$-direction links constitute $\hat{H}_E$.
That is,
\begin{equation}
  \begin{split}
    \hat{H}_E = \frac{g_0^2}{4} \sum_{x}\sum_{j=1}^{2}& \left[ \tfrac{1}{2} \Nj({x^\prime}) \left(\tfrac{1}{2} \Nj({x^\prime}) + 1 \right)  \right. \\
    & + \left.  \tfrac{1}{2} \mathcal{N}_{\bar{j}} ({\bar{x}^{\prime}}) \left(\tfrac{1}{2} \mathcal{N}_{\bar{j}} ({\bar{x}^{\prime}}) + 1 \right)  \right] \ .
  \end{split}
  \label{eq:2dLSHHamiltonian}
\end{equation}
Note that this $d>1$ expression for $\hat{H}_E$ only involves number operators from gluonic sites.

For one dimension, $\hat{H}_M$ was translated in (\ref{eq:HMLoopString}).
The translation of $\hat{H}_M$ carries over identically to $d>1$:
\begin{equation}
  \hat{H}_M = m_0 \sum_{x} (-)^{x} (\Ni(x) + \No (x)) \ .
\end{equation}

The hopping terms in $\hat{H}_I$ were factored for one dimension in (\ref{eq:hoppingLoopString}) and (\ref{eq:denominatorsLoopString}).
The hopping terms for two dimensions are translated as follows:
The links are naturally oriented such that a typical hopping term takes the schematic form $\psi^\dagger U^\dagger U U^\dagger \psi$, where the middle $U$ comes from the original square lattice;
these orientations can be seen from the cutout of a point-split plaquette shown in Fig.\ \ref{fig:2dPlaquetteXY}.
\newcommand{\plaqscaletwodim}{5}
\newcommand{\twositedraw}[3]
{
    \begin{scope}[decoration={markings, mark=at position 0.5 with {\arrow{latex}}}]
      \coordinate (O) at (#1 , #2);
      \coordinate (A) at ($(O) + (0.75, 0.75)$);
      \coordinate (D) at ($(O) - (0.75, 0.75)$);
      \draw (A) -- ($(A) + (1,0)$) node[above left] {$1$};
      \draw (A) -- ($(A) + (0,1)$) node[below right] {$2$};
      \draw ($(D) - (1,0)$) node[above right] {$\bar{1}$} -- (D);
      \draw ($(D) - (0,1)$) node[above right] {$\bar{2}$} -- (D);
      \draw[postaction={decorate}] (O) node [shift={(-0.35,-0.1)}] {$o$}
      -- (D) node [shift={(0.1,0.4)}] {$\bar{3}$};
      \draw[postaction={decorate}] (A) node [shift={(-0.3,-0.05)}] {$3$}
      -- (O) node [shift={(0.15,0.4)}] {$i$};
      \filldraw [fill=white, draw=black] (O) circle [radius=0.2]
      node {$\psi$};
      \draw ($(O) + (0, -0.5)$) node [right] {#3};
      \filldraw [fill=gray, draw=gray] (D) circle [radius=0.06125];
      \filldraw [fill=gray, draw=gray] (A) circle [radius=0.06125];
    \end{scope}
}
  \begin{figure}[!t]
    \centering
    %
    %
    \scalebox{0.9}{
  \begin{tikzpicture}
    \begin{scope}[decoration={markings, mark=at position 0.5 with {\arrow{latex}}}]
    \twositedraw{0}{0}{$x$};
    \draw[postaction={decorate}] ($(0.75,0.75)+(1,0)$) .. controls ($(0.75,0.75)+(1,0)+(0.5,0)$) and ($(\plaqscaletwodim,0)-(0.75,0.75)-(1,0)-(0.5,0)$) .. ($(\plaqscaletwodim,0)-(0.75,0.75)-(1,0)$);
    \twositedraw{\plaqscaletwodim}{0}{$x+e_{1}$};
    \draw[postaction={decorate}] ($(\plaqscaletwodim,0)+(0.75,0.75)+(0,1)$) .. controls ($(\plaqscaletwodim,0)+(0.75,0.75)+(0,1)+(0,0.5)$) and ($(\plaqscaletwodim,\plaqscaletwodim)-(0.75,0.75)-(0,1)-(0,0.5)$) .. ($(\plaqscaletwodim,\plaqscaletwodim)-(0.75,0.75)-(0,1)$);
    \twositedraw{\plaqscaletwodim}{\plaqscaletwodim}{$x+e_{1}+e_{2}$};
    \draw[postaction={decorate}] ($(0,\plaqscaletwodim)+(0.75,0.75)+(1,0)$) .. controls ($(0,\plaqscaletwodim)+(0.75,0.75)+(1,0)+(0.5,0)$) and ($(\plaqscaletwodim,\plaqscaletwodim)-(0.75,0.75)-(1,0)-(0.5,0)$) .. ($(\plaqscaletwodim,\plaqscaletwodim)-(0.75,0.75)-(1,0)$);
    \twositedraw{0}{\plaqscaletwodim}{$x+e_{2}$};
    \draw[postaction={decorate}] ($(0,0)+(0.75,0.75)+(0,1)$) .. controls ($(0,0)+(0.75,0.75)+(0,1)+(0,0.5)$) and ($(0,\plaqscaletwodim)-(0.75,0.75)-(0,1)-(0,0.5)$) .. ($(0,\plaqscaletwodim)-(0.75,0.75)-(0,1)$);
    \end{scope}
  \end{tikzpicture}
}
\caption{Connectivity of a point-split plaquette in two dimensions. Arrows indicated flow from the ``left'' end of a link to its ``right'' end.}
    \label{fig:2dPlaquetteXY}
  \end{figure}
In a 2D Schwinger boson framework, the hopping term in the $j$ direction would be expanded as
\begin{align*}
  \psi^\dagger(x) U(x,x+e_j) \psi & (x+e_j) \rightarrow \\*
  &\psi^\dagger(x) U^{R\dagger}_{3} (x) \times \\*
  &\times U^{L\dagger}_{3} (x^\prime) U^L_{j}(x^\prime) \times \\*
  &\times U^R_{j} ( \overline{x+e_j}^{\, \prime} ) U^{R\dagger}_{3} ( \overline{x+e_j}^{\, \prime}) \times \\*
  &\times U^{L\dagger}_{3} (x+e_j) \psi(x+e_j) \ .
\end{align*}
This same object is realized in the LSH framework by stringing together the vertex factors from Sec. \ref{sec:vertex}.
The translation of the right-hand side into LSH operators is
\begin{align*}
  & \left[ \inverseRoot[\mathcal{N}_{R}+1]
  \left(\begin{matrix} \Sinmp & \Sinpp \end{matrix}\right) \right]_{x}  \times \\*
  & \times \left[ \inverseRoot[\Nj +1]
    \left( \begin{matrix} \Lmp[3j] & -\Lmm[3j]\\ \Lpp[3j] & \Lpm[3j] \end{matrix} \right)
  \inverseRoot[\mathcal{N}_3+1]
\right]_{x^\prime} \times \\*
  & \times \left[ \inverseRoot[\mathcal{N}_{\bar{3}}  +1]
    \left( \begin{matrix} \Lpm[\bar{j}\bar{3}] & -\Lpp[\bar{j}\bar{3}]\\ \Lmm[\bar{j}\bar{3}] & \Lmp[\bar{j}\bar{3}] \end{matrix} \right)
  \inverseRoot[\mathcal{N}_{\bar{j}} +1] \right]_{\overline{x+e_j}^{\,\prime}} \times \\*
  & \times \left[ \left(\begin{matrix}\Soutmm\\ \Soutmp\end{matrix}\right)
  \inverseRoot[\mathcal{N}_{L}+1] \right]_{x+e_j} \ .
\end{align*}
To express the result of matrix multiplication, it is helpful to introduce a sign function $\eta^{(2{\mathrm D})}_h$ to carry overall signs:
\begin{align}
  \eta^{(2{\mathrm D})}_h(\vec{\sigma}) & \equiv \eta^{(2{\mathrm D})}_h(\sigma_1, \sigma_2, \sigma_3) \nonumber \\
  & = (-1)^{\delta_{(\sigma_1,\sigma_2),(-,-)}} (-1)^{\delta_{(\sigma_2,\sigma_3),(+,+)}} \ .
\end{align}
Therefore, the translation into loop-string-hadron operators is
\begin{align}
  &\psi^\dagger  (x) U(x,x+e_j) \psi(x+e_j) \rightarrow \nonumber \\*
  & \sum_{\sigma_1, \sigma_2, \sigma_3 }
  \eta^{(2{\mathrm D})}_h(\vec{\sigma})
  \left[
    \inverseRoot[\mathcal{N}_{R}+1]
    \inverseRoot[\Nj+1]
  \right]_{x}
  \left[
    \inverseRoot[\mathcal{N}_{\bar{3}}+1]
  \right]_{x+e_j}
  \nonumber \\*
  & \qquad \times \left[
    \Sin^{\sigma_1, +}
    \mathcal{L}_{3j}^{\sigma_1, \sigma_2}
  \right]_{x}
  \left[
    \mathcal{L}_{\bar{j}\bar{3}}^{\sigma_2, \sigma_3}
    \Sout^{-, \sigma_3} \right]_{x+e_j} \times \nonumber \\*
    & \qquad \times
    \left[
      \inverseRoot[\mathcal{N}_{3}+1]
    \right]_{x}
    \left[
      \inverseRoot[\mathcal{N}_{\bar{j}}+1]
      \inverseRoot[\mathcal{N}_{L}+1]
    \right]_{x+e_j} \ .
    \label{eq:hopping2dInLoopString}
\end{align}

The final piece of the Hamiltonian is $\hat{H}_B$.
The plaquette operators can be translated by following the gluonic-site vertex contractions around a plaquette as described in Sec. \ref{sec:vertex}.
A generic plaquette is depicted in Fig.\ \ref{fig:2dPlaquetteXY}.
Similar to hopping terms, the result is given in terms of plaquette signs $\eta^{(2{\mathrm D})}_{p}$ stemming from the vertex contractions:
\begin{widetext}
\begin{align}
  \label{eq:2dplaquetteSigns}
  & \qquad \eta^{(2{\mathrm D})}_{p} (\vec{\sigma} ) \equiv \eta^{(2{\mathrm D})}_p ( \sigma_{1}, \sigma_{2}, \cdots , \sigma_{8} ) \nonumber \\*
  & \qquad \qquad \ \ \ \ \ =
  (-1)^{\delta_{( \sigma_{1}, \sigma_{2} ), (+,-)}}
  (-1)^{\delta_{ ( \sigma_{2}, \sigma_{3} ), (-,-)}} (-1)^{\delta_{( \sigma_{3}, \sigma_{4} ) , (+,+)}}
  (-1)^{\delta_{( \sigma_{4}, \sigma_{5} ) , (-,-)}} \times \nonumber \\*
  & \qquad \qquad \qquad \ \  \times
  (-1)^{\delta_{( \sigma_{5}, \sigma_{6} ) , (-,+)}}
  (-1)^{\delta_{( \sigma_{6}, \sigma_{7} ) , (+,+)}} (-1)^{\delta_{( \sigma_{7}, \sigma_{8} ) , (-,-)}}
  (-1)^{\delta_{( \sigma_{8}, \sigma_{1} ) , (+,+)}}\\
\label{eq:plaquette2dInLoopString}
  & -\text{tr} \left( U_{\square}^{} (x) \right) \rightarrow \nonumber \\
  &\sum_{\sigma_{1},\cdots,\sigma_{8}}\eta^{(2{\mathrm D})}_p (\vec{\sigma})
  \left[
  \inverseRoot[\mathcal{N}_{2}+1]
    \mathcal{L}_{12}^{\sigma_{7} \sigma_{8}}
  \inverseRoot[\mathcal{N}_{1}+1]
  \right]_{x}
  \times \nonumber \\
  &
  \times
  \left[
  \inverseRoot[\mathcal{N}_{\bar{3}}+1]
  \inverseRoot[\mathcal{N}_{i}+1]
  \inverseRoot[\mathcal{N}_{1}+1]
    \mathcal{L}_{\bar{2} \bar{3}}^{\sigma_{8} \sigma_{1}}
    \mathcal{L}_{o i}^{\sigma_{1} \sigma_{2}}
    \mathcal{L}_{3 1}^{\sigma_{2} \sigma_{3}}
  \inverseRoot[\mathcal{N}_{\bar{2}}+1]
  \inverseRoot[\mathcal{N}_{o}+1]
  \inverseRoot[\mathcal{N}_{3}+1]
  \right]_{x+e_2} \times \nonumber \\
  &
  \times \left[
  \inverseRoot[\mathcal{N}_{\bar{2}}+1]
  \mathcal{L}_{\bar{1}\bar{2}}^{\sigma_{3} \sigma_{4}}
  \inverseRoot[\mathcal{N}_{\bar{1}}+1]
  \right]_{x+e_1+e_2}
  \left[
  \inverseRoot[\mathcal{N}_{3}+1]
  \inverseRoot[\mathcal{N}_{o}+1]
  \inverseRoot[\mathcal{N}_{\bar{1}}+1]
    \mathcal{L}_{2 3}^{\sigma_{4} \sigma_{5}}
    \mathcal{L}_{i o}^{\sigma_{5} \sigma_{6}}
    \mathcal{L}_{\bar{3} \bar{1}}^{\sigma_{6} \sigma_{7}}
  \inverseRoot[\mathcal{N}_{2}+1]
  \inverseRoot[\mathcal{N}_{i}+1]
  \inverseRoot[\mathcal{N}_{\bar{3}}+1]
  \right]_{
x+e_1}
\end{align}
\end{widetext}

\subsection{2D dynamics on an orthonormal basis}
Following the development for one dimension, we have identified all SU(2)-invariant operators and used them to express the loop-string-hadron Hamiltonian.
Now, we introduce a basis and factorize all loop-string-hadron operators for convenience in that basis.
We then arrive at the Hamiltonian terms in their factorized form.

\subsubsection{On-site gluonic Hilbert space}
Here we summarize the local Hilbert space structure that has been studied in Ref.\,\cite{raychowdhuryLowEnergy19}.

The local vacant state is again characterized as a normalized state $\ket{0}_{x_g}$ that is annihilated by any $\mathcal{L}^{\sigma' \sigma}_{ij}$ carrying at least one minus sign.
Acting on $\ket{0}_{x_g}$, only the $\Lpp[ij]$ are nonzero and will build up the local loop Hilbert space.
A local loop state basis can be constructed following steps in analogy to the matter sites in Sec. \ref{sec:LSHOnSite}.
This local loop space is characterized by three independent linking numbers $l_{ij}$ denoting the flux flowing along three $(ij)$ directions [$(pq)$, $(qr)$, and $(rp)$].
The orthonormal basis is given by
\begin{equation}
  \ket{\ell_{pq},\ell_{qr},\ell_{rp}} \equiv \frac{(\Lpp[pq] )^{\ell_{pq}} (\Lpp[qr])^{\ell_{qr}} (\Lpp[rp])^{\ell_{rp}}}{\sqrt{\ell_{pq}!\ell_{qr}!\ell_{rp}!(\ell_{pq}+\ell_{qr}+\ell_{rp}+1)!}} \ket{0}_{x_g}
\label{lij}
\end{equation}
The number operators analogous to (\ref{eq:LSHNumberOps}) are
\begin{subequations}
  \begin{align}
    \N[pq] &\equiv \frac{1}{2}(\N[p]+\N[q]-\N[r]) \ , \\*
    \N[qr] &\equiv \frac{1}{2}(\N[q]+\N[r]-\N[p]) \ , \\*
    \N[rp] &\equiv \frac{1}{2}(\N[r]+\N[p]-\N[q]) \ .
  \end{align}
  \label{eq:loopNumberOps}
\end{subequations}
It will also be convenient to introduce
\begin{equation}
  \N[\Sigma] \equiv \N[pq]+\N[qr]+\N[rp]+1 \ .
\end{equation}

\subsubsection{Operator factorization}
Now we will factor operators at gluonic sites in such a way that their actions in the loop basis are transparent.

We have the following normalized ladder operators:
\begin{subequations}
  \begin{align}
    \hat{\Lambda}^{+}_{ij} &= \Lpp[ij] \inverseRoot[{(\N[ij] + 1)(\N[\Sigma] + 1)}]\\
    \hat{\Lambda}^{-}_{ij} &= \inverseRoot[{(\N[ij] + 1)(\N[\Sigma] + 1)}] \Lmm[ij]
  \end{align}
\end{subequations}
The operator factorizations for gluonic sites are given in terms of these normalized shift operators in Table \ref{tab:loopOpFactorizations}.

These simple local loop operators, contracted together along the links consistent with the AGL (\ref{eq:AGL}), reproduce the nonlocal loops and strings of the original theory.
Moreover, these loop operators now act more like their U(1) counterparts;
loop operators in U(1) theories shift $E$ by unit increments along an infinite tower of states, but in U(1), the normalization factor is always trivial.

\begin{table}[H]
  \centering
  \begin{tabular}{m{7cm}}
    \hline
    \begin{center}\textsc{Loop operator factorizations} \end{center}\\
    \hline
    {
      \begin{subequations}
        \label{eq:loopEffectiveOps}
        \begin{align}
          \Lpp[ij] &= \ \ \hat{\Lambda}^{+}_{ij} \sqrt{(\N[ij] + 1)(\N[\Sigma] + 1)}\\
          \Lmm[ij] &= \ \ \hat{\Lambda}^{-}_{ij} \sqrt{\N[ij]\N[\Sigma]}\\
          \Lpm[ij] &= - \hat{\Lambda}^{+}_{ki} \hat{\Lambda}^{-}_{jk} \sqrt{(\N[ki]+1)\N[jk]}\\
          \Lmp[ij] &= - \hat{\Lambda}^{-}_{ki} \hat{\Lambda}^{+}_{jk} \sqrt{\N[ki](\N[jk]+1)}\\
          ijk &= \text{$pqr$, $qrp$, or $rpq$} \nonumber
        \end{align}
      \end{subequations}
    } \\
    \hline
  \end{tabular}
  \caption{Factorization of all SU(2) singlet operators at a gluonic site.}
  \label{tab:loopOpFactorizations}
\end{table}

\subsubsection{Global Hilbert space construction in two dimensions}
As in one dimension, the 2D lattice vacant state is characterized as that state on which
\begin{align*}
  \Ni(x) \ket{0} = \No(x) \ket{0} = \Nl(x) \ket{0} &= 0 \quad \text{for all $x$},\\
  \N[12](x_g) \ket{0} = \N[23](x_g) \ket{0} = \N[31](x_g) \ket{0} &= 0 \quad \text{for all $x_g$}.
\end{align*}
In two dimensions, we call a site $x$ even (odd) if $x_1+x_2$ is even (odd).
The strong-coupling vacuum $\ket{v}$ is then defined analogously to one dimension:
\begin{align*}
  \Nl(x) \ket{v} &= 0\\
  (\Ni(x)+\No(x))\ket{v} &= 0 \quad \text{for even $x$}\\
  (\Ni(x)+\No(x))\ket{v} &= 2 \ket{v} \quad \text{for odd $x$}\\
  \Sin^{\pm, -}(x) \ket{v} = \Sout^{-, \pm}(x) \ket{v} &= 0 \quad \text{for even $x$}\\
  \Sin^{\pm, +}(x) \ket{v} = \Sout^{+, \pm}(x) \ket{v} &= 0 \quad \text{for odd $x$}\\
  (\N[12](x_g) + \N[23](x_g) + \N[31](x_g) ) \ket{v} &= 0 \ .
\end{align*}

The 2D lattice Hilbert space structure is as follows:
\begin{enumerate}
  \item The gluonic sites $x^\prime,\bar{x}^{\prime}$ have only loop states $|l_{12},l_{23},l_{31}\rangle_{x'/\bar{x}^{\prime}}$, being treated identically as in pure gauge theory.
  \item The matter sites $x$ have loop and quark states $\ket{n_l,n_i,n_o}$, being structurally identical to sites with matter in one dimension.
    Physical quark degrees of freedom still require an ordering in order to treat lattice basis states as tensor products:
    We denote the physical quark modes associated with $x$ by
    \begin{equation*}
    \begin{split}
      \chi^\dagger_{q}(x_1,x_2) \quad q=& \ 0,1\\
      0 \sim \text{in}, &\quad 1\sim \text{out}
    \end{split}
    \end{equation*}
    and order fermions by the map
    \begin{equation*}
      f(q,x_1,x_2) \rightarrow q + 2(x_1 + L_x  x_2)
    \end{equation*}
    Generalizing the 1D convention, basis states are defined to have $\chi^\dagger_{q}(x)$'s being applied on $\ket{0}$ from greatest $f(q,x_1,x_2)$ to least.
  \item The Abelian Gauss laws along the three directions of the hexagonal lattice are
\begin{subequations}
\begin{align}
  n_1(x^{\prime}) &= n_{\bar 1}(\overline{x+e_1}^{\,\prime}) \ , \\
  n_2(x^{\prime}) &= n_{\bar 2}(\overline{x+e_2}^{\,\prime}) \ , \\
  n_3 (x^\prime)  &=  n_l(x) + n_i(x) [1-n_o(x)] \\
  n_{\bar{3}} (\bar{x}^{\prime}) &= n_l(x) + n_o(x)[1-n_i(x)]
\end{align}
\end{subequations}
\end{enumerate}

\subsubsection{2D dynamics of loop-string-hadron states}
Matrix elements of the Hamiltonian with respect to the global basis described above are straightforward to obtain by using the operator factorizations at gluonic sites \eqref{eq:loopEffectiveOps} and at matter sites \eqref{eq:LSHEffectiveOps} in place of the LSH operators appearing in the various parts of the 2D Hamiltonian given in \eqref{eq:2dLSHHamiltonian}--\eqref{eq:plaquette2dInLoopString}.

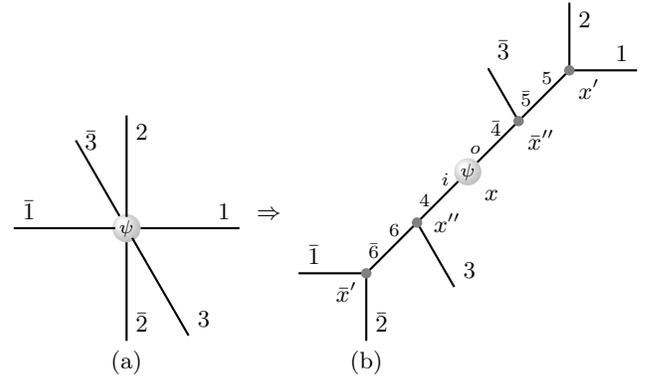
\begin{figure}[H]
\begin{center}
  \begin{tikzpicture}[scale=1.5, thick]
    %
    %
    \begin{scope}
      \draw (0,0) -- (1,0) node[above left] {$1$};
      \draw (0,-1) node[below] {(a)} node[above right] {$\bar{2}$} -- (0,0);
      \draw (0,0) -- ({0.9 * cos(120)}, {0.9 * sin(120)}) node[right] {$\bar{3}$};
      \fill [fill=white] (0,0) circle [radius=0.125];
      \shade[ball color = gray!25, opacity = 0.4] (0,0) circle (.125);
      \draw (0,0) node {\scriptsize{$\psi$}};
      \draw (-1,0) node[above right] {$\bar{1}$} -- (-0.1,0);
      \draw (0,0.1) -- (0,1) node[below right] {$2$};
      \draw ({1.1 * cos(300)}, {1.1 * sin(300)})  node[above right] {$3$} -- ({0.1 * cos(300)}, {0.1 * sin(300)});
    \end{scope}
  \end{tikzpicture}
    %
    %
  \begin{tikzpicture}
    \draw (0,0) node {$\Rightarrow$};
    \path (0,-2) node [below] {\, };
  \end{tikzpicture}
    %
    %
  \begin{tikzpicture}[scale=0.9, thick]
    \begin{scope}
      \coordinate (A) at (1.5, 1.5);
      \coordinate (B) at (0.75, 0.75);
      \coordinate (C) at (-0.75, -0.75);
      \coordinate (D) at (-1.5, -1.5);
      \draw (A) -- ($(A)+(1,0)$) node[above left] {$1$};
      \draw (A) -- ($(A)+(0,1)$) node[below right] {$2$};
      \draw ($(D)+(-1,0)$) node[above right] {$\bar{1}$} -- (D);
      \draw ($(D)+(0,-1)$) node[above right] {$\bar{2}$} node[below] {(b)} -- (D);
      \draw (D) node [shift={(0.1,0.3)}] {\scriptsize $\bar{6}$}
      -- (C) node [shift={(-0.3,-0.1)}] {\scriptsize $6$};
      \draw (C) node [shift={(0.1,0.3)}] {\scriptsize $4$}
      -- (0,0) node [shift={(-0.3,-0.1)}] {\scriptsize $i$};
      \draw (0,0) node [shift={(0.1,0.3)}] {\scriptsize $o$}
      -- (B) node [shift={(-0.3,-0.1)}] {\scriptsize $\bar{4}$};
      \draw (B) node [shift={(0.1,0.3)}] {\scriptsize $\bar{5}$}
      -- (A) node [shift={(-0.3,-0.1)}] {\scriptsize $5$};
      \draw ($(C)+({1.1 * cos(300)}, {1.1 * sin(300)})$) node[above right] {$3$} -- (C);
      \draw ($(B)+({0.9 * cos(120)}, {0.9 * sin(120)})$) node[above right] {$\bar{3}$} -- (B);
      \fill [fill=white] (0, 0) circle [radius=0.2];
      \shade[ball color = gray!25, opacity = 0.4] (0,0) circle (0.2);
      \draw (0,0) node {\scriptsize{$\psi$}};
      \draw (0.35,-0.35) node {$x$};
      \filldraw [fill=gray, draw=gray] (D) circle [radius=0.06125] node[below left] {$\bar{x}^{\prime}$};
      \filldraw [fill=gray, draw=gray] (C) circle [radius=0.06125] node [shift={(0.1,0)}, right] {$x^{\prime\prime}$};
      \filldraw [fill=gray, draw=gray] (B) circle [radius=0.06125] node [below right] {$\bar{x}^{\prime\prime}$};
      \filldraw [fill=gray, draw=gray] (A) circle [radius=0.06125] node [below right] {$x^{\prime}$};
    \end{scope}
  \end{tikzpicture}
\end{center}
\caption{(a) A site $x$ from a 3D cubic lattice. (b) The lattice site is virtually split into sites $x, x', \bar{x}', x'', \bar{x}''$ connected by intermediate links along internal directions. Matter continues to live at site $x$.}
\label{fig:3dPointSplit}
\end{figure}
\section{\label{sec:LSH3dim}3D lattice with matter}
The same scheme of point splitting used in two dimensions can be continued up to arbitrary spatial dimensionality $d$.
As shown in Fig. \ref{fig:3dPointSplit}, point splitting in three dimensions results in four gluonic three-point vertices, while matter is accommodated by creating a fifth virtual site along one of the internal lines.

As in two dimensions, the local loop Hilbert space at gluonic vertices remains identical (three linking numbers) to the pure gauge theory.
Matter is incorporated by dividing one virtual link (the $4-\bar{4}$ in Fig. \ref{fig:3dPointSplit}) into two;
the Hilbert space at the virtual matter site has two string numbers and one loop number, again with the same structure used in one dimension.
The modified Abelian Gauss laws on the three-dimensional (3D) lattice are
\begin{subequations}
\begin{align}
  (\Nj (x_g) - \mathcal{N}_{\bar{j}} (x_g+e_j) )\ket{\text{phys}} &= 0 \ , \ \ (j=1,2,3)\\*
  (\N[5](x') - \N[\bar{5}](\bar{x}'') ) \ket{\text{phys}} &= 0 \ , \quad \\*
  (\N[6](x'') - \N[\bar{6}](\bar{x}') ) \ket{\text{phys}} &= 0 \ , \quad \\*
  (\NL (x) - \mathcal{N}_{\bar{4}} (\bar{x}'') ) \ket{\text{phys}} &= 0  \ , \\*
  (\NR (x) - \mathcal{N}_{4} (x'') ) \ket{\text{phys}} &= 0  \ .
\end{align}
\end{subequations}

The Hamiltonian for three dimensions has no conceptually new objects---the terms present in two dimensions are just more numerous.
The explicit decompositions in the LSH framework do, however, have more operator factors and there is less notational symmetry shared by all three spatial directions.
Below, we provide a summary of the operators in the Hamiltonian of 3D SU(2) gauge theory with one staggered quark flavor.

For the interaction $H_I$, the hopping terms are given below, using sign factors $\eta^{(3{\mathrm D})}_{h,j}$ analogous to two dimensions:
  \begin{widetext}
\begin{align}
    \label{eq:hopping3dSignsXY}
    \eta^{(3{\mathrm D})}_{h,j} (\vec{\sigma} ) \equiv\ \hspace{1.1cm} & \hspace{-1.1cm}  \eta^{(3{\mathrm D})}_{h,j} ( \sigma_{1}, \sigma_{2}, \cdots , \sigma_{5} )  \qquad \qquad \qquad \qquad \qquad \qquad \qquad \qquad \qquad \qquad \quad\ (j=1,2) \nonumber \\*
  =\ \hspace{1.1cm} & \hspace{-1.1cm} (-1)^{\delta_{( \sigma_{1}, \sigma_{2} ), (+,+)}}
  (-1)^{\delta_{ ( \sigma_{2}, \sigma_{3} ), (-,-)}}
  (-1)^{\delta_{( \sigma_{3}, \sigma_{4} ) , (+,+)}}
  (-1)^{\delta_{( \sigma_{4}, \sigma_{5} ) , (-,-)}}\\*
    \label{eq:hopping3dSignsZ}
    \eta^{(3{\mathrm D})}_{h,3} (\vec{\sigma} ) \equiv\ \hspace{1.1cm} & \hspace{-1.1cm}  \eta^{(3{\mathrm D})}_{h,3} ( \sigma_{1}, \sigma_{2}, \sigma_{3} ) \nonumber  \\*
  =\ \hspace{1.1cm} & \hspace{-1.1cm} (-1)^{\delta_{( \sigma_{1}, \sigma_{2} ), (-,-)}}
  (-1)^{\delta_{ ( \sigma_{2}, \sigma_{3} ), (+,+)}} \\
  \psi^{\dagger} (x) U(x,x+e_j) \psi(x+e_j)  \rightarrow \sum_{\sigma_1,\cdots,\sigma_5} & \eta^{(3{\mathrm D})}_{h,j} (\sigma_1, \sigma_2, \sigma_3, \sigma_4, \sigma_5) \times \quad \quad \quad \quad \quad \quad \quad \quad \qquad \qquad \qquad \qquad (j=1,2) \nonumber \\
  \times & \left[
  \inverseRoot[\mathcal{N}_{L}+1]
  \inverseRoot[\mathcal{N}_{\bar{5}}+1]
  \inverseRoot[\mathcal{N}_{j}+1]
    \Sout^{+,\sigma_1}
    \mathcal{L}_{\bar{4} \bar{5}}^{\sigma_1 \sigma_2 }
    \mathcal{L}_{5 j}^{\sigma_2 \sigma_3}
  \inverseRoot[\mathcal{N}_{\bar{4}}+1]
  \inverseRoot[\mathcal{N}_{5}+1]
  \right]_{x} \times \nonumber \\
  \times & \left[
  \inverseRoot[\mathcal{N}_{\bar{6}}+1]
  \inverseRoot[\mathcal{N}_{4}+1]
    \mathcal{L}_{\bar{j} \bar{6}}^{\sigma_3 \sigma_4}
    \mathcal{L}_{6 4}^{\sigma_4 \sigma_5 }
    \Sin^{\sigma_5,-}
  \inverseRoot[\mathcal{N}_{\bar{j}}+1]
  \inverseRoot[\mathcal{N}_{6}+1]
  \inverseRoot[\mathcal{N}_{R}+1] \right]_{x+e_j} \\
  & \nonumber \\
  \psi^{\dagger} (x) U(x,x+e_3) \psi(x+e_3)  \rightarrow \sum_{\sigma_1,\sigma_2 ,\sigma_3} & \eta^{(3{\mathrm D})}_{h,3} (\sigma_1, \sigma_2, \sigma_3 )  \left[
  \inverseRoot[\mathcal{N}_{R}+1]
  \inverseRoot[\mathcal{N}_{3}+1]
    \Sin^{\sigma_1,+}
    \mathcal{L}_{4 3}^{\sigma_1 \sigma_2 }
  \inverseRoot[\mathcal{N}_{4}+1]
  \right]_{x} \times \nonumber \\*
  & \quad \quad \quad \quad \quad \ \times \left[
  \inverseRoot[\mathcal{N}_{4}+1]
    \mathcal{L}_{\bar{3} 4}^{\sigma_4 \sigma_5 }
    \Sout^{-,\sigma_5}
  \inverseRoot[\mathcal{N}_{\bar{3}}+1]
  \inverseRoot[\mathcal{N}_{L}+1] \right]_{x+e_3}
\end{align}
\end{widetext}

Turning to the magnetic energy $\hat{H}_B$, each plaquette trace is a contraction of LSH operators, with their three orientations being displayed in Figs. \ref{fig:3dPlaquetteXY}--\ref{fig:3dPlaquetteZX}.
As in two dimensions, there are sign factors to keep track of from the vertex contractions.
All three plaquette operators can be expressed using a single sign function $\eta^{(3{\mathrm D})}_p (\vec{\sigma} )$, as given in (\ref{eq:PlaquetteSigns3d}),
and the formulas for the $\hat{H}_B$ contributions are given in \eqref{eq:PlaquetteDecomp3dXY} and \eqref{eq:PlaquetteDecomp3djZ}:
\begin{widetext}
\begin{align}
  \label{eq:PlaquetteSigns3d}
  \hspace{3.75cm} & \hspace{-3.75cm} \eta^{(3{\mathrm D})}_p (\vec{\sigma} ) \equiv  \eta^{(3{\mathrm D})}_p ( \sigma_{1}, \sigma_{2}, \cdots ,\sigma_{12} ) \nonumber \\*
  \hspace{3.75cm} & \hspace{-2.5cm}=
  (-1)^{\delta_{( \sigma_{1}, \sigma_{2} ), (-,+)}}
  (-1)^{\delta_{ ( \sigma_{2}, \sigma_{3} ), (+,+)}}
  (-1)^{\delta_{( \sigma_{3}, \sigma_{4} ) , (-,-)}}
  (-1)^{\delta_{( \sigma_{4}, \sigma_{5} ) , (+,+)}}
  (-1)^{\delta_{( \sigma_{5}, \sigma_{6} ) , (-,-)}}
  (-1)^{\delta_{( \sigma_{6}, \sigma_{7} ) , (+,+)}} \times \nonumber \\*
  \hspace{3.75cm} & \hspace{-2.5cm}
  \quad \times
  (-1)^{\delta_{( \sigma_{7}, \sigma_{8} ) , (+,-)}}
  (-1)^{\delta_{( \sigma_{8}, \sigma_{9} ) , (-,-)}}
  (-1)^{\delta_{( \sigma_{9}, \sigma_{10} ) , (+,+)}}
  (-1)^{\delta_{( \sigma_{10}, \sigma_{11} ) , (-,-)}}
  (-1)^{\delta_{( \sigma_{11}, \sigma_{12} ) , (+,+)}}
  (-1)^{\delta_{( \sigma_{12}, \sigma_{1} ) , (-,-)} } \\
  -\text{tr} \left( U_{\square}^{(12)} (x) \right) \rightarrow  \sum_{\sigma_{1},\cdots,\sigma_{12}} & \eta^{(3{\mathrm D})}_{p} (\vec{\sigma}) \left[
  \inverseRoot[\mathcal{N}_{2}+1]
    \mathcal{L}_{1 2}^{\sigma_{10} \sigma_{11}}
  \inverseRoot[\mathcal{N}_{1}+1]
  \right]_{x}
  \times \nonumber \\*
  & \times \left[
  \inverseRoot[\mathcal{N}_{\bar{6}}+1]
  \inverseRoot[\mathcal{N}_{4}+1]
  \inverseRoot[\mathcal{N}_{o}+1]
  \inverseRoot[\mathcal{N}_{\bar{5}}+1]
  \inverseRoot[\mathcal{N}_{1}+1]
    \mathcal{L}_{\bar{2} \bar{6}}^{\sigma_{11} \sigma_{12}}
    \mathcal{L}_{6 4}^{\sigma_{12} \sigma_{1}}
    \mathcal{L}_{i o}^{\sigma_{1} \sigma_{2}}
    \mathcal{L}_{\bar{4} \bar{5}}^{\sigma_{2} \sigma_{3}}
    \mathcal{L}_{5 1}^{\sigma_{3} \sigma_{4}}
  \right.
  \nonumber \\*
  & \qquad \qquad \qquad \qquad \qquad \qquad \qquad \quad \quad \left.
  \inverseRoot[\mathcal{N}_{\bar{2}}+1]
  \inverseRoot[\mathcal{N}_{6}+1]
  \inverseRoot[\mathcal{N}_{i}+1]
  \inverseRoot[\mathcal{N}_{\bar{4}}+1]
  \inverseRoot[\mathcal{N}_{5}+1]
  \right]_{x+e_2}
  \times \nonumber \\*
  &
  \times \left[
  \inverseRoot[\mathcal{N}_{\bar{2}}+1]
    \mathcal{L}_{\bar{1} \bar{2}}^{\sigma_{4} \sigma_{5}}
  \inverseRoot[\mathcal{N}_{\bar{1}}+1]
  \right]_{x+e_1+e_2}
  \times \nonumber \\
  & \times \left[
  \inverseRoot[\mathcal{N}_{5}+1]
  \inverseRoot[\mathcal{N}_{\bar{4}}+1]
  \inverseRoot[\mathcal{N}_{i}+1]
  \inverseRoot[\mathcal{N}_{6}+1]
  \inverseRoot[\mathcal{N}_{\bar{1}}+1]
    \mathcal{L}_{2 5}^{\sigma_{5} \sigma_{6}}
    \mathcal{L}_{\bar{5} \bar{4}}^{\sigma_{6} \sigma_{7}}
    \mathcal{L}_{o i}^{\sigma_{7} \sigma_{8}}
    \mathcal{L}_{4 6}^{\sigma_{8} \sigma_{9}}
    \mathcal{L}_{\bar{6} \bar{1}}^{\sigma_{9} \sigma_{10}}
  \right.
  \nonumber \\
  & \qquad \qquad \qquad \qquad \qquad \qquad \qquad \quad \quad \left.
  \inverseRoot[\mathcal{N}_{2}+1]
  \inverseRoot[\mathcal{N}_{\bar{5}}+1]
  \inverseRoot[\mathcal{N}_{o}+1]
  \inverseRoot[\mathcal{N}_{4}+1]
  \inverseRoot[\mathcal{N}_{\bar{6}}+1]
  \right]_{x+e_1}
  \label{eq:PlaquetteDecomp3dXY}
\end{align}
\begin{align}
  & -\text{tr} \left( U_{\square}^{(j3)} (x) \right) \rightarrow \qquad \qquad \qquad \qquad \qquad \qquad \qquad \qquad \qquad \qquad \qquad \qquad \qquad \qquad \qquad \qquad \qquad \quad (j=1,2) \nonumber \\*
  & \sum_{\sigma_{1},\cdots,\sigma_{12}} \eta^{(3{\mathrm D})}_p (\vec{\sigma}) \times \nonumber \\*
  & \times
  \left[
  \inverseRoot[\mathcal{N}_{\bar{6}}+1]
  \inverseRoot[\mathcal{N}_{3}+1]
    \mathcal{L}_{\bar{j} \bar{6}}^{\sigma_{4} \sigma_{5}}
    \mathcal{L}_{6 3}^{\sigma_{5} \sigma_{6}}
  \inverseRoot[\mathcal{N}_{\bar{j}}+1]
  \inverseRoot[\mathcal{N}_{6}+1]
  \right]_{x+e_j}
  \times \nonumber \\
  & \times
  \left[
  \inverseRoot[\mathcal{N}_{\bar{4}}+1]
  \inverseRoot[\mathcal{N}_{i}+1]
  \inverseRoot[\mathcal{N}_{6}+1]
  \inverseRoot[\mathcal{N}_{\bar{j}}+1]
    \mathcal{L}_{\bar{3} \bar{4}}^{\sigma_{6} \sigma_{7}}
    \mathcal{L}_{o i}^{\sigma_{7} \sigma_{8}}
    \mathcal{L}_{4 6}^{\sigma_{8} \sigma_{9}}
    \mathcal{L}_{\bar{6} \bar{j}}^{\sigma_{9} \sigma_{10}}
  \inverseRoot[\mathcal{N}_{\bar{3}}+1]
  \inverseRoot[\mathcal{N}_{o}+1]
  \inverseRoot[\mathcal{N}_{4}+1]
  \inverseRoot[\mathcal{N}_{\bar{6}}+1]
  \right]_{x+e_j+e_3}
  \times \nonumber \\
  &
  \times \left[
  \inverseRoot[\mathcal{N}_{5}+1]
  \inverseRoot[\mathcal{N}_{\bar{3}}+1]
    \mathcal{L}_{j 5}^{\sigma_{10} \sigma_{11}}
    \mathcal{L}_{\bar{5} \bar{3}}^{\sigma_{11} \sigma_{12}}
  \inverseRoot[\mathcal{N}_{j}+1]
  \inverseRoot[\mathcal{N}_{\bar{5}}+1]
  \right]_{x+e_3}
  \times \nonumber \\
  & \times \left[
  \inverseRoot[\mathcal{N}_{4}+1]
  \inverseRoot[\mathcal{N}_{o}+1]
  \inverseRoot[\mathcal{N}_{\bar{5}}+1]
  \inverseRoot[\mathcal{N}_{j}+1]
    \mathcal{L}_{3 4}^{\sigma_{12} \sigma_{1}}
    \mathcal{L}_{i o}^{\sigma_{1} \sigma_{2}}
    \mathcal{L}_{\bar{4} \bar{5}}^{\sigma_{2} \sigma_{3}}
    \mathcal{L}_{5 j}^{\sigma_{3} \sigma_{4}}
  \inverseRoot[\mathcal{N}_{3}+1]
  \inverseRoot[\mathcal{N}_{i}+1]
  \inverseRoot[\mathcal{N}_{\bar{4}}+1]
  \inverseRoot[\mathcal{N}_{5}+1]
  \right]_{x}
  \label{eq:PlaquetteDecomp3djZ}
\end{align}
\end{widetext}
\newcommand{\threesitedraw}[3]
{
    \begin{scope}[decoration={markings, mark=at position 0.5 with {\arrow{latex}}}]
      \coordinate (A) at (#1 + 1.5, #2 + 1.5);
      \coordinate (B) at (#1 + 0.75, #2 + 0.75);
      \coordinate (C) at (#1 + -0.75, #2 + -0.75);
      \coordinate (D) at (#1 + -1.5, #2 + -1.5);
      \draw (A) -- ($(A)+(1,0)$) node[above left] {$1$};
      \draw (A) -- ($(A)+(0,1)$) node[below right] {$2$};
      \draw ($(D)+(-1,0)$) node[above right] {$\bar{1}$} -- (D);
      \draw ($(D)+(0,-1)$) node[above right] {$\bar{2}$} -- (D);
      \draw[postaction={decorate}] (C) node [shift={(-0.25,-0.05)}] {\scriptsize $6$}
      -- (D) node [shift={(0.05,0.3)}] {\scriptsize $\bar{6}$};
      \draw[postaction={decorate}] (C) node [shift={(0.05,0.3)}] {\scriptsize $4$}
      -- (#1 + 0, #2 + 0) node [shift={(-0.3,-0.1)}] {\scriptsize $i$};
      \draw[postaction={decorate}] (#1 + 0, #2 + 0) node [shift={(0.1,0.3)}] {\scriptsize $o$}
      -- (B) node [shift={(-0.25,-0.05)}] {\scriptsize $\bar{4}$};
      \draw[postaction={decorate}] (A) node [shift={(-0.25,-0.05)}] {\scriptsize $5$}
      -- (B) node [shift={(0.05,0.3)}] {\scriptsize $\bar{5}$};
      \draw ($(C)+({1.0 * cos(300)}, {1.0 * sin(300)})$) node[above right] {$3$} -- (C);
      \draw ($(B)+({1.0 * cos(120)}, {1.0 * sin(120)})$) node[above right] {$\bar{3}$} -- (B);
      \filldraw [fill=white, draw=black] (#1 + 0, #2 + 0) circle [radius=0.2]
      node {\scriptsize $\psi$};
      \draw (#1 + 0.0, #2 + -0.5) node [right] {#3};
      \filldraw [fill=gray, draw=gray] (D) circle [radius=0.06125];
      \filldraw [fill=gray, draw=gray] (C) circle [radius=0.06125];
      \filldraw [fill=gray, draw=gray] (B) circle [radius=0.06125];
      \filldraw [fill=gray, draw=gray] (A) circle [radius=0.06125];
    \end{scope}
}
\newcommand{\plaqscalethreedim}{8}
  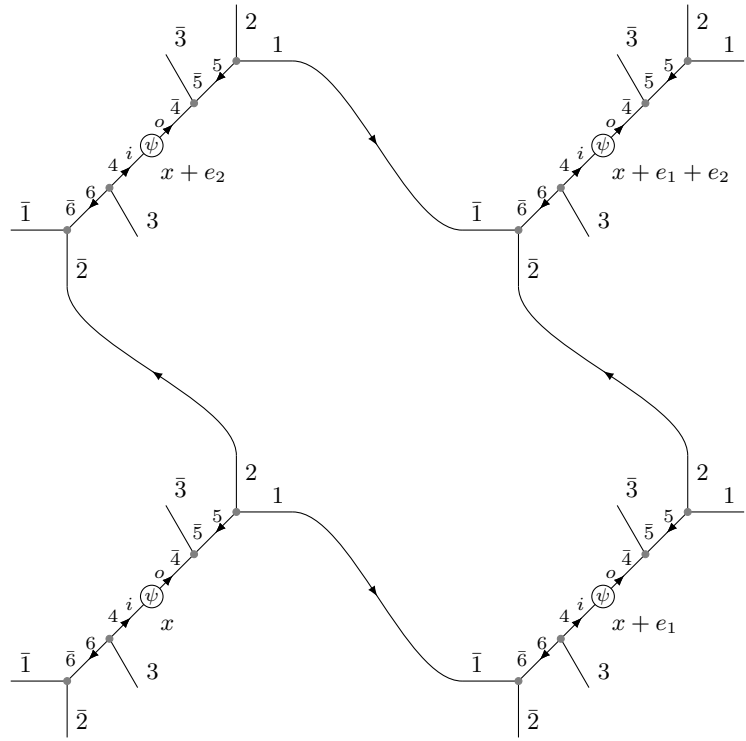
\begin{figure*}[!h]
    \centering
    %
    %
  \begin{tikzpicture}[scale=0.75]
    \begin{scope}[decoration={markings, mark=at position 0.5 with {\arrow{latex}}}]
    \threesitedraw{0}{0}{$x$};
    \draw[postaction={decorate}] (2.5, 1.5) .. controls (3.5, 1.5) and (\plaqscalethreedim -3.5, -1.5) .. (\plaqscalethreedim -2.5, -1.5);
    \threesitedraw{\plaqscalethreedim}{0}{$x+e_{1}$};
    \draw[postaction={decorate}] (\plaqscalethreedim +1.5, 2.5) .. controls (\plaqscalethreedim +1.5, 3.5) and (\plaqscalethreedim -1.5,\plaqscalethreedim - 3.5) .. (\plaqscalethreedim -1.5,\plaqscalethreedim - 2.5);
    \threesitedraw{\plaqscalethreedim}{\plaqscalethreedim}{$x+e_{1}+e_{2}$};
    \draw[postaction={decorate}] (2.5, \plaqscalethreedim + 1.5) .. controls (3.5, \plaqscalethreedim + 1.5) and (\plaqscalethreedim -3.5, \plaqscalethreedim + -1.5) .. (\plaqscalethreedim -2.5, \plaqscalethreedim + -1.5);
    \threesitedraw{0}{\plaqscalethreedim}{$x+e_{2}$};
    \draw[postaction={decorate}] (1.5, 2.5) .. controls (1.5, 3.5) and (-1.5,\plaqscalethreedim - 3.5) .. (-1.5,\plaqscalethreedim - 2.5);
    \end{scope}
  \end{tikzpicture}
  \caption{Connectivity of a $xy$-plaquette in three dimensions.}
    \label{fig:3dPlaquetteXY}
  \end{figure*}
  \begin{figure*}[htb]
    \centering
    %
    %
    \begin{tikzpicture}[scale=.75]
    \begin{scope}[decoration={markings, mark=at position 0.5 with {\arrow{latex}}}]
    \threesitedraw{0}{0}{$x$};
    \draw[postaction={decorate}]
    ($(0,0) - (0.75, 0.75) - ({cos(120)}, {sin(120)})$)
    .. controls ($(0,0) - (0.75, 0.75) - ({2.5 * cos(120)}, {2.5 * sin(120)})$)
    and ($(\plaqscalethreedim ,0) + (0.75, 0.75) +({2.5 * cos(120)}, {2.5 * sin(120)})$)
    .. ($(\plaqscalethreedim ,0) + (0.75, 0.75) + ({cos(120)}, {sin(120)})$);
    \threesitedraw{\plaqscalethreedim}{0}{$x+e_{3}$};
    \draw[postaction={decorate}] (\plaqscalethreedim +1.5, 2.5) .. controls (\plaqscalethreedim +1.5, 3.5) and (\plaqscalethreedim -1.5,\plaqscalethreedim - 3.5) .. (\plaqscalethreedim -1.5,\plaqscalethreedim - 2.5);
    \threesitedraw{\plaqscalethreedim}{\plaqscalethreedim}{$x+e_{2}+e_{3}$};
    \draw[postaction={decorate}]
    ($(0,\plaqscalethreedim) - (0.75, 0.75) - ({cos(120)}, {sin(120)})$)
    .. controls ($(0,\plaqscalethreedim) - (0.75, 0.75) - ({2.5 * cos(120)}, {2.5 * sin(120)})$)
    and ($(\plaqscalethreedim ,\plaqscalethreedim) + (0.75, 0.75) +({2.5 * cos(120)}, {2.5 * sin(120)})$)
    .. ($(\plaqscalethreedim ,\plaqscalethreedim) + (0.75, 0.75) + ({cos(120)}, {sin(120)})$);
    \threesitedraw{0}{\plaqscalethreedim}{$x+e_{2}$};
    \draw[postaction={decorate}] (1.5, 2.5) .. controls (1.5, 3.5) and (-1.5,\plaqscalethreedim - 3.5) .. (-1.5,\plaqscalethreedim - 2.5);
  \end{scope}
  \end{tikzpicture}
  \caption{Connectivity of a $yz$-plaquette in three dimensions.}
    \label{fig:3dPlaquetteYZ}
  \end{figure*}
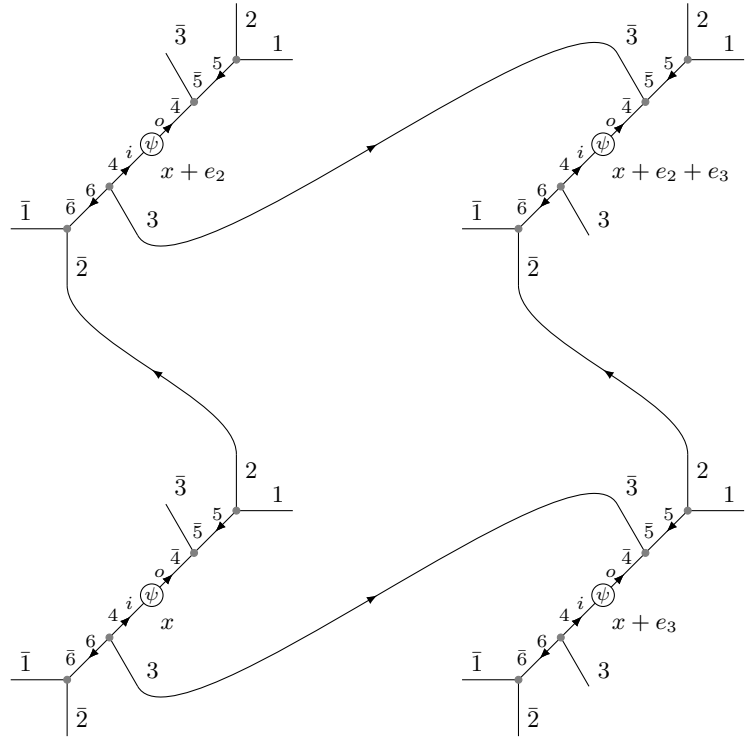
  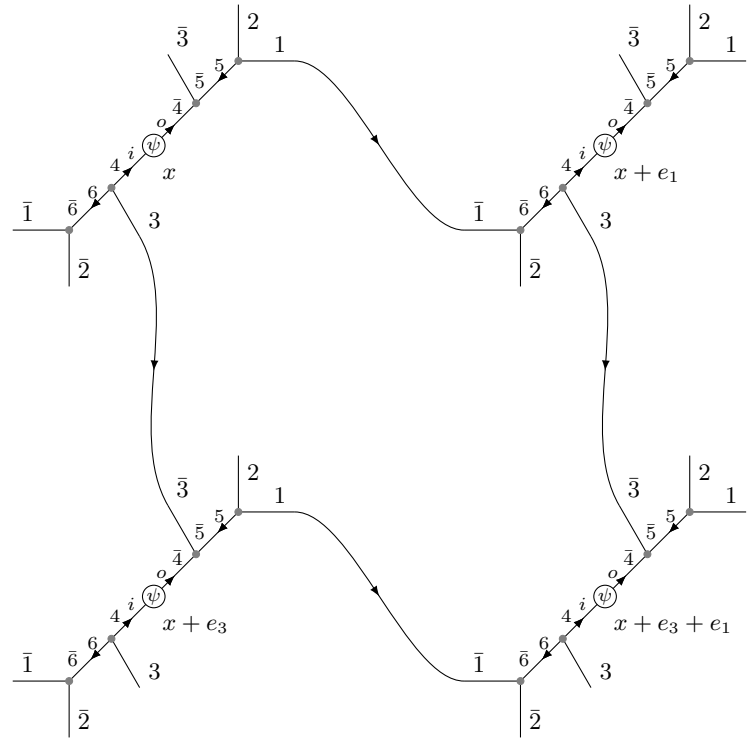
\begin{figure*}[!h]
    \centering
    %
    %
  \begin{tikzpicture}[scale=.75]
    \begin{scope}[decoration={markings, mark=at position 0.5 with {\arrow{latex}}}]
    \threesitedraw{0}{0}{$x+e_3$};
    \draw[postaction={decorate}] (2.5, 1.5) .. controls (3.5, 1.5) and (\plaqscalethreedim -3.5, -1.5) .. (\plaqscalethreedim -2.5, -1.5);
    \threesitedraw{\plaqscalethreedim}{0}{$x+e_{3}+e_{1}$};
    \draw[postaction={decorate}] ($(\plaqscalethreedim, \plaqscalethreedim) - (0.75, 0.75) - ({cos(120)}, {sin(120)})$)
    .. controls ($(\plaqscalethreedim, \plaqscalethreedim) - (0.75, 0.75) - ({2.5 * cos(120)}, {2.5 * sin(120)})$)
    and ($(\plaqscalethreedim ,0) + (0.75, 0.75) +({2.5 * cos(120)}, {2.5 * sin(120)})$)
    .. ($(\plaqscalethreedim ,0) + (0.75, 0.75) + ({cos(120)}, {sin(120)})$);
    \threesitedraw{\plaqscalethreedim}{\plaqscalethreedim}{$x+e_{1}$};
    \draw[postaction={decorate}] (2.5, \plaqscalethreedim + 1.5) .. controls (3.5, \plaqscalethreedim + 1.5) and (\plaqscalethreedim -3.5, \plaqscalethreedim + -1.5) .. (\plaqscalethreedim -2.5, \plaqscalethreedim + -1.5);
    \threesitedraw{0}{\plaqscalethreedim}{$x$};
    \draw[postaction={decorate}] ($(0, \plaqscalethreedim) - (0.75, 0.75) - ({cos(120)}, {sin(120)})$)
    .. controls ($(0, \plaqscalethreedim) - (0.75, 0.75) - ({2.5 * cos(120)}, {2.5 * sin(120)})$)
    and ($(0 ,0) + (0.75, 0.75) +({2.5 * cos(120)}, {2.5 * sin(120)})$)
    .. ($(0 ,0) + (0.75, 0.75) + ({cos(120)}, {sin(120)})$);
  \end{scope}
  \end{tikzpicture}
  \caption{Connectivity of a $zx$-plaquette in three dimensions.}
    \label{fig:3dPlaquetteZX}
  \end{figure*}

\section{\label{sec:LSH-KSComparison}Comparison of loop-string-hadron and Kogut-Susskind}

From the perspective of quantum computation and simulation, the LSH framework exhibits the following benefits:
\begin{itemize}
  \item \emph{Abelian constraints}.---%
    The AGLs are the only remnant constraints in the LSH framework.
    A LSH basis naturally diagonalizes these constraints.
    Any basis state in the AGL-satisfying subspace is physically observable.
  \item \emph{Simple quantum numbers}.---%
    The LSH Hilbert space is naturally characterized by integers.
    Quark quantum numbers are bounded by the Pauli principle, while loop quantum numbers ($n_l$ or $\ell_{ij}$) can be any non-negative integer.
    Additionally, all elementary operators are 1-sparse in the LSH basis.
  \item \emph{No Clebsch-Gordon coefficients}.---%
    Like the prepotential formulation, the LSH treatment avoids the need for Clebsch-Gordon coefficients.
    What makes the theory describe SU(2) is the available set of operators and their algebra.
    Establishing the same for SU(3) is the subject of ongoing work.
  \item \emph{Gauge redundancy in preliminary simulations}.---%
    In $d=1$, the LSH formulation gives a clear advantage over Kogut-Susskind in terms of qubit requirements---simulating the gauge degrees of freedom takes half the number of qubits.
    The qubit requirements are also less for pure gauge theory in $d=2$.
    By bringing down the qubit costs in these cases, one can learn how to deal with the implementation of LSH structures sooner.
\end{itemize}

At the same time, there are some caveats:
\begin{itemize}
  \item \emph{Proliferation of Hamiltonian terms}.---%
    Introducing virtual links causes the hopping and plaquette terms to grow in size and number.
    In $d=3$, the number of terms is formidable.
    This would pose a problem for quantum simulation methods such as Trotter-Suzuki decompositions \cite{trotterProductSemigroups59,suzukiGeneralizedTrotter76}.
  \item \emph{Qubit costs without solving the Abelian Gauss law}.---%
    During the onset of scientific quantum computing, each and every qubit is important to count.
    This framework would have direct benefit on qubit costs for the 1+1-dimensional theory and the 2+1-dimensional pure gauge theory, and those are important to study on their own.
    Beyond pure gauge theory in two dimensions, qubit requirements of directly simulating the multidimensional LSH formulation outpace those of simulating the Kogut-Susskind formulation.
    However, it is possible to push the LSH framework further by actually solving the Abelian Gauss law.
    If this is done, then simulating the gluons will cost fewer qubits in any number of dimensions.
\end{itemize}
These drawbacks are resource oriented, rather than theoretical.
Given that the LSH framework is just being introduced, we can hope that novel algorithmic solutions will alleviate the practical issues.

\section{Conclusion}

In this paper we have provided a complete Hamiltonian for SU(2) gauge theory coupled to staggered fermions in $1+1$, $2+1$, and $3+1$ dimensions.
Dynamics is described in terms of physical and local observables: hadrons and segments of flux loops and meson strings.
By using a staggered fermion prescription, the matter field carried only a color (no spinor) index, allowing the LSH dynamics to be formulated without unnecessary complications.
Studying adaptations of the LSH framework to other fermion discretizations or to more flavors will be of future interest.

We also point out that, while the focus of this paper was limited to SU(2) for concreteness, the geometric approach makes no explicit use of the SU(2) angular momentum characterization of states or of SU(2) Clebsch-Gordon coefficients.
The prepotential formulation from which this LSH formulation was derived has already been generalized to SU(3)
\cite{anishetty.mathur.eaPrepotentialFormulation10}
and even SU(N)
\cite{mathur.raychowdhury.eaSUIrreducible10,raychowdhuryPrepotentialFormulation13}.
Generalization to SU(3) preserves the local loop Hilbert space construction, this time with two Abelian Gauss laws for every link and each of the AGLs of the same form as in SU(2).
However, finding a suitable point splitting scheme to describe the dynamics using only physical degrees of freedom is not done yet and is of significant interest to work out in the future.

We have illustrated in this paper how the present scheme translates the dynamics of all possible irreps of the gauge group into the dynamics of many local towers of states characterized by single integers.
This is a major gain of this formalism over the Kogut-Susskind one.
The LSH framework makes non-Abelian gauge theory dynamics more similar to that of Schwinger model by completely solving the non-Abelian Gauss law constraint.
This particular formalism, which is structurally closer to U(1) gauge theories, stands to directly benefit from  algorithms developed for Abelian theories.

The major price paid is the introduction of more lattice links and a new AGL on each virtual link.
It turns out, however, that half or more of the bosonic degrees of freedom can be removed by solving the Abelian Gauss law.
Solving the Abelian Gauss law would render the qubit cost of LSH simulation less than that of the Kogut-Susskind formulation in any dimension and will be the subject of future work.
Nonetheless, even before solving the AGL, the truncated LSH framework costs fewer qubits than the truncated Kogut-Susskind formulation would for theories that will be important milestones along the way to three-dimensional simulations.

Combining all the above benefits, the LSH framework may take us one step closer to quantum-simulating theories that model fundamental interactions of Nature.

\section*{Acknowledgments}
I.R.\,would like to thank Z. Davoudi for many useful discussions and comments on the manuscript.
J.R.S.\,would like to thank D.B. Kaplan and Stephan Caspar for helpful discussions, and Z. Davoudi and the Maryland Center for Fundamental Physics (MCFP) at the University of Maryland at College Park for hosting him while part of this research was carried out.
I.R.\,is supported by the U.S. Department of Energy (DOE), Office of Science, Office of Advanced Scientific Computing Research (ASCR) Quantum Computing Application Teams program, under fieldwork proposal number ERKJ347, and by the MCFP.
J.R.S.\,was supported by DOE Grant No.~DE-FG02-00ER41132 and by the National Science Foundation Graduate Research Fellowship under Grant No.~1256082.
Part of this research was carried out with support from a Thomas L.~and Margo G.~Wyckoff Endowed Faculty Fellowship.

\appendix

\section{\label{app:norm}Normalizing the loop-string-hadron basis}
The norm of a pure-gauge-flux state $\ket{|\ell ,0,0}$ at a matter site is readily obtained by recursion and repeated use of the fact that $\Lmm\ket{0}=0$.
One starts with
\begin{align*}
  \braket{\ell ,0,0||\ell,0,0} &= \bra{0}(\Lmm)^\ell (\Lpp)^\ell \ket{0}\\
  &=  \bra{0}\left[ (\Lmm)^\ell , (\Lpp)^\ell \right] \ket{0}
\end{align*}
The commutator is then expanded with a product rule:
\begin{align*}
  \begin{split}
  \left[ (\Lmm)^\ell , (\Lpp)^\ell \right] = & (\Lmm)^{\ell-1} \left[ (\Lmm , (\Lpp)^\ell \right]\\
  & + (\Lmm)^{\ell-2} \left[ \Lmm , (\Lpp)^\ell \right] \Lmm \\
  & + \cdots \\
  & + \left[ \Lmm , (\Lpp)^\ell \right] (\Lmm)^{\ell-1}
\end{split}
\end{align*}
All terms except the first annihilate $\ket{0}$, so we have
\begin{align*}
  \bra{0} (\Lmm)^\ell  (\Lpp)^\ell \ket{0} &= \bra{0} (\Lmm)^{\ell-1} \left[ \Lmm , (\Lpp)^\ell \right] \ket{0} \ .
\end{align*}
Now, it is straightforward to show that $\left[ \Lmm , (\Lpp)^\ell \right]= \ell\left(\NL +\NR +2 -(\ell-1)\right) (\Lpp)^{\ell-1}$.
Evaluating the previous result then leads to
\begin{align*}
  \bra{0} (\Lmm)^\ell  (\Lpp)^\ell \ket{0} &= (\ell+1)\ell \bra{0} (\Lmm)^{\ell-1}  (\Lpp)^{\ell-1} \ket{0}
\end{align*}
By recursion, we arrive at
\begin{align*}
\bra{0} (\Lmm)^\ell  (\Lpp)^\ell \ket{0} &= (\ell+1)!\ \ell!
\end{align*}
Thus, the states
\begin{equation*}
  \ket{n_l,0,0}=\frac{\ket{|n_l,0,0}}{\sqrt{(n_l+1)!\ n_l! }}
\end{equation*}
are normalized.

Using similar tactics, one can also show that
\begin{align*}
  \braket{n_l,1,0||n_l,1,0}&=(n_l+2) \braket{n_l,0,0||n_l,0,0}\\
  &= (n_l+2)!\ n_l! \ ,
\end{align*}
which shows that
\begin{equation*}
  \ket{n_l,1,0}=\frac{\ket{|n_l,1,0}}{\sqrt{(n_l+2)!\ n_l! }}
\end{equation*}
is normalized.
The remaining states can again be normalized using the same techniques and all are encapsulated by (\ref{eq:normalSiteBasis}).

\bibliography{lshQuantum.bib}

\begin{thebibliography}{56}%
\makeatletter
\providecommand \@ifxundefined [1]{%
 \@ifx{#1\undefined}
}%
\providecommand \@ifnum [1]{%
 \ifnum #1\expandafter \@firstoftwo
 \else \expandafter \@secondoftwo
 \fi
}%
\providecommand \@ifx [1]{%
 \ifx #1\expandafter \@firstoftwo
 \else \expandafter \@secondoftwo
 \fi
}%
\providecommand \natexlab [1]{#1}%
\providecommand \enquote  [1]{``#1''}%
\providecommand \bibnamefont  [1]{#1}%
\providecommand \bibfnamefont [1]{#1}%
\providecommand \citenamefont [1]{#1}%
\providecommand \href@noop [0]{\@secondoftwo}%
\providecommand \href [0]{\begingroup \@sanitize@url \@href}%
\providecommand \@href[1]{\@@startlink{#1}\@@href}%
\providecommand \@@href[1]{\endgroup#1\@@endlink}%
\providecommand \@sanitize@url [0]{\catcode `\\12\catcode `\$12\catcode
  `\&12\catcode `\#12\catcode `\^12\catcode `\_12\catcode `\%12\relax}%
\providecommand \@@startlink[1]{}%
\providecommand \@@endlink[0]{}%
\providecommand \url  [0]{\begingroup\@sanitize@url \@url }%
\providecommand \@url [1]{\endgroup\@href {#1}{\urlprefix }}%
\providecommand \urlprefix  [0]{URL }%
\providecommand \Eprint [0]{\href }%
\providecommand \doibase [0]{http://dx.doi.org/}%
\providecommand \selectlanguage [0]{\@gobble}%
\providecommand \bibinfo  [0]{\@secondoftwo}%
\providecommand \bibfield  [0]{\@secondoftwo}%
\providecommand \translation [1]{[#1]}%
\providecommand \BibitemOpen [0]{}%
\providecommand \bibitemStop [0]{}%
\providecommand \bibitemNoStop [0]{.\EOS\space}%
\providecommand \EOS [0]{\spacefactor3000\relax}%
\providecommand \BibitemShut  [1]{\csname bibitem#1\endcsname}%
\let\auto@bib@innerbib\@empty
\bibitem [{\citenamefont {Gross}\ and\ \citenamefont
  {Wilczek}(1973)}]{gross.wilczekUltravioletBehavior73}%
  \BibitemOpen
  \bibfield  {author} {\bibinfo {author} {\bibfnamefont {D.~J.}\ \bibnamefont
  {Gross}}\ and\ \bibinfo {author} {\bibfnamefont {F.}~\bibnamefont
  {Wilczek}},\ }\href {\doibase 10.1103/PhysRevLett.30.1343} {\bibfield
  {journal} {\bibinfo  {journal} {Phys. Rev. Lett.}\ }\textbf {\bibinfo
  {volume} {30}},\ \bibinfo {pages} {1343} (\bibinfo {year}
  {1973})}\BibitemShut {NoStop}%
\bibitem [{\citenamefont {Politzer}(1973)}]{politzerReliablePerturbative73}%
  \BibitemOpen
  \bibfield  {author} {\bibinfo {author} {\bibfnamefont {H.~D.}\ \bibnamefont
  {Politzer}},\ }\href {\doibase 10.1103/PhysRevLett.30.1346} {\bibfield
  {journal} {\bibinfo  {journal} {Phys. Rev. Lett.}\ }\textbf {\bibinfo
  {volume} {30}},\ \bibinfo {pages} {1346} (\bibinfo {year}
  {1973})}\BibitemShut {NoStop}%
\bibitem [{\citenamefont {{Gell-Mann}}(1961)}]{gell-mannEightfoldWay61}%
  \BibitemOpen
  \bibfield  {author} {\bibinfo {author} {\bibfnamefont {M.}~\bibnamefont
  {{Gell-Mann}}},\ }\href {\doibase 10.2172/4008239} {\emph {\bibinfo {title}
  {The Eightfold Way: {{A}} Theory of Strong Interaction Symmetry}}},\ \bibinfo
  {type} {Tech. Rep.}\ \bibinfo {number} {TID-12608; CTSL-20}\ (\bibinfo
  {institution} {{Caltech, Pasadena. Synchrotron Lab.}},\ \bibinfo {year}
  {1961})\BibitemShut {NoStop}%
\bibitem [{\citenamefont {Leurs}(2007)}]{leursNonabelianGauge07}%
  \BibitemOpen
  \bibfield  {author} {\bibinfo {author} {\bibfnamefont {B.~W.}\ \bibnamefont
  {Leurs}},\ }\emph {\bibinfo {title} {Non-Abelian Gauge Theories: Analogies in
  Condensed Matter Systems}},\ \href@noop {} {\bibinfo {type} {Book}},\
  \bibinfo  {school} {Leiden University}, \bibinfo {address} {{Leiden,
  Netherlands}} (\bibinfo {year} {2007})\BibitemShut {NoStop}%
\bibitem [{\citenamefont {Lee}\ \emph {et~al.}(2006)\citenamefont {Lee},
  \citenamefont {Nagaosa},\ and\ \citenamefont
  {Wen}}]{lee.nagaosa.eaDopingMott06}%
  \BibitemOpen
  \bibfield  {author} {\bibinfo {author} {\bibfnamefont {P.~A.}\ \bibnamefont
  {Lee}}, \bibinfo {author} {\bibfnamefont {N.}~\bibnamefont {Nagaosa}}, \ and\
  \bibinfo {author} {\bibfnamefont {X.-G.}\ \bibnamefont {Wen}},\ }\href
  {\doibase 10.1103/RevModPhys.78.17} {\bibfield  {journal} {\bibinfo
  {journal} {Rev. Mod. Phys.}\ }\textbf {\bibinfo {volume} {78}},\ \bibinfo
  {pages} {17} (\bibinfo {year} {2006})}\BibitemShut {NoStop}%
\bibitem [{\citenamefont {Wilson}(1974)}]{wilsonConfinementQuarks74}%
  \BibitemOpen
  \bibfield  {author} {\bibinfo {author} {\bibfnamefont {K.~G.}\ \bibnamefont
  {Wilson}},\ }\href {\doibase 10.1103/PhysRevD.10.2445} {\bibfield  {journal}
  {\bibinfo  {journal} {Phys. Rev. D}\ }\textbf {\bibinfo {volume} {10}},\
  \bibinfo {pages} {2445} (\bibinfo {year} {1974})}\BibitemShut {NoStop}%
\bibitem [{\citenamefont {Bowler}\ \emph {et~al.}(1996)\citenamefont {Bowler},
  \citenamefont {Kenway}, \citenamefont {Oliveira}, \citenamefont {Richards},
  \citenamefont {Ueberholz}, \citenamefont {{UKQCD Collaboration}},
  \citenamefont {Lellouch}, \citenamefont {Nieves}, \citenamefont {Sachrajda},
  \citenamefont {Stella},\ and\ \citenamefont
  {Wittig}}]{bowler.kenway.eaHeavyBaryon96}%
  \BibitemOpen
  \bibfield  {author} {\bibinfo {author} {\bibfnamefont {K.~C.}\ \bibnamefont
  {Bowler}}, \bibinfo {author} {\bibfnamefont {R.~D.}\ \bibnamefont {Kenway}},
  \bibinfo {author} {\bibfnamefont {O.}~\bibnamefont {Oliveira}}, \bibinfo
  {author} {\bibfnamefont {D.~G.}\ \bibnamefont {Richards}}, \bibinfo {author}
  {\bibfnamefont {P.}~\bibnamefont {Ueberholz}}, \bibinfo {author}
  {\bibnamefont {{UKQCD Collaboration}}}, \bibinfo {author} {\bibfnamefont
  {L.}~\bibnamefont {Lellouch}}, \bibinfo {author} {\bibfnamefont
  {J.}~\bibnamefont {Nieves}}, \bibinfo {author} {\bibfnamefont {C.~T.}\
  \bibnamefont {Sachrajda}}, \bibinfo {author} {\bibfnamefont {N.}~\bibnamefont
  {Stella}}, \ and\ \bibinfo {author} {\bibfnamefont {H.}~\bibnamefont
  {Wittig}},\ }\href {\doibase 10.1103/PhysRevD.54.3619} {\bibfield  {journal}
  {\bibinfo  {journal} {Phys. Rev. D}\ }\textbf {\bibinfo {volume} {54}},\
  \bibinfo {pages} {3619} (\bibinfo {year} {1996})}\BibitemShut {NoStop}%
\bibitem [{\citenamefont {{PACS-CS Collaboration}}\ \emph
  {et~al.}(2009)\citenamefont {{PACS-CS Collaboration}}, \citenamefont {Aoki},
  \citenamefont {Ishikawa}, \citenamefont {Ishizuka}, \citenamefont {Izubuchi},
  \citenamefont {Kadoh}, \citenamefont {Kanaya}, \citenamefont {Kuramashi},
  \citenamefont {Namekawa}, \citenamefont {Okawa}, \citenamefont {Taniguchi},
  \citenamefont {Ukawa}, \citenamefont {Ukita},\ and\ \citenamefont
  {Yoshi{\'e}}}]{pacs-cscollaboration.aoki.eaFlavorLattice09}%
  \BibitemOpen
  \bibfield  {author} {\bibinfo {author} {\bibnamefont {{PACS-CS
  Collaboration}}}, \bibinfo {author} {\bibfnamefont {S.}~\bibnamefont {Aoki}},
  \bibinfo {author} {\bibfnamefont {K.-I.}\ \bibnamefont {Ishikawa}}, \bibinfo
  {author} {\bibfnamefont {N.}~\bibnamefont {Ishizuka}}, \bibinfo {author}
  {\bibfnamefont {T.}~\bibnamefont {Izubuchi}}, \bibinfo {author}
  {\bibfnamefont {D.}~\bibnamefont {Kadoh}}, \bibinfo {author} {\bibfnamefont
  {K.}~\bibnamefont {Kanaya}}, \bibinfo {author} {\bibfnamefont
  {Y.}~\bibnamefont {Kuramashi}}, \bibinfo {author} {\bibfnamefont
  {Y.}~\bibnamefont {Namekawa}}, \bibinfo {author} {\bibfnamefont
  {M.}~\bibnamefont {Okawa}}, \bibinfo {author} {\bibfnamefont
  {Y.}~\bibnamefont {Taniguchi}}, \bibinfo {author} {\bibfnamefont
  {A.}~\bibnamefont {Ukawa}}, \bibinfo {author} {\bibfnamefont
  {N.}~\bibnamefont {Ukita}}, \ and\ \bibinfo {author} {\bibfnamefont
  {T.}~\bibnamefont {Yoshi{\'e}}},\ }\href {\doibase
  10.1103/PhysRevD.79.034503} {\bibfield  {journal} {\bibinfo  {journal} {Phys.
  Rev. D}\ }\textbf {\bibinfo {volume} {79}},\ \bibinfo {pages} {034503}
  (\bibinfo {year} {2009})}\BibitemShut {NoStop}%
\bibitem [{\citenamefont {Bazavov}\ \emph {et~al.}(2009)\citenamefont
  {Bazavov}, \citenamefont {Bhattacharya}, \citenamefont {Cheng}, \citenamefont
  {Christ}, \citenamefont {DeTar}, \citenamefont {Ejiri}, \citenamefont
  {Gottlieb}, \citenamefont {Gupta}, \citenamefont {Heller}, \citenamefont
  {Huebner}, \citenamefont {Jung}, \citenamefont {Karsch}, \citenamefont
  {Laermann}, \citenamefont {Levkova}, \citenamefont {Miao}, \citenamefont
  {Mawhinney}, \citenamefont {Petreczky}, \citenamefont {Schmidt},
  \citenamefont {Soltz}, \citenamefont {Soeldner}, \citenamefont {Sugar},
  \citenamefont {Toussaint},\ and\ \citenamefont
  {Vranas}}]{bazavov.bhattacharya.eaEquationState09}%
  \BibitemOpen
  \bibfield  {author} {\bibinfo {author} {\bibfnamefont {A.}~\bibnamefont
  {Bazavov}}, \bibinfo {author} {\bibfnamefont {T.}~\bibnamefont
  {Bhattacharya}}, \bibinfo {author} {\bibfnamefont {M.}~\bibnamefont {Cheng}},
  \bibinfo {author} {\bibfnamefont {N.~H.}\ \bibnamefont {Christ}}, \bibinfo
  {author} {\bibfnamefont {C.}~\bibnamefont {DeTar}}, \bibinfo {author}
  {\bibfnamefont {S.}~\bibnamefont {Ejiri}}, \bibinfo {author} {\bibfnamefont
  {S.}~\bibnamefont {Gottlieb}}, \bibinfo {author} {\bibfnamefont
  {R.}~\bibnamefont {Gupta}}, \bibinfo {author} {\bibfnamefont {U.~M.}\
  \bibnamefont {Heller}}, \bibinfo {author} {\bibfnamefont {K.}~\bibnamefont
  {Huebner}}, \bibinfo {author} {\bibfnamefont {C.}~\bibnamefont {Jung}},
  \bibinfo {author} {\bibfnamefont {F.}~\bibnamefont {Karsch}}, \bibinfo
  {author} {\bibfnamefont {E.}~\bibnamefont {Laermann}}, \bibinfo {author}
  {\bibfnamefont {L.}~\bibnamefont {Levkova}}, \bibinfo {author} {\bibfnamefont
  {C.}~\bibnamefont {Miao}}, \bibinfo {author} {\bibfnamefont {R.~D.}\
  \bibnamefont {Mawhinney}}, \bibinfo {author} {\bibfnamefont {P.}~\bibnamefont
  {Petreczky}}, \bibinfo {author} {\bibfnamefont {C.}~\bibnamefont {Schmidt}},
  \bibinfo {author} {\bibfnamefont {R.~A.}\ \bibnamefont {Soltz}}, \bibinfo
  {author} {\bibfnamefont {W.}~\bibnamefont {Soeldner}}, \bibinfo {author}
  {\bibfnamefont {R.}~\bibnamefont {Sugar}}, \bibinfo {author} {\bibfnamefont
  {D.}~\bibnamefont {Toussaint}}, \ and\ \bibinfo {author} {\bibfnamefont
  {P.}~\bibnamefont {Vranas}},\ }\href {\doibase 10.1103/PhysRevD.80.014504}
  {\bibfield  {journal} {\bibinfo  {journal} {Phys. Rev. D}\ }\textbf {\bibinfo
  {volume} {80}},\ \bibinfo {pages} {014504} (\bibinfo {year}
  {2009})}\BibitemShut {NoStop}%
\bibitem [{\citenamefont {Aoki}\ \emph {et~al.}(2009)\citenamefont {Aoki},
  \citenamefont {Bors{\'a}nyi}, \citenamefont {D{\"u}rr}, \citenamefont
  {Fodor}, \citenamefont {Katz}, \citenamefont {Krieg},\ and\ \citenamefont
  {Szabo}}]{aoki.borsanyi.eaQCDTransition09}%
  \BibitemOpen
  \bibfield  {author} {\bibinfo {author} {\bibfnamefont {Y.}~\bibnamefont
  {Aoki}}, \bibinfo {author} {\bibfnamefont {S.}~\bibnamefont {Bors{\'a}nyi}},
  \bibinfo {author} {\bibfnamefont {S.}~\bibnamefont {D{\"u}rr}}, \bibinfo
  {author} {\bibfnamefont {Z.}~\bibnamefont {Fodor}}, \bibinfo {author}
  {\bibfnamefont {S.~D.}\ \bibnamefont {Katz}}, \bibinfo {author}
  {\bibfnamefont {S.}~\bibnamefont {Krieg}}, \ and\ \bibinfo {author}
  {\bibfnamefont {K.}~\bibnamefont {Szabo}},\ }\href {\doibase
  10.1088/1126-6708/2009/06/088} {\bibfield  {journal} {\bibinfo  {journal} {J.
  High Energy Phys.}\ }\textbf {\bibinfo {volume} {2009}},\ \bibinfo {pages}
  {088} (\bibinfo {year} {2009})}\BibitemShut {NoStop}%
\bibitem [{\citenamefont {Bors{\'a}nyi}\ \emph {et~al.}(2013)\citenamefont
  {Bors{\'a}nyi}, \citenamefont {Fodor}, \citenamefont {Katz}, \citenamefont
  {Krieg}, \citenamefont {Ratti},\ and\ \citenamefont
  {Szab{\'o}}}]{borsanyi.fodor.eaFreezeOutParameters13}%
  \BibitemOpen
  \bibfield  {author} {\bibinfo {author} {\bibfnamefont {S.}~\bibnamefont
  {Bors{\'a}nyi}}, \bibinfo {author} {\bibfnamefont {Z.}~\bibnamefont {Fodor}},
  \bibinfo {author} {\bibfnamefont {S.~D.}\ \bibnamefont {Katz}}, \bibinfo
  {author} {\bibfnamefont {S.}~\bibnamefont {Krieg}}, \bibinfo {author}
  {\bibfnamefont {C.}~\bibnamefont {Ratti}}, \ and\ \bibinfo {author}
  {\bibfnamefont {K.~K.}\ \bibnamefont {Szab{\'o}}},\ }\href {\doibase
  10.1103/PhysRevLett.111.062005} {\bibfield  {journal} {\bibinfo  {journal}
  {Phys. Rev. Lett.}\ }\textbf {\bibinfo {volume} {111}},\ \bibinfo {pages}
  {062005} (\bibinfo {year} {2013})}\BibitemShut {NoStop}%
\bibitem [{\citenamefont {{NPLQCD Collaboration}}\ \emph
  {et~al.}(2017{\natexlab{a}})\citenamefont {{NPLQCD Collaboration}},
  \citenamefont {Tiburzi}, \citenamefont {Wagman}, \citenamefont {Winter},
  \citenamefont {Chang}, \citenamefont {Davoudi}, \citenamefont {Detmold},
  \citenamefont {Orginos}, \citenamefont {Savage},\ and\ \citenamefont
  {Shanahan}}]{nplqcdcollaboration.tiburzi.eaDoubleBeta17}%
  \BibitemOpen
  \bibfield  {author} {\bibinfo {author} {\bibnamefont {{NPLQCD
  Collaboration}}}, \bibinfo {author} {\bibfnamefont {B.~C.}\ \bibnamefont
  {Tiburzi}}, \bibinfo {author} {\bibfnamefont {M.~L.}\ \bibnamefont {Wagman}},
  \bibinfo {author} {\bibfnamefont {F.}~\bibnamefont {Winter}}, \bibinfo
  {author} {\bibfnamefont {E.}~\bibnamefont {Chang}}, \bibinfo {author}
  {\bibfnamefont {Z.}~\bibnamefont {Davoudi}}, \bibinfo {author} {\bibfnamefont
  {W.}~\bibnamefont {Detmold}}, \bibinfo {author} {\bibfnamefont
  {K.}~\bibnamefont {Orginos}}, \bibinfo {author} {\bibfnamefont {M.~J.}\
  \bibnamefont {Savage}}, \ and\ \bibinfo {author} {\bibfnamefont {P.~E.}\
  \bibnamefont {Shanahan}},\ }\href {\doibase 10.1103/PhysRevD.96.054505}
  {\bibfield  {journal} {\bibinfo  {journal} {Phys. Rev. D}\ }\textbf {\bibinfo
  {volume} {96}},\ \bibinfo {pages} {054505} (\bibinfo {year}
  {2017}{\natexlab{a}})}\BibitemShut {NoStop}%
\bibitem [{\citenamefont {{NPLQCD Collaboration}}\ \emph
  {et~al.}(2017{\natexlab{b}})\citenamefont {{NPLQCD Collaboration}},
  \citenamefont {Savage}, \citenamefont {Shanahan}, \citenamefont {Tiburzi},
  \citenamefont {Wagman}, \citenamefont {Winter}, \citenamefont {Beane},
  \citenamefont {Chang}, \citenamefont {Davoudi}, \citenamefont {Detmold},\
  and\ \citenamefont
  {Orginos}}]{nplqcdcollaboration.savage.eaProtonProtonFusion17}%
  \BibitemOpen
  \bibfield  {author} {\bibinfo {author} {\bibnamefont {{NPLQCD
  Collaboration}}}, \bibinfo {author} {\bibfnamefont {M.~J.}\ \bibnamefont
  {Savage}}, \bibinfo {author} {\bibfnamefont {P.~E.}\ \bibnamefont
  {Shanahan}}, \bibinfo {author} {\bibfnamefont {B.~C.}\ \bibnamefont
  {Tiburzi}}, \bibinfo {author} {\bibfnamefont {M.~L.}\ \bibnamefont {Wagman}},
  \bibinfo {author} {\bibfnamefont {F.}~\bibnamefont {Winter}}, \bibinfo
  {author} {\bibfnamefont {S.~R.}\ \bibnamefont {Beane}}, \bibinfo {author}
  {\bibfnamefont {E.}~\bibnamefont {Chang}}, \bibinfo {author} {\bibfnamefont
  {Z.}~\bibnamefont {Davoudi}}, \bibinfo {author} {\bibfnamefont
  {W.}~\bibnamefont {Detmold}}, \ and\ \bibinfo {author} {\bibfnamefont
  {K.}~\bibnamefont {Orginos}},\ }\href {\doibase
  10.1103/PhysRevLett.119.062002} {\bibfield  {journal} {\bibinfo  {journal}
  {Phys. Rev. Lett.}\ }\textbf {\bibinfo {volume} {119}},\ \bibinfo {pages}
  {062002} (\bibinfo {year} {2017}{\natexlab{b}})}\BibitemShut {NoStop}%
\bibitem [{\citenamefont {Kronfeld}(2012)}]{kronfeldTwentyFirstCentury12}%
  \BibitemOpen
  \bibfield  {author} {\bibinfo {author} {\bibfnamefont {A.~S.}\ \bibnamefont
  {Kronfeld}},\ }\href {\doibase 10.1146/annurev-nucl-102711-094942} {\bibfield
   {journal} {\bibinfo  {journal} {Annu. Rev. Nucl. Part. Sci.}\ }\textbf
  {\bibinfo {volume} {62}},\ \bibinfo {pages} {265} (\bibinfo {year}
  {2012})}\BibitemShut {NoStop}%
\bibitem [{\citenamefont {Kogut}\ and\ \citenamefont
  {Susskind}(1975)}]{kogut.susskindHamiltonianFormulation75}%
  \BibitemOpen
  \bibfield  {author} {\bibinfo {author} {\bibfnamefont {J.}~\bibnamefont
  {Kogut}}\ and\ \bibinfo {author} {\bibfnamefont {L.}~\bibnamefont
  {Susskind}},\ }\href {\doibase 10.1103/PhysRevD.11.395} {\bibfield  {journal}
  {\bibinfo  {journal} {Phys. Rev. D}\ }\textbf {\bibinfo {volume} {11}},\
  \bibinfo {pages} {395} (\bibinfo {year} {1975})}\BibitemShut {NoStop}%
\bibitem [{\citenamefont {Feynman}(1982)}]{feynmanSimulatingPhysics82}%
  \BibitemOpen
  \bibfield  {author} {\bibinfo {author} {\bibfnamefont {R.~P.}\ \bibnamefont
  {Feynman}},\ }\href {\doibase 10.1007/BF02650179} {\bibfield  {journal}
  {\bibinfo  {journal} {Int. J. Theor. Phys.}\ }\textbf {\bibinfo {volume}
  {21}},\ \bibinfo {pages} {467} (\bibinfo {year} {1982})}\BibitemShut
  {NoStop}%
\bibitem [{\citenamefont {Arute}\ \emph {et~al.}(2019)\citenamefont {Arute},
  \citenamefont {Arya}, \citenamefont {Babbush}, \citenamefont {Bacon},
  \citenamefont {Bardin}, \citenamefont {Barends}, \citenamefont {Biswas},
  \citenamefont {Boixo}, \citenamefont {Brandao}, \citenamefont {Buell},
  \citenamefont {Burkett}, \citenamefont {Chen}, \citenamefont {Chen},
  \citenamefont {Chiaro}, \citenamefont {Collins}, \citenamefont {Courtney},
  \citenamefont {Dunsworth}, \citenamefont {Farhi}, \citenamefont {Foxen},
  \citenamefont {Fowler}, \citenamefont {Gidney}, \citenamefont {Giustina},
  \citenamefont {Graff}, \citenamefont {Guerin}, \citenamefont {Habegger},
  \citenamefont {Harrigan}, \citenamefont {Hartmann}, \citenamefont {Ho},
  \citenamefont {Hoffmann}, \citenamefont {Huang}, \citenamefont {Humble},
  \citenamefont {Isakov}, \citenamefont {Jeffrey}, \citenamefont {Jiang},
  \citenamefont {Kafri}, \citenamefont {Kechedzhi}, \citenamefont {Kelly},
  \citenamefont {Klimov}, \citenamefont {Knysh}, \citenamefont {Korotkov},
  \citenamefont {Kostritsa}, \citenamefont {Landhuis}, \citenamefont
  {Lindmark}, \citenamefont {Lucero}, \citenamefont {Lyakh}, \citenamefont
  {Mandr{\`a}}, \citenamefont {McClean}, \citenamefont {McEwen}, \citenamefont
  {Megrant}, \citenamefont {Mi}, \citenamefont {Michielsen}, \citenamefont
  {Mohseni}, \citenamefont {Mutus}, \citenamefont {Naaman}, \citenamefont
  {Neeley}, \citenamefont {Neill}, \citenamefont {Niu}, \citenamefont {Ostby},
  \citenamefont {Petukhov}, \citenamefont {Platt}, \citenamefont {Quintana},
  \citenamefont {Rieffel}, \citenamefont {Roushan}, \citenamefont {Rubin},
  \citenamefont {Sank}, \citenamefont {Satzinger}, \citenamefont {Smelyanskiy},
  \citenamefont {Sung}, \citenamefont {Trevithick}, \citenamefont
  {Vainsencher}, \citenamefont {Villalonga}, \citenamefont {White},
  \citenamefont {Yao}, \citenamefont {Yeh}, \citenamefont {Zalcman},
  \citenamefont {Neven},\ and\ \citenamefont
  {Martinis}}]{arute.arya.eaQuantumSupremacy19}%
  \BibitemOpen
  \bibfield  {author} {\bibinfo {author} {\bibfnamefont {F.}~\bibnamefont
  {Arute}}, \bibinfo {author} {\bibfnamefont {K.}~\bibnamefont {Arya}},
  \bibinfo {author} {\bibfnamefont {R.}~\bibnamefont {Babbush}}, \bibinfo
  {author} {\bibfnamefont {D.}~\bibnamefont {Bacon}}, \bibinfo {author}
  {\bibfnamefont {J.~C.}\ \bibnamefont {Bardin}}, \bibinfo {author}
  {\bibfnamefont {R.}~\bibnamefont {Barends}}, \bibinfo {author} {\bibfnamefont
  {R.}~\bibnamefont {Biswas}}, \bibinfo {author} {\bibfnamefont
  {S.}~\bibnamefont {Boixo}}, \bibinfo {author} {\bibfnamefont {F.~G. S.~L.}\
  \bibnamefont {Brandao}}, \bibinfo {author} {\bibfnamefont {D.~A.}\
  \bibnamefont {Buell}}, \bibinfo {author} {\bibfnamefont {B.}~\bibnamefont
  {Burkett}}, \bibinfo {author} {\bibfnamefont {Y.}~\bibnamefont {Chen}},
  \bibinfo {author} {\bibfnamefont {Z.}~\bibnamefont {Chen}}, \bibinfo {author}
  {\bibfnamefont {B.}~\bibnamefont {Chiaro}}, \bibinfo {author} {\bibfnamefont
  {R.}~\bibnamefont {Collins}}, \bibinfo {author} {\bibfnamefont
  {W.}~\bibnamefont {Courtney}}, \bibinfo {author} {\bibfnamefont
  {A.}~\bibnamefont {Dunsworth}}, \bibinfo {author} {\bibfnamefont
  {E.}~\bibnamefont {Farhi}}, \bibinfo {author} {\bibfnamefont
  {B.}~\bibnamefont {Foxen}}, \bibinfo {author} {\bibfnamefont
  {A.}~\bibnamefont {Fowler}}, \bibinfo {author} {\bibfnamefont
  {C.}~\bibnamefont {Gidney}}, \bibinfo {author} {\bibfnamefont
  {M.}~\bibnamefont {Giustina}}, \bibinfo {author} {\bibfnamefont
  {R.}~\bibnamefont {Graff}}, \bibinfo {author} {\bibfnamefont
  {K.}~\bibnamefont {Guerin}}, \bibinfo {author} {\bibfnamefont
  {S.}~\bibnamefont {Habegger}}, \bibinfo {author} {\bibfnamefont {M.~P.}\
  \bibnamefont {Harrigan}}, \bibinfo {author} {\bibfnamefont {M.~J.}\
  \bibnamefont {Hartmann}}, \bibinfo {author} {\bibfnamefont {A.}~\bibnamefont
  {Ho}}, \bibinfo {author} {\bibfnamefont {M.}~\bibnamefont {Hoffmann}},
  \bibinfo {author} {\bibfnamefont {T.}~\bibnamefont {Huang}}, \bibinfo
  {author} {\bibfnamefont {T.~S.}\ \bibnamefont {Humble}}, \bibinfo {author}
  {\bibfnamefont {S.~V.}\ \bibnamefont {Isakov}}, \bibinfo {author}
  {\bibfnamefont {E.}~\bibnamefont {Jeffrey}}, \bibinfo {author} {\bibfnamefont
  {Z.}~\bibnamefont {Jiang}}, \bibinfo {author} {\bibfnamefont
  {D.}~\bibnamefont {Kafri}}, \bibinfo {author} {\bibfnamefont
  {K.}~\bibnamefont {Kechedzhi}}, \bibinfo {author} {\bibfnamefont
  {J.}~\bibnamefont {Kelly}}, \bibinfo {author} {\bibfnamefont {P.~V.}\
  \bibnamefont {Klimov}}, \bibinfo {author} {\bibfnamefont {S.}~\bibnamefont
  {Knysh}}, \bibinfo {author} {\bibfnamefont {A.}~\bibnamefont {Korotkov}},
  \bibinfo {author} {\bibfnamefont {F.}~\bibnamefont {Kostritsa}}, \bibinfo
  {author} {\bibfnamefont {D.}~\bibnamefont {Landhuis}}, \bibinfo {author}
  {\bibfnamefont {M.}~\bibnamefont {Lindmark}}, \bibinfo {author}
  {\bibfnamefont {E.}~\bibnamefont {Lucero}}, \bibinfo {author} {\bibfnamefont
  {D.}~\bibnamefont {Lyakh}}, \bibinfo {author} {\bibfnamefont
  {S.}~\bibnamefont {Mandr{\`a}}}, \bibinfo {author} {\bibfnamefont {J.~R.}\
  \bibnamefont {McClean}}, \bibinfo {author} {\bibfnamefont {M.}~\bibnamefont
  {McEwen}}, \bibinfo {author} {\bibfnamefont {A.}~\bibnamefont {Megrant}},
  \bibinfo {author} {\bibfnamefont {X.}~\bibnamefont {Mi}}, \bibinfo {author}
  {\bibfnamefont {K.}~\bibnamefont {Michielsen}}, \bibinfo {author}
  {\bibfnamefont {M.}~\bibnamefont {Mohseni}}, \bibinfo {author} {\bibfnamefont
  {J.}~\bibnamefont {Mutus}}, \bibinfo {author} {\bibfnamefont
  {O.}~\bibnamefont {Naaman}}, \bibinfo {author} {\bibfnamefont
  {M.}~\bibnamefont {Neeley}}, \bibinfo {author} {\bibfnamefont
  {C.}~\bibnamefont {Neill}}, \bibinfo {author} {\bibfnamefont {M.~Y.}\
  \bibnamefont {Niu}}, \bibinfo {author} {\bibfnamefont {E.}~\bibnamefont
  {Ostby}}, \bibinfo {author} {\bibfnamefont {A.}~\bibnamefont {Petukhov}},
  \bibinfo {author} {\bibfnamefont {J.~C.}\ \bibnamefont {Platt}}, \bibinfo
  {author} {\bibfnamefont {C.}~\bibnamefont {Quintana}}, \bibinfo {author}
  {\bibfnamefont {E.~G.}\ \bibnamefont {Rieffel}}, \bibinfo {author}
  {\bibfnamefont {P.}~\bibnamefont {Roushan}}, \bibinfo {author} {\bibfnamefont
  {N.~C.}\ \bibnamefont {Rubin}}, \bibinfo {author} {\bibfnamefont
  {D.}~\bibnamefont {Sank}}, \bibinfo {author} {\bibfnamefont {K.~J.}\
  \bibnamefont {Satzinger}}, \bibinfo {author} {\bibfnamefont {V.}~\bibnamefont
  {Smelyanskiy}}, \bibinfo {author} {\bibfnamefont {K.~J.}\ \bibnamefont
  {Sung}}, \bibinfo {author} {\bibfnamefont {M.~D.}\ \bibnamefont
  {Trevithick}}, \bibinfo {author} {\bibfnamefont {A.}~\bibnamefont
  {Vainsencher}}, \bibinfo {author} {\bibfnamefont {B.}~\bibnamefont
  {Villalonga}}, \bibinfo {author} {\bibfnamefont {T.}~\bibnamefont {White}},
  \bibinfo {author} {\bibfnamefont {Z.~J.}\ \bibnamefont {Yao}}, \bibinfo
  {author} {\bibfnamefont {P.}~\bibnamefont {Yeh}}, \bibinfo {author}
  {\bibfnamefont {A.}~\bibnamefont {Zalcman}}, \bibinfo {author} {\bibfnamefont
  {H.}~\bibnamefont {Neven}}, \ and\ \bibinfo {author} {\bibfnamefont {J.~M.}\
  \bibnamefont {Martinis}},\ }\href {\doibase 10.1038/s41586-019-1666-5}
  {\bibfield  {journal} {\bibinfo  {journal} {Nature}\ }\textbf {\bibinfo
  {volume} {574}},\ \bibinfo {pages} {505} (\bibinfo {year}
  {2019})}\BibitemShut {NoStop}%
\bibitem [{\citenamefont {Banerjee}\ \emph {et~al.}(2012)\citenamefont
  {Banerjee}, \citenamefont {Dalmonte}, \citenamefont {M{\"u}ller},
  \citenamefont {Rico}, \citenamefont {Stebler}, \citenamefont {Wiese},\ and\
  \citenamefont {Zoller}}]{banerjee.dalmonte.eaAtomicQuantum12}%
  \BibitemOpen
  \bibfield  {author} {\bibinfo {author} {\bibfnamefont {D.}~\bibnamefont
  {Banerjee}}, \bibinfo {author} {\bibfnamefont {M.}~\bibnamefont {Dalmonte}},
  \bibinfo {author} {\bibfnamefont {M.}~\bibnamefont {M{\"u}ller}}, \bibinfo
  {author} {\bibfnamefont {E.}~\bibnamefont {Rico}}, \bibinfo {author}
  {\bibfnamefont {P.}~\bibnamefont {Stebler}}, \bibinfo {author} {\bibfnamefont
  {U.-J.}\ \bibnamefont {Wiese}}, \ and\ \bibinfo {author} {\bibfnamefont
  {P.}~\bibnamefont {Zoller}},\ }\href {\doibase
  10.1103/PhysRevLett.109.175302} {\bibfield  {journal} {\bibinfo  {journal}
  {Phys. Rev. Lett.}\ }\textbf {\bibinfo {volume} {109}},\ \bibinfo {pages}
  {175302} (\bibinfo {year} {2012})}\BibitemShut {NoStop}%
\bibitem [{\citenamefont {Ba{\~n}uls}\ and\ \citenamefont
  {Cichy}(2020)}]{banuls.cichyReviewNovel20}%
  \BibitemOpen
  \bibfield  {author} {\bibinfo {author} {\bibfnamefont {M.~C.}\ \bibnamefont
  {Ba{\~n}uls}}\ and\ \bibinfo {author} {\bibfnamefont {K.}~\bibnamefont
  {Cichy}},\ }\href {\doibase 10.1088/1361-6633/ab6311} {\bibfield  {journal}
  {\bibinfo  {journal} {Rep. Prog. Phys.}\ }\textbf {\bibinfo {volume} {83}},\
  \bibinfo {pages} {024401} (\bibinfo {year} {2020})}\BibitemShut {NoStop}%
\bibitem [{\citenamefont {Ba{\~n}uls}\ \emph {et~al.}(2019)\citenamefont
  {Ba{\~n}uls}, \citenamefont {Blatt}, \citenamefont {Catani}, \citenamefont
  {Celi}, \citenamefont {Cirac}, \citenamefont {Dalmonte}, \citenamefont
  {Fallani}, \citenamefont {Jansen}, \citenamefont {Lewenstein}, \citenamefont
  {Montangero}, \citenamefont {Muschik}, \citenamefont {Reznik}, \citenamefont
  {Rico}, \citenamefont {Tagliacozzo}, \citenamefont {Van~Acoleyen},
  \citenamefont {Verstraete}, \citenamefont {Wiese}, \citenamefont {Wingate},
  \citenamefont {Zakrzewski},\ and\ \citenamefont
  {Zoller}}]{banuls.blatt.eaSimulatingLattice19}%
  \BibitemOpen
  \bibfield  {author} {\bibinfo {author} {\bibfnamefont {M.~C.}\ \bibnamefont
  {Ba{\~n}uls}}, \bibinfo {author} {\bibfnamefont {R.}~\bibnamefont {Blatt}},
  \bibinfo {author} {\bibfnamefont {J.}~\bibnamefont {Catani}}, \bibinfo
  {author} {\bibfnamefont {A.}~\bibnamefont {Celi}}, \bibinfo {author}
  {\bibfnamefont {J.~I.}\ \bibnamefont {Cirac}}, \bibinfo {author}
  {\bibfnamefont {M.}~\bibnamefont {Dalmonte}}, \bibinfo {author}
  {\bibfnamefont {L.}~\bibnamefont {Fallani}}, \bibinfo {author} {\bibfnamefont
  {K.}~\bibnamefont {Jansen}}, \bibinfo {author} {\bibfnamefont
  {M.}~\bibnamefont {Lewenstein}}, \bibinfo {author} {\bibfnamefont
  {S.}~\bibnamefont {Montangero}}, \bibinfo {author} {\bibfnamefont {C.~A.}\
  \bibnamefont {Muschik}}, \bibinfo {author} {\bibfnamefont {B.}~\bibnamefont
  {Reznik}}, \bibinfo {author} {\bibfnamefont {E.}~\bibnamefont {Rico}},
  \bibinfo {author} {\bibfnamefont {L.}~\bibnamefont {Tagliacozzo}}, \bibinfo
  {author} {\bibfnamefont {K.}~\bibnamefont {Van~Acoleyen}}, \bibinfo {author}
  {\bibfnamefont {F.}~\bibnamefont {Verstraete}}, \bibinfo {author}
  {\bibfnamefont {U.-J.}\ \bibnamefont {Wiese}}, \bibinfo {author}
  {\bibfnamefont {M.}~\bibnamefont {Wingate}}, \bibinfo {author} {\bibfnamefont
  {J.}~\bibnamefont {Zakrzewski}}, \ and\ \bibinfo {author} {\bibfnamefont
  {P.}~\bibnamefont {Zoller}},\ }\href@noop {} {\bibfield  {journal} {\bibinfo
  {journal} {arXiv:1911.00003 [cond-mat, physics:hep-lat, physics:hep-th,
  physics:quant-ph]}\ } (\bibinfo {year} {2019})},\ \Eprint
  {http://arxiv.org/abs/1911.00003} {arXiv:1911.00003 [cond-mat,
  physics:hep-lat, physics:hep-th, physics:quant-ph]} \BibitemShut {NoStop}%
\bibitem [{\citenamefont {Zohar}\ \emph {et~al.}(2017)\citenamefont {Zohar},
  \citenamefont {Farace}, \citenamefont {Reznik},\ and\ \citenamefont
  {Cirac}}]{zohar.farace.eaDigitalQuantum17}%
  \BibitemOpen
  \bibfield  {author} {\bibinfo {author} {\bibfnamefont {E.}~\bibnamefont
  {Zohar}}, \bibinfo {author} {\bibfnamefont {A.}~\bibnamefont {Farace}},
  \bibinfo {author} {\bibfnamefont {B.}~\bibnamefont {Reznik}}, \ and\ \bibinfo
  {author} {\bibfnamefont {J.~I.}\ \bibnamefont {Cirac}},\ }\href {\doibase
  10.1103/PhysRevLett.118.070501} {\bibfield  {journal} {\bibinfo  {journal}
  {Phys. Rev. Lett.}\ }\textbf {\bibinfo {volume} {118}},\ \bibinfo {pages}
  {070501} (\bibinfo {year} {2017})}\BibitemShut {NoStop}%
\bibitem [{\citenamefont {Schweizer}\ \emph {et~al.}(2019)\citenamefont
  {Schweizer}, \citenamefont {Grusdt}, \citenamefont {Berngruber},
  \citenamefont {Barbiero}, \citenamefont {Demler}, \citenamefont {Goldman},
  \citenamefont {Bloch},\ and\ \citenamefont
  {Aidelsburger}}]{schweizer.grusdt.eaFloquetApproach19}%
  \BibitemOpen
  \bibfield  {author} {\bibinfo {author} {\bibfnamefont {C.}~\bibnamefont
  {Schweizer}}, \bibinfo {author} {\bibfnamefont {F.}~\bibnamefont {Grusdt}},
  \bibinfo {author} {\bibfnamefont {M.}~\bibnamefont {Berngruber}}, \bibinfo
  {author} {\bibfnamefont {L.}~\bibnamefont {Barbiero}}, \bibinfo {author}
  {\bibfnamefont {E.}~\bibnamefont {Demler}}, \bibinfo {author} {\bibfnamefont
  {N.}~\bibnamefont {Goldman}}, \bibinfo {author} {\bibfnamefont
  {I.}~\bibnamefont {Bloch}}, \ and\ \bibinfo {author} {\bibfnamefont
  {M.}~\bibnamefont {Aidelsburger}},\ }\href {\doibase
  10.1038/s41567-019-0649-7} {\bibfield  {journal} {\bibinfo  {journal} {Nat.
  Phys.}\ }\textbf {\bibinfo {volume} {15}},\ \bibinfo {pages} {1168} (\bibinfo
  {year} {2019})}\BibitemShut {NoStop}%
\bibitem [{\citenamefont {Rico}\ \emph {et~al.}(2014)\citenamefont {Rico},
  \citenamefont {Pichler}, \citenamefont {Dalmonte}, \citenamefont {Zoller},\
  and\ \citenamefont {Montangero}}]{rico.pichler.eaTensorNetworks14}%
  \BibitemOpen
  \bibfield  {author} {\bibinfo {author} {\bibfnamefont {E.}~\bibnamefont
  {Rico}}, \bibinfo {author} {\bibfnamefont {T.}~\bibnamefont {Pichler}},
  \bibinfo {author} {\bibfnamefont {M.}~\bibnamefont {Dalmonte}}, \bibinfo
  {author} {\bibfnamefont {P.}~\bibnamefont {Zoller}}, \ and\ \bibinfo {author}
  {\bibfnamefont {S.}~\bibnamefont {Montangero}},\ }\href {\doibase
  10.1103/PhysRevLett.112.201601} {\bibfield  {journal} {\bibinfo  {journal}
  {Phys. Rev. Lett.}\ }\textbf {\bibinfo {volume} {112}},\ \bibinfo {pages}
  {201601} (\bibinfo {year} {2014})}\BibitemShut {NoStop}%
\bibitem [{\citenamefont {Muschik}\ \emph {et~al.}(2017)\citenamefont
  {Muschik}, \citenamefont {Heyl}, \citenamefont {Martinez}, \citenamefont
  {Monz}, \citenamefont {Schindler}, \citenamefont {Vogell}, \citenamefont
  {Dalmonte}, \citenamefont {Hauke}, \citenamefont {Blatt},\ and\ \citenamefont
  {Zoller}}]{muschik.heyl.eaWilsonLattice17}%
  \BibitemOpen
  \bibfield  {author} {\bibinfo {author} {\bibfnamefont {C.}~\bibnamefont
  {Muschik}}, \bibinfo {author} {\bibfnamefont {M.}~\bibnamefont {Heyl}},
  \bibinfo {author} {\bibfnamefont {E.}~\bibnamefont {Martinez}}, \bibinfo
  {author} {\bibfnamefont {T.}~\bibnamefont {Monz}}, \bibinfo {author}
  {\bibfnamefont {P.}~\bibnamefont {Schindler}}, \bibinfo {author}
  {\bibfnamefont {B.}~\bibnamefont {Vogell}}, \bibinfo {author} {\bibfnamefont
  {M.}~\bibnamefont {Dalmonte}}, \bibinfo {author} {\bibfnamefont
  {P.}~\bibnamefont {Hauke}}, \bibinfo {author} {\bibfnamefont
  {R.}~\bibnamefont {Blatt}}, \ and\ \bibinfo {author} {\bibfnamefont
  {P.}~\bibnamefont {Zoller}},\ }\href {\doibase 10.1088/1367-2630/aa89ab}
  {\bibfield  {journal} {\bibinfo  {journal} {New J. Phys.}\ }\textbf {\bibinfo
  {volume} {19}},\ \bibinfo {pages} {103020} (\bibinfo {year}
  {2017})}\BibitemShut {NoStop}%
\bibitem [{\citenamefont {Klco}\ \emph {et~al.}(2018)\citenamefont {Klco},
  \citenamefont {Dumitrescu}, \citenamefont {McCaskey}, \citenamefont {Morris},
  \citenamefont {Pooser}, \citenamefont {Sanz}, \citenamefont {Solano},
  \citenamefont {Lougovski},\ and\ \citenamefont
  {Savage}}]{klco.dumitrescu.eaQuantumclassicalComputation18}%
  \BibitemOpen
  \bibfield  {author} {\bibinfo {author} {\bibfnamefont {N.}~\bibnamefont
  {Klco}}, \bibinfo {author} {\bibfnamefont {E.~F.}\ \bibnamefont
  {Dumitrescu}}, \bibinfo {author} {\bibfnamefont {A.~J.}\ \bibnamefont
  {McCaskey}}, \bibinfo {author} {\bibfnamefont {T.~D.}\ \bibnamefont
  {Morris}}, \bibinfo {author} {\bibfnamefont {R.~C.}\ \bibnamefont {Pooser}},
  \bibinfo {author} {\bibfnamefont {M.}~\bibnamefont {Sanz}}, \bibinfo {author}
  {\bibfnamefont {E.}~\bibnamefont {Solano}}, \bibinfo {author} {\bibfnamefont
  {P.}~\bibnamefont {Lougovski}}, \ and\ \bibinfo {author} {\bibfnamefont
  {M.~J.}\ \bibnamefont {Savage}},\ }\href {\doibase
  10.1103/PhysRevA.98.032331} {\bibfield  {journal} {\bibinfo  {journal} {Phys.
  Rev. A}\ }\textbf {\bibinfo {volume} {98}},\ \bibinfo {pages} {032331}
  (\bibinfo {year} {2018})}\BibitemShut {NoStop}%
\bibitem [{\citenamefont {Davoudi}\ \emph {et~al.}(2020)\citenamefont
  {Davoudi}, \citenamefont {Hafezi}, \citenamefont {Monroe}, \citenamefont
  {Pagano}, \citenamefont {Seif},\ and\ \citenamefont
  {Shaw}}]{davoudi.hafezi.eaAnalogQuantum20}%
  \BibitemOpen
  \bibfield  {author} {\bibinfo {author} {\bibfnamefont {Z.}~\bibnamefont
  {Davoudi}}, \bibinfo {author} {\bibfnamefont {M.}~\bibnamefont {Hafezi}},
  \bibinfo {author} {\bibfnamefont {C.}~\bibnamefont {Monroe}}, \bibinfo
  {author} {\bibfnamefont {G.}~\bibnamefont {Pagano}}, \bibinfo {author}
  {\bibfnamefont {A.}~\bibnamefont {Seif}}, \ and\ \bibinfo {author}
  {\bibfnamefont {A.}~\bibnamefont {Shaw}},\ }\href {\doibase
  10.1103/PhysRevResearch.2.023015} {\bibfield  {journal} {\bibinfo  {journal}
  {Phys. Rev. Research}\ }\textbf {\bibinfo {volume} {2}},\ \bibinfo {pages}
  {023015} (\bibinfo {year} {2020})}\BibitemShut {NoStop}%
\bibitem [{\citenamefont {Martinez}\ \emph {et~al.}(2016)\citenamefont
  {Martinez}, \citenamefont {Muschik}, \citenamefont {Schindler}, \citenamefont
  {Nigg}, \citenamefont {Erhard}, \citenamefont {Heyl}, \citenamefont {Hauke},
  \citenamefont {Dalmonte}, \citenamefont {Monz}, \citenamefont {Zoller},\ and\
  \citenamefont {Blatt}}]{martinez.muschik.eaRealtimeDynamics16}%
  \BibitemOpen
  \bibfield  {author} {\bibinfo {author} {\bibfnamefont {E.~A.}\ \bibnamefont
  {Martinez}}, \bibinfo {author} {\bibfnamefont {C.~A.}\ \bibnamefont
  {Muschik}}, \bibinfo {author} {\bibfnamefont {P.}~\bibnamefont {Schindler}},
  \bibinfo {author} {\bibfnamefont {D.}~\bibnamefont {Nigg}}, \bibinfo {author}
  {\bibfnamefont {A.}~\bibnamefont {Erhard}}, \bibinfo {author} {\bibfnamefont
  {M.}~\bibnamefont {Heyl}}, \bibinfo {author} {\bibfnamefont {P.}~\bibnamefont
  {Hauke}}, \bibinfo {author} {\bibfnamefont {M.}~\bibnamefont {Dalmonte}},
  \bibinfo {author} {\bibfnamefont {T.}~\bibnamefont {Monz}}, \bibinfo {author}
  {\bibfnamefont {P.}~\bibnamefont {Zoller}}, \ and\ \bibinfo {author}
  {\bibfnamefont {R.}~\bibnamefont {Blatt}},\ }\href {\doibase
  10.1038/nature18318} {\bibfield  {journal} {\bibinfo  {journal} {Nature}\
  }\textbf {\bibinfo {volume} {534}},\ \bibinfo {pages} {516} (\bibinfo {year}
  {2016})}\BibitemShut {NoStop}%
\bibitem [{\citenamefont {Byrnes}\ and\ \citenamefont
  {Yamamoto}(2006)}]{byrnes.yamamotoSimulatingLattice06}%
  \BibitemOpen
  \bibfield  {author} {\bibinfo {author} {\bibfnamefont {T.}~\bibnamefont
  {Byrnes}}\ and\ \bibinfo {author} {\bibfnamefont {Y.}~\bibnamefont
  {Yamamoto}},\ }\href {\doibase 10.1103/PhysRevA.73.022328} {\bibfield
  {journal} {\bibinfo  {journal} {Phys. Rev. A}\ }\textbf {\bibinfo {volume}
  {73}},\ \bibinfo {pages} {022328} (\bibinfo {year} {2006})}\BibitemShut
  {NoStop}%
\bibitem [{\citenamefont {Zohar}\ \emph {et~al.}(2013)\citenamefont {Zohar},
  \citenamefont {Cirac},\ and\ \citenamefont
  {Reznik}}]{zohar.cirac.eaColdAtomQuantum13}%
  \BibitemOpen
  \bibfield  {author} {\bibinfo {author} {\bibfnamefont {E.}~\bibnamefont
  {Zohar}}, \bibinfo {author} {\bibfnamefont {J.~I.}\ \bibnamefont {Cirac}}, \
  and\ \bibinfo {author} {\bibfnamefont {B.}~\bibnamefont {Reznik}},\ }\href
  {\doibase 10.1103/PhysRevLett.110.125304} {\bibfield  {journal} {\bibinfo
  {journal} {Phys. Rev. Lett.}\ }\textbf {\bibinfo {volume} {110}},\ \bibinfo
  {pages} {125304} (\bibinfo {year} {2013})}\BibitemShut {NoStop}%
\bibitem [{\citenamefont {Tagliacozzo}\ \emph {et~al.}(2013)\citenamefont
  {Tagliacozzo}, \citenamefont {Celi}, \citenamefont {Orland}, \citenamefont
  {Mitchell},\ and\ \citenamefont
  {Lewenstein}}]{tagliacozzo.celi.eaSimulationNonAbelian13}%
  \BibitemOpen
  \bibfield  {author} {\bibinfo {author} {\bibfnamefont {L.}~\bibnamefont
  {Tagliacozzo}}, \bibinfo {author} {\bibfnamefont {A.}~\bibnamefont {Celi}},
  \bibinfo {author} {\bibfnamefont {P.}~\bibnamefont {Orland}}, \bibinfo
  {author} {\bibfnamefont {M.~W.}\ \bibnamefont {Mitchell}}, \ and\ \bibinfo
  {author} {\bibfnamefont {M.}~\bibnamefont {Lewenstein}},\ }\href {\doibase
  10.1038/ncomms3615} {\bibfield  {journal} {\bibinfo  {journal} {Nat.
  Commun.}\ }\textbf {\bibinfo {volume} {4}},\ \bibinfo {pages} {2615}
  (\bibinfo {year} {2013})}\BibitemShut {NoStop}%
\bibitem [{\citenamefont {Banerjee}\ \emph {et~al.}(2013)\citenamefont
  {Banerjee}, \citenamefont {B{\"o}gli}, \citenamefont {Dalmonte},
  \citenamefont {Rico}, \citenamefont {Stebler}, \citenamefont {Wiese},\ and\
  \citenamefont {Zoller}}]{banerjee.bogli.eaAtomicQuantum13}%
  \BibitemOpen
  \bibfield  {author} {\bibinfo {author} {\bibfnamefont {D.}~\bibnamefont
  {Banerjee}}, \bibinfo {author} {\bibfnamefont {M.}~\bibnamefont {B{\"o}gli}},
  \bibinfo {author} {\bibfnamefont {M.}~\bibnamefont {Dalmonte}}, \bibinfo
  {author} {\bibfnamefont {E.}~\bibnamefont {Rico}}, \bibinfo {author}
  {\bibfnamefont {P.}~\bibnamefont {Stebler}}, \bibinfo {author} {\bibfnamefont
  {U.-J.}\ \bibnamefont {Wiese}}, \ and\ \bibinfo {author} {\bibfnamefont
  {P.}~\bibnamefont {Zoller}},\ }\href {\doibase
  10.1103/PhysRevLett.110.125303} {\bibfield  {journal} {\bibinfo  {journal}
  {Phys. Rev. Lett.}\ }\textbf {\bibinfo {volume} {110}},\ \bibinfo {pages}
  {125303} (\bibinfo {year} {2013})}\BibitemShut {NoStop}%
\bibitem [{\citenamefont {Wiese}(2013)}]{wieseUltracoldQuantum13}%
  \BibitemOpen
  \bibfield  {author} {\bibinfo {author} {\bibfnamefont {U.-J.}\ \bibnamefont
  {Wiese}},\ }\href {\doibase 10.1002/andp.201300104} {\bibfield  {journal}
  {\bibinfo  {journal} {Ann. Phys.}\ }\textbf {\bibinfo {volume} {525}},\
  \bibinfo {pages} {777} (\bibinfo {year} {2013})}\BibitemShut {NoStop}%
\bibitem [{\citenamefont {Wiese}(2014)}]{wieseQuantumSimulating14}%
  \BibitemOpen
  \bibfield  {author} {\bibinfo {author} {\bibfnamefont {U.-J.}\ \bibnamefont
  {Wiese}},\ }\href {\doibase 10.1016/j.nuclphysa.2014.09.102} {\bibfield
  {journal} {\bibinfo  {journal} {Nucl. Phys. A}\ }\bibinfo {series} {Quark
  {{Matter}} 2014},\ \textbf {\bibinfo {volume} {931}},\ \bibinfo {pages} {246}
  (\bibinfo {year} {2014})}\BibitemShut {NoStop}%
\bibitem [{\citenamefont {Mezzacapo}\ \emph {et~al.}(2015)\citenamefont
  {Mezzacapo}, \citenamefont {Rico}, \citenamefont {Sab{\'i}n}, \citenamefont
  {Egusquiza}, \citenamefont {Lamata},\ and\ \citenamefont
  {Solano}}]{mezzacapo.rico.eaNonAbelianSU15}%
  \BibitemOpen
  \bibfield  {author} {\bibinfo {author} {\bibfnamefont {A.}~\bibnamefont
  {Mezzacapo}}, \bibinfo {author} {\bibfnamefont {E.}~\bibnamefont {Rico}},
  \bibinfo {author} {\bibfnamefont {C.}~\bibnamefont {Sab{\'i}n}}, \bibinfo
  {author} {\bibfnamefont {I.~L.}\ \bibnamefont {Egusquiza}}, \bibinfo {author}
  {\bibfnamefont {L.}~\bibnamefont {Lamata}}, \ and\ \bibinfo {author}
  {\bibfnamefont {E.}~\bibnamefont {Solano}},\ }\href {\doibase
  10.1103/PhysRevLett.115.240502} {\bibfield  {journal} {\bibinfo  {journal}
  {Phys. Rev. Lett.}\ }\textbf {\bibinfo {volume} {115}},\ \bibinfo {pages}
  {240502} (\bibinfo {year} {2015})}\BibitemShut {NoStop}%
\bibitem [{\citenamefont {{NuQS Collaboration}}\ \emph
  {et~al.}(2019{\natexlab{a}})\citenamefont {{NuQS Collaboration}},
  \citenamefont {Lamm}, \citenamefont {Lawrence},\ and\ \citenamefont
  {Yamauchi}}]{nuqscollaboration.lamm.eaGeneralMethods19}%
  \BibitemOpen
  \bibfield  {author} {\bibinfo {author} {\bibnamefont {{NuQS Collaboration}}},
  \bibinfo {author} {\bibfnamefont {H.}~\bibnamefont {Lamm}}, \bibinfo {author}
  {\bibfnamefont {S.}~\bibnamefont {Lawrence}}, \ and\ \bibinfo {author}
  {\bibfnamefont {Y.}~\bibnamefont {Yamauchi}},\ }\href {\doibase
  10.1103/PhysRevD.100.034518} {\bibfield  {journal} {\bibinfo  {journal}
  {Phys. Rev. D}\ }\textbf {\bibinfo {volume} {100}},\ \bibinfo {pages}
  {034518} (\bibinfo {year} {2019}{\natexlab{a}})}\BibitemShut {NoStop}%
\bibitem [{\citenamefont {{NuQS Collaboration}}\ \emph
  {et~al.}(2019{\natexlab{b}})\citenamefont {{NuQS Collaboration}},
  \citenamefont {Alexandru}, \citenamefont {Bedaque}, \citenamefont
  {Harmalkar}, \citenamefont {Lamm}, \citenamefont {Lawrence},\ and\
  \citenamefont {Warrington}}]{nuqscollaboration.alexandru.eaGluonField19}%
  \BibitemOpen
  \bibfield  {author} {\bibinfo {author} {\bibnamefont {{NuQS Collaboration}}},
  \bibinfo {author} {\bibfnamefont {A.}~\bibnamefont {Alexandru}}, \bibinfo
  {author} {\bibfnamefont {P.~F.}\ \bibnamefont {Bedaque}}, \bibinfo {author}
  {\bibfnamefont {S.}~\bibnamefont {Harmalkar}}, \bibinfo {author}
  {\bibfnamefont {H.}~\bibnamefont {Lamm}}, \bibinfo {author} {\bibfnamefont
  {S.}~\bibnamefont {Lawrence}}, \ and\ \bibinfo {author} {\bibfnamefont
  {N.~C.}\ \bibnamefont {Warrington}},\ }\href {\doibase
  10.1103/PhysRevD.100.114501} {\bibfield  {journal} {\bibinfo  {journal}
  {Phys. Rev. D}\ }\textbf {\bibinfo {volume} {100}},\ \bibinfo {pages}
  {114501} (\bibinfo {year} {2019}{\natexlab{b}})}\BibitemShut {NoStop}%
\bibitem [{\citenamefont {Klco}\ \emph {et~al.}(2020)\citenamefont {Klco},
  \citenamefont {Savage},\ and\ \citenamefont
  {Stryker}}]{klco.savage.eaSUNonAbelian20}%
  \BibitemOpen
  \bibfield  {author} {\bibinfo {author} {\bibfnamefont {N.}~\bibnamefont
  {Klco}}, \bibinfo {author} {\bibfnamefont {M.~J.}\ \bibnamefont {Savage}}, \
  and\ \bibinfo {author} {\bibfnamefont {J.~R.}\ \bibnamefont {Stryker}},\
  }\href {\doibase 10.1103/PhysRevD.101.074512} {\bibfield  {journal} {\bibinfo
   {journal} {Phys. Rev. D}\ }\textbf {\bibinfo {volume} {101}},\ \bibinfo
  {pages} {074512} (\bibinfo {year} {2020})}\BibitemShut {NoStop}%
\bibitem [{\citenamefont {Raychowdhury}\ and\ \citenamefont
  {Stryker}(2020)}]{raychowdhury.strykerSolvingGauss20}%
  \BibitemOpen
  \bibfield  {author} {\bibinfo {author} {\bibfnamefont {I.}~\bibnamefont
  {Raychowdhury}}\ and\ \bibinfo {author} {\bibfnamefont {J.~R.}\ \bibnamefont
  {Stryker}},\ }\href@noop {} {\bibfield  {journal} {\bibinfo  {journal} {to
  appear in Phys. Rev. Res.}\ } (\bibinfo {year} {2020})},\ \Eprint
  {http://arxiv.org/abs/1812.07554} {arXiv:1812.07554} \BibitemShut {NoStop}%
\bibitem [{\citenamefont {Schwinger}(1952)}]{schwingerAngularMomentum52}%
  \BibitemOpen
  \bibfield  {author} {\bibinfo {author} {\bibfnamefont {J.}~\bibnamefont
  {Schwinger}},\ }\href@noop {} {\emph {\bibinfo {title} {On Angular
  Momentum}}},\ \bibinfo {type} {Tech. Rep.}\ \bibinfo {number} {NYO-3071}\
  (\bibinfo  {institution} {{U.S. Atomic Energy Commission}},\ \bibinfo
  {address} {{Oak Ridge, Tennessee}},\ \bibinfo {year} {1952})\BibitemShut
  {NoStop}%
\bibitem [{\citenamefont {Mathur}(2005)}]{mathurHarmonicOscillator05}%
  \BibitemOpen
  \bibfield  {author} {\bibinfo {author} {\bibfnamefont {M.}~\bibnamefont
  {Mathur}},\ }\href {\doibase 10.1088/0305-4470/38/46/008} {\bibfield
  {journal} {\bibinfo  {journal} {J. Phys. A: Math. Gen.}\ }\textbf {\bibinfo
  {volume} {38}},\ \bibinfo {pages} {10015} (\bibinfo {year}
  {2005})}\BibitemShut {NoStop}%
\bibitem [{\citenamefont {Mathur}(2006)}]{mathurLoopStates06}%
  \BibitemOpen
  \bibfield  {author} {\bibinfo {author} {\bibfnamefont {M.}~\bibnamefont
  {Mathur}},\ }\href {\doibase 10.1016/j.physletb.2006.08.022} {\bibfield
  {journal} {\bibinfo  {journal} {Physics Letters B}\ }\textbf {\bibinfo
  {volume} {640}},\ \bibinfo {pages} {292} (\bibinfo {year}
  {2006})}\BibitemShut {NoStop}%
\bibitem [{\citenamefont {Mathur}(2007)}]{mathurLoopApproach07}%
  \BibitemOpen
  \bibfield  {author} {\bibinfo {author} {\bibfnamefont {M.}~\bibnamefont
  {Mathur}},\ }\href {\doibase 10.1016/j.nuclphysb.2007.04.031} {\bibfield
  {journal} {\bibinfo  {journal} {Nucl. Phys. B}\ }\textbf {\bibinfo {volume}
  {779}},\ \bibinfo {pages} {32} (\bibinfo {year} {2007})}\BibitemShut
  {NoStop}%
\bibitem [{\citenamefont {Anishetty}\ \emph {et~al.}(2010)\citenamefont
  {Anishetty}, \citenamefont {Mathur},\ and\ \citenamefont
  {Raychowdhury}}]{anishetty.mathur.eaPrepotentialFormulation10}%
  \BibitemOpen
  \bibfield  {author} {\bibinfo {author} {\bibfnamefont {R.}~\bibnamefont
  {Anishetty}}, \bibinfo {author} {\bibfnamefont {M.}~\bibnamefont {Mathur}}, \
  and\ \bibinfo {author} {\bibfnamefont {I.}~\bibnamefont {Raychowdhury}},\
  }\href {\doibase 10.1088/1751-8113/43/3/035403} {\bibfield  {journal}
  {\bibinfo  {journal} {J. Phys. A: Math. Theor.}\ }\textbf {\bibinfo {volume}
  {43}},\ \bibinfo {pages} {035403} (\bibinfo {year} {2010})}\BibitemShut
  {NoStop}%
\bibitem [{\citenamefont {Mathur}\ \emph {et~al.}(2010)\citenamefont {Mathur},
  \citenamefont {Raychowdhury},\ and\ \citenamefont
  {Anishetty}}]{mathur.raychowdhury.eaSUIrreducible10}%
  \BibitemOpen
  \bibfield  {author} {\bibinfo {author} {\bibfnamefont {M.}~\bibnamefont
  {Mathur}}, \bibinfo {author} {\bibfnamefont {I.}~\bibnamefont
  {Raychowdhury}}, \ and\ \bibinfo {author} {\bibfnamefont {R.}~\bibnamefont
  {Anishetty}},\ }\href {\doibase 10.1063/1.3464267} {\bibfield  {journal}
  {\bibinfo  {journal} {Journal of Mathematical Physics}\ }\textbf {\bibinfo
  {volume} {51}},\ \bibinfo {pages} {093504} (\bibinfo {year}
  {2010})}\BibitemShut {NoStop}%
\bibitem [{\citenamefont
  {Raychowdhury}(2013)}]{raychowdhuryPrepotentialFormulation13}%
  \BibitemOpen
  \bibfield  {author} {\bibinfo {author} {\bibfnamefont {I.}~\bibnamefont
  {Raychowdhury}},\ }\emph {\bibinfo {title} {Prepotential {{Formulation}} of
  {{Lattice Gauge Theories}}}},\ \href@noop {} {Ph.D. thesis},\ \bibinfo
  {school} {Calcutta U.} (\bibinfo {year} {2013})\BibitemShut {NoStop}%
\bibitem [{\citenamefont {Anishetty}\ and\ \citenamefont
  {Raychowdhury}(2014)}]{anishetty.raychowdhurySULattice14}%
  \BibitemOpen
  \bibfield  {author} {\bibinfo {author} {\bibfnamefont {R.}~\bibnamefont
  {Anishetty}}\ and\ \bibinfo {author} {\bibfnamefont {I.}~\bibnamefont
  {Raychowdhury}},\ }\href {\doibase 10.1103/PhysRevD.90.114503} {\bibfield
  {journal} {\bibinfo  {journal} {Phys. Rev. D}\ }\textbf {\bibinfo {volume}
  {90}},\ \bibinfo {pages} {114503} (\bibinfo {year} {2014})}\BibitemShut
  {NoStop}%
\bibitem [{\citenamefont {Anishetty}\ and\ \citenamefont
  {Sreeraj}(2018)}]{anishetty.sreerajMassGap18}%
  \BibitemOpen
  \bibfield  {author} {\bibinfo {author} {\bibfnamefont {R.}~\bibnamefont
  {Anishetty}}\ and\ \bibinfo {author} {\bibfnamefont {T.~P.}\ \bibnamefont
  {Sreeraj}},\ }\href {\doibase 10.1103/PhysRevD.97.074511} {\bibfield
  {journal} {\bibinfo  {journal} {Phys. Rev. D}\ }\textbf {\bibinfo {volume}
  {97}},\ \bibinfo {pages} {074511} (\bibinfo {year} {2018})}\BibitemShut
  {NoStop}%
\bibitem [{\citenamefont
  {Sharatchandra}(1982)}]{sharatchandraLocalObservables82}%
  \BibitemOpen
  \bibfield  {author} {\bibinfo {author} {\bibfnamefont {H.~S.}\ \bibnamefont
  {Sharatchandra}},\ }\href {\doibase 10.1016/0550-3213(82)90302-9} {\bibfield
  {journal} {\bibinfo  {journal} {Nuclear Physics B}\ }\textbf {\bibinfo
  {volume} {196}},\ \bibinfo {pages} {62} (\bibinfo {year} {1982})}\BibitemShut
  {NoStop}%
\bibitem [{\citenamefont
  {Mandelstam}(1979)}]{mandelstamChargemonopoleDuality79}%
  \BibitemOpen
  \bibfield  {author} {\bibinfo {author} {\bibfnamefont {S.}~\bibnamefont
  {Mandelstam}},\ }\href {\doibase 10.1103/PhysRevD.19.2391} {\bibfield
  {journal} {\bibinfo  {journal} {Phys. Rev. D}\ }\textbf {\bibinfo {volume}
  {19}},\ \bibinfo {pages} {2391} (\bibinfo {year} {1979})}\BibitemShut
  {NoStop}%
\bibitem [{\citenamefont {Loll}(1993)}]{lollYangMillsTheory93}%
  \BibitemOpen
  \bibfield  {author} {\bibinfo {author} {\bibfnamefont {R.}~\bibnamefont
  {Loll}},\ }\href {\doibase 10.1016/0550-3213(93)90400-J} {\bibfield
  {journal} {\bibinfo  {journal} {Nucl. Phys. B}\ }\textbf {\bibinfo {volume}
  {400}},\ \bibinfo {pages} {126} (\bibinfo {year} {1993})}\BibitemShut
  {NoStop}%
\bibitem [{\citenamefont {Watson}(1994)}]{watsonSolutionSU94}%
  \BibitemOpen
  \bibfield  {author} {\bibinfo {author} {\bibfnamefont {N.~J.}\ \bibnamefont
  {Watson}},\ }\href {\doibase 10.1016/0370-2693(94)91236-X} {\bibfield
  {journal} {\bibinfo  {journal} {Phys. Lett. B}\ }\textbf {\bibinfo {volume}
  {323}},\ \bibinfo {pages} {385} (\bibinfo {year} {1994})}\BibitemShut
  {NoStop}%
\bibitem [{\citenamefont {Zohar}\ and\ \citenamefont
  {Burrello}(2015)}]{zohar.burrelloFormulationLattice15}%
  \BibitemOpen
  \bibfield  {author} {\bibinfo {author} {\bibfnamefont {E.}~\bibnamefont
  {Zohar}}\ and\ \bibinfo {author} {\bibfnamefont {M.}~\bibnamefont
  {Burrello}},\ }\href {\doibase 10.1103/PhysRevD.91.054506} {\bibfield
  {journal} {\bibinfo  {journal} {Phys. Rev. D}\ }\textbf {\bibinfo {volume}
  {91}},\ \bibinfo {pages} {054506} (\bibinfo {year} {2015})}\BibitemShut
  {NoStop}%
\bibitem [{\citenamefont {Anishetty}\ \emph {et~al.}(2009)\citenamefont
  {Anishetty}, \citenamefont {Mathur},\ and\ \citenamefont
  {Raychowdhury}}]{anishetty.mathur.eaIrreducibleSU09}%
  \BibitemOpen
  \bibfield  {author} {\bibinfo {author} {\bibfnamefont {R.}~\bibnamefont
  {Anishetty}}, \bibinfo {author} {\bibfnamefont {M.}~\bibnamefont {Mathur}}, \
  and\ \bibinfo {author} {\bibfnamefont {I.}~\bibnamefont {Raychowdhury}},\
  }\href {\doibase 10.1063/1.3122666} {\bibfield  {journal} {\bibinfo
  {journal} {Journal of Mathematical Physics}\ }\textbf {\bibinfo {volume}
  {50}},\ \bibinfo {pages} {053503} (\bibinfo {year} {2009})}\BibitemShut
  {NoStop}%
\bibitem [{\citenamefont {Raychowdhury}(2019)}]{raychowdhuryLowEnergy19}%
  \BibitemOpen
  \bibfield  {author} {\bibinfo {author} {\bibfnamefont {I.}~\bibnamefont
  {Raychowdhury}},\ }\href {\doibase 10.1140/epjc/s10052-019-6753-0} {\bibfield
   {journal} {\bibinfo  {journal} {Eur. Phys. J. C}\ }\textbf {\bibinfo
  {volume} {79}},\ \bibinfo {pages} {235} (\bibinfo {year} {2019})}\BibitemShut
  {NoStop}%
\bibitem [{\citenamefont {Trotter}(1959)}]{trotterProductSemigroups59}%
  \BibitemOpen
  \bibfield  {author} {\bibinfo {author} {\bibfnamefont {H.~F.}\ \bibnamefont
  {Trotter}},\ }\href {\doibase 10.1090/S0002-9939-1959-0108732-6} {\bibfield
  {journal} {\bibinfo  {journal} {Proc. Amer. Math. Soc.}\ }\textbf {\bibinfo
  {volume} {10}},\ \bibinfo {pages} {545} (\bibinfo {year} {1959})}\BibitemShut
  {NoStop}%
\bibitem [{\citenamefont {Suzuki}(1976)}]{suzukiGeneralizedTrotter76}%
  \BibitemOpen
  \bibfield  {author} {\bibinfo {author} {\bibfnamefont {M.}~\bibnamefont
  {Suzuki}},\ }\href {\doibase 10.1007/BF01609348} {\bibfield  {journal}
  {\bibinfo  {journal} {Commun.Math. Phys.}\ }\textbf {\bibinfo {volume}
  {51}},\ \bibinfo {pages} {183} (\bibinfo {year} {1976})}\BibitemShut
  {NoStop}%
\end{thebibliography}%

\end{document}